\documentclass[11pt]{article}
\pdfoutput=1 
\usepackage[utf8]{inputenc}
\usepackage{jheppub}
\usepackage{graphicx}
\usepackage{caption}
\usepackage{subcaption}
\usepackage{hyperref}
\usepackage[T1]{fontenc}

\newcommand{\be}{\begin{equation}}
\newcommand{\ee}{\end{equation}}
\newcommand{\bea}{\begin{eqnarray}}
\newcommand{\eea}{\end{eqnarray}}

\newcommand{\nn}{\nonumber\\}
\def\CB{\mathcal{B}}
\def\CC{\mathcal{C}}

\def\CI{\mathcal{I}}

\def\CN{\mathcal{N}}

\def\CP{\mathcal{P}}

\def\qfr{\mathfrak{q}}
\def\wfr{\mathfrak{w}}
\def\kfr{\ell}
\def\Ofr{\mathfrak{O}}

\def\q{{\bf q}}
\def\x{{\bf x}}

\def\re{\text{Re}}

\title{The complex life of hydrodynamic modes}

\author[a]{Sa\v{s}o Grozdanov,}
\author[b]{Pavel K. Kovtun,}
\author[c]{Andrei O. Starinets}
\author[d]{and Petar Tadi\'{c}}

\affiliation[a]{Center for Theoretical Physics, Massachusetts Institute of Technology, \\Cambridge, MA 02139, USA}

\affiliation[b]{Department of Physics \& Astronomy, University of Victoria, \\ PO Box 1700 STN CSC, Victoria,
BC, V8W 2Y2, Canada}

\affiliation[c]{Rudolf Peierls Centre for Theoretical Physics, University of Oxford, \\ Parks Road,  
Oxford, OX1 3PU, United Kingdom }

\affiliation[d]{School of Mathematics, Trinity College Dublin, Dublin, D02 W272, Ireland}

\emailAdd{saso@mit.edu}

\emailAdd{pkovtun@uvic.ca}

\emailAdd{andrei.starinets@physics.ox.ac.uk}

\emailAdd{tadicp@tcd.ie}

\abstract{
We study analytic properties of the dispersion relations in classical hydrodynamics by treating them as Puiseux series in complex momentum. The radii of convergence of the series are determined by the critical points of the associated complex spectral curves. For theories that admit a dual gravitational description through holography, the critical points correspond to level-crossings in the quasinormal spectrum of the dual black hole. We illustrate these methods in ${\cal N}=4$ supersymmetric Yang-Mills theory in 3+1 dimensions, in a holographic model with broken translation symmetry in 2+1 dimensions, and in conformal field theory in 1+1 dimensions. We comment on the pole-skipping phenomenon in thermal correlation functions, and show that it is not specific to energy density correlations.
}

\preprint{MIT-CTP/5101\\ \hspace*{\fill} OUTP-19-02P}

\begin{document} 
\maketitle
\flushbottom

\section{Introduction}
\label{sec:intro}
Hydrodynamics is an effective theory of fluids  valid at sufficiently long times and sufficiently large distances. Classical hydrodynamics is formulated by combining conservation laws for energy, momentum, and other charges (such as the particle number in non-relativistic systems) with the constitutive relations expressing the conserved fluxes in terms of the hydrodynamic variables: local temperature, fluid velocity, and charge density~\cite{landau}. The constitutive relations are written down on the basis of symmetry, using the derivative expansion. The constitutive relations which contain no derivatives of the hydrodynamic variables are said to describe ``perfect fluids'', corresponding to ``zeroth-order'' hydrodynamics. The constitutive relations which contain up to one derivative of the hydrodynamic variables are commonly said to describe ``viscous fluids'', corresponding to ``first-order'' or ``Navier-Stokes'' hydrodynamics. One can proceed by adding terms with more derivatives of the hydrodynamic variables to the constitutive relations, hoping to improve the hydrodynamic description of  the actual physical fluids. The constitutive relations which contain up to $n$ derivatives of the hydrodynamic variables then give rise to $n^{\rm th}$-order hydrodynamics. In this paper, we will use simple facts from the theory of complex curves in order to study some aspects of  convergence of the above derivative expansion. Our focus will be on relativistic fluids, and we shall illustrate general results with the examples of strongly interacting quantum field theories possessing a dual holographic description. 

A neutral homogeneous and isotropic relativistic fluid whose energy-momentum tensor is conserved supports collective excitations in the form of shear and sound hydrodynamic modes \cite{Kovtun:2012rj}. The collective modes arise from the analysis of linearised fluctuations of energy and momentum densities around the equilibrium state at temperature $T$. The mode's frequency $\omega$ is related to the wave vector $\q$ by the dispersion relation $\omega=\omega(\q)$. In hydrodynamics, the dispersion relations are given by
\begin{align}
\mbox{Shear mode:} & &  &\omega_{\rm shear}(\q) = -i D \q^2 + \cdots\,, 
\label{hydro-modes-shear} \\
\mbox{Sound mode:}& & &\omega_{\rm sound}^{\pm}(\q) = \pm v_s |\q| - i \frac{\Gamma_{\!s}}{2}\, \q^2 + \cdots\,, 
\label{hydro-modes-sound}
\end{align}
where $v_s$ is the speed of sound, and $D$ and $\Gamma_{\!s}$ can be expressed through relevant transport coefficients. In $d_s$ spatial dimensions, we have
\begin{align}
\,& D = \frac{\eta}{\epsilon +p}\,, \label{hydro-modes-coeffi-D}\\
\,& \Gamma_{\!s} = \frac{1}{\epsilon +p}\left[ \zeta + \frac{2d_s-2}{d_s} \, \eta\right],
\label{hydro-modes-coeffi-Gamma}
\end{align}
where $\epsilon$ and $p$ are the equilibrium energy density and pressure, and $\eta$ and $\zeta$ are the shear and bulk viscosities. The shear mode \eqref{hydro-modes-shear} describes  diffusion of the transverse (to the wave vector) velocity fluctuations which are damped by the shear viscosity. The sound mode \eqref{hydro-modes-sound} describes propagation of the longitudinal velocity fluctuations together with the fluctuations in the energy density and pressure. The terms written down in  \eqref{hydro-modes-shear} and \eqref{hydro-modes-sound} represent the contributions from first-order hydrodynamics, while the ellipses  denote terms with higher powers of $\q$, which can be matched to transport coefficients in second- and higher-order hydrodynamics \cite{Baier:2007ix, Bhattacharyya:2012nq, Grozdanov:2015kqa}. To all orders in the hydrodynamic derivative expansion, the dispersion relations \eqref{hydro-modes-shear}, \eqref{hydro-modes-sound} are represented by infinite series in $\q$.

Are these hydrodynamic series convergent? If so, what are their radii of convergence and what determines them? If the series are only asymptotic, can they be resummed?  The shear mode \eqref{hydro-modes-shear} {\it appears} to be an expansion in powers of $\q^2$, whereas the sound mode \eqref{hydro-modes-sound} seems to contain odd and even powers of 
$|\q|=\sqrt{\q^2}$ in its real and imaginary part, respectively. Can we prove that no other power of $\q^2$ or non-analytic terms appears in the hydrodynamic expansions at {\it any} order?%
\footnote{\label{fn1}
In this paper, we consider classical hydrodynamics only, ignoring the effects of statistical fluctuations such as those discussed for example in refs.~\cite{Pomeau:1974hg,Ernst1975}. Such fluctuation effects are suppressed in the holographic models we study below.} These questions were considered in our recent short paper~\cite{Grozdanov:2019kge}, and we shall provide more details here.

In general, proving  convergence of a perturbative series and finding the corresponding radius of convergence may constitute rather difficult (and rather different) problems. For example, the convergence of the so called $1/Z$ series representing the lowest energy eigenvalue of the two-electron atom with the nucleus of charge $Z$ was rigorously 
proven by Kato in 1951 \cite{kato} but  reliably computing the actual value of the radius of convergence  from the series coefficients and their Pad\'{e} extensions  remained a controversial problem for many decades \cite{1-Z-exp,peck}. Yet another example is the series solution to Kepler's equation  whose ``mysterious'' convergence 
properties were discussed by the giants such as Laplace and Cauchy and were finally understood as arising from the critical points of the associated curve in the complex eccentricity plane (see Appendix \ref{kepler-app}).

The problems involving re-summing all-order hydrodynamic expansions and finding the radius of convergence of hydrodynamic series have been previously addressed in the literature. All-order linearised hydrodynamics was investigated by Bu and Lublinsky \cite{Bu:2014sia, Bu:2014ena} using fluid-gravity correspondence. Re-summations of the hydrodynamic derivative expansion have been also considered in the framework of kinetic theory \cite{Kurkela:2017xis} and in cosmological models \cite{Buchel:2016cbj}. By far the largest body of work on the subject has been done in the setting of the boost-invariant flow, where the gradient expansion in the position space typically produces asymptotic rather than convergent series, and the Borel-Pad\'{e} and ``resurgence'' methods \cite{Aniceto:2018bis} can be used to re-sum the series and extract information about gapped modes in the spectrum from the hydrodynamic series \cite{Heller:2013fn,Heller:2015dha,Basar:2015ava,Aniceto:2015mto,Florkowski:2016zsi,Denicol:2016bjh,Florkowski:2017olj,Spalinski:2017mel,Casalderrey-Solana:2017zyh,Heller:2018qvh}. In ref.~\cite{Withers:2018srf}, Withers studied the convergence properties and analytic continuation of a dispersion relation for the shear-diffusion mode in a holographic model involving a dual Reissner-Nordstrom-AdS${}_4$ black brane. There, a finite radius of convergence resulted from a branch point singularity at a certain value of the purely imaginary momentum. From the point of view of the quasinormal spectrum, this point corresponds to the collision of the modes or level-crossing, very similar to the discussion in ref.~\cite{Grozdanov:2019kge} and in the present paper. 

The prediction of classical hydrodynamics is that the above frequencies $\omega(\q)$ appear as poles of the retarded two-point functions\footnote{ We shall often call the retarded two-point functions ``response functions'', as they form the basis of the linear response theory.} of the energy-momentum tensor in thermal equilibrium, as $\q\to0$~\cite{KM,Kovtun:2012rj}. Assuming that the prediction of classical hydrodynamics is correct and that the actual response functions computed from quantum field theory (viewed as functions of~$\omega$) indeed have isolated poles at $\omega=\omega(\q)$ with $\omega(\q \to 0)\to0$, we can define the function $\omega(\q)$ as the location of the relevant pole. In what follows, we shall discuss several models where the poles of the full response functions can be readily analysed, both analytically and numerically, for generic values of $\q$, either real or complex, small or large. This will allow us to study the analytic properties of the derivative expansion in hydrodynamics by studying the dispersion relations $\omega(\q)$ obtained from the poles of the relevant retarded functions in thermal equilibrium.

In general, the hydrodynamic dispersion relations arise as solutions to 
\begin{align}
 P(\q^2,\omega)=0\,,
\label{al-curve}
\end{align}
where $P$ is proportional to the inverse of the corresponding retarded two-point function.\footnote{ Here, the proportionality is assumed to be trivial in the sense that the equations $G^{-1}_R=0$ and \eqref{al-curve} are equivalent.} The hydrodynamic dispersion relations $\omega(\q)$ are solutions to \eqref{al-curve} obeying $\omega(\q \to 0) \to 0$, where $\q^2$ and $\omega$ are treated as complex variables. We shall refer to $P(\q^2,\omega)$ as the hydrodynamic spectral curve. In order to obtain $P(\q^2,\omega)$ from classical hydrodynamics, we choose a set of hydrodynamic variables $\varphi_a$ (such as the fluid velocity and temperature), and linearise the hydrodynamic equations about the equilibrium state, $\varphi_a = \varphi_a^{( {\rm eq})} + \delta\varphi_a$. In the absence of external sources, the hydrodynamic equations are homogeneous and, upon Fourier transforming, can be written as a set of linear algebraic equations, $K_{ab}\delta\varphi_b = 0$. The hydrodynamic spectral curve is then simply $P=\det K$. 

In $n^{\rm th}$-order (classical) hydrodynamics, $P(\q^2,\omega)$ is a polynomial of a finite order, and Eq.~\eqref{al-curve} defines a complex algebraic curve. The theorems of analysis such as the implicit function theorem and the theorem of Puiseux then determine the structure and properties of $\omega(\q)$. In particular, these theorems guarantee that for any finite order of the derivative expansion, the dispersion relations $\omega(\q)$ are given by series converging in some vicinity of the origin $(\q^2,\omega)=(0,0)$, and the same is true as $n\to\infty$, provided $P(\q^2,\omega)$ is an analytic function at $(0,0)$. 
We discuss the spectral curve and the associated dispersion relations of the hydrodynamic modes in section~\ref{rel-hydro-det} of the paper.

In addition to the spectral curve, we shall also study the retarded functions of the energy-momentum tensor. As a simple example, the prediction of $1^{\rm st}$-order hydrodynamics for the retarded function of the transverse momentum density is~\cite{Kovtun:2012rj}
\begin{align}
\label{eq:GR-1}
  G^R_{{\scriptscriptstyle\perp},{\scriptscriptstyle\perp}}(\q^2,\omega) = \frac{(\epsilon{+}p) D \q^2}{i\omega - D \q^2}\,,
\end{align}
where $D=\eta/(\epsilon+p)$, as before. When viewed as a function of $\omega$, there is a simple pole given by the shear mode dispersion relation \eqref{hydro-modes-shear}. When viewed as a function of two variables $\q^2$ and $\omega$, one can immediately see that the point $(\q^2_*,\omega_*)=(0,0)$ is a singular point of $G^R(\q^2,\omega)$, such that the value of the response function at $(\q^2_*,\omega_*)$ is undefined. This is commonly expressed by saying that the limits $\omega\to0$ and $\q\to0$ do not commute. Such indeterminacy-type singularities in physical response functions can also exist at finite non-zero $(\q^2_*,\omega_*)$. This was recently explored for the sound mode (retarded function of the energy density) in refs.~\cite{Grozdanov:2017ajz, Blake:2017ris, Blake:2018leo}, where the phenomenon of the indeterminacy-type singularities at non-zero $(\q^2_*,\omega_*)$ was called ``pole-skipping''. Put differently, pole-skipping happens when $P(\q^2_*,\omega_*) = 0$, and the residue of the corresponding pole of $G^R(\q^2,\omega)$ vanishes at $(\q^2_*,\omega_*)$. In the example of eq.~(\ref{eq:GR-1}), the ``skipping'' of the shear pole happens at the origin. A study of pole-skipping at non-zero $(\q^2_*,\omega_*)$ will be another focus of our paper.%
\footnote{
The connection between the pole-skipping values  $(\q^2_*,\omega_*)$  and the growth of the out-of-time-ordered four-point correlation functions (OTOC) of local operators in quantum field theory has been studied 
in refs.~\cite{Grozdanov:2017ajz,Blake:2017ris,Blake:2018leo,Grozdanov:2018kkt}.
}
In fact, the original motivation for our work was to see whether the pole-skipping singularities in response functions and the non-zero radius of convergence of the hydrodynamic derivative expansion might be related to each other.

To illustrate our approach in a simple exactly solvable example, in section \ref{DG-model} we discuss  the holographic bottom-up model studied, in particular, by Davison and Gout\'{e}r\-aux in ref.~\cite{Davison:2014lua}.  The advantage of the model is that the effects of translation symmetry breaking on hydrodynamics can be studied in a controlled manner, and that the hydrodynamic and non-hydrodynamic degrees of freedom can be easily separated. The explicit breaking of the translation symmetry means that momentum is no longer conserved, and the sound mode is absent from the spectrum as $\q\to0$. Thus, the only modes with $\omega(\q \to0) = 0$ are the diffusive modes of the energy and charge densities. For a certain special value of the translation symmetry breaking parameter in the model, the response functions of the energy and charge densities can be found exactly for all (not just small) momenta \cite{Davison:2014lua}. One then finds the following dispersion relation for the diffusive modes:
\begin{align}
\label{DG-FullDisp}
\wfr (\qfr)  = - \frac{i}{2} \left( 1 - \sqrt{1 - 4 \qfr^2 }\right) = - i \qfr^2 - i \qfr^4 + \cdots\,.
\end{align} 
It is clear that the corresponding hydrodynamic series converges in the circle $|\qfr|<|\qfr_{\rm c}|=1/2$ due to the branch point singularities at $\qfr_{\rm c}=\pm 1/2$. We shall study the exact and approximate spectral curves and obstruction to convergence in this  model in section~\ref{DG-model}.

\begin{figure}[t!]
\centering
\includegraphics[width=0.45\textwidth]{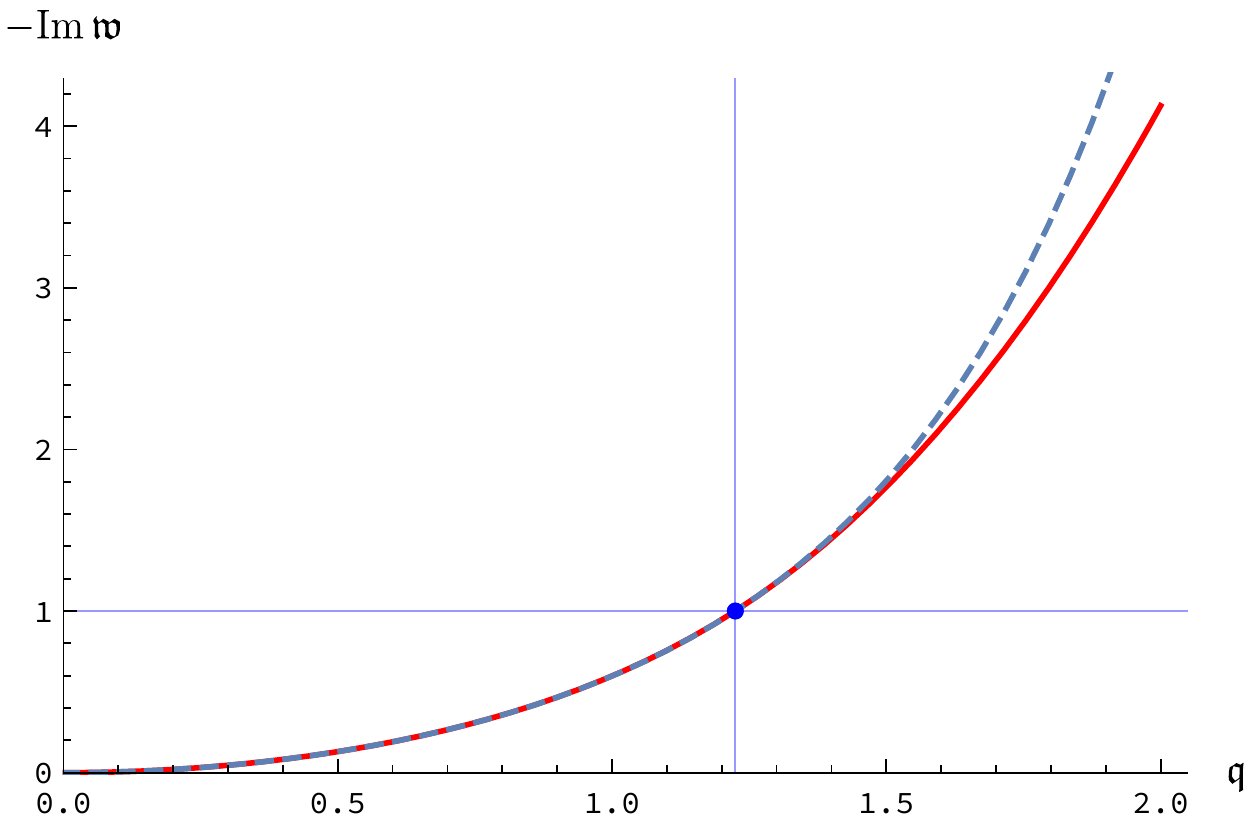}
\hspace{0.05\textwidth}
\includegraphics[width=0.45\textwidth]{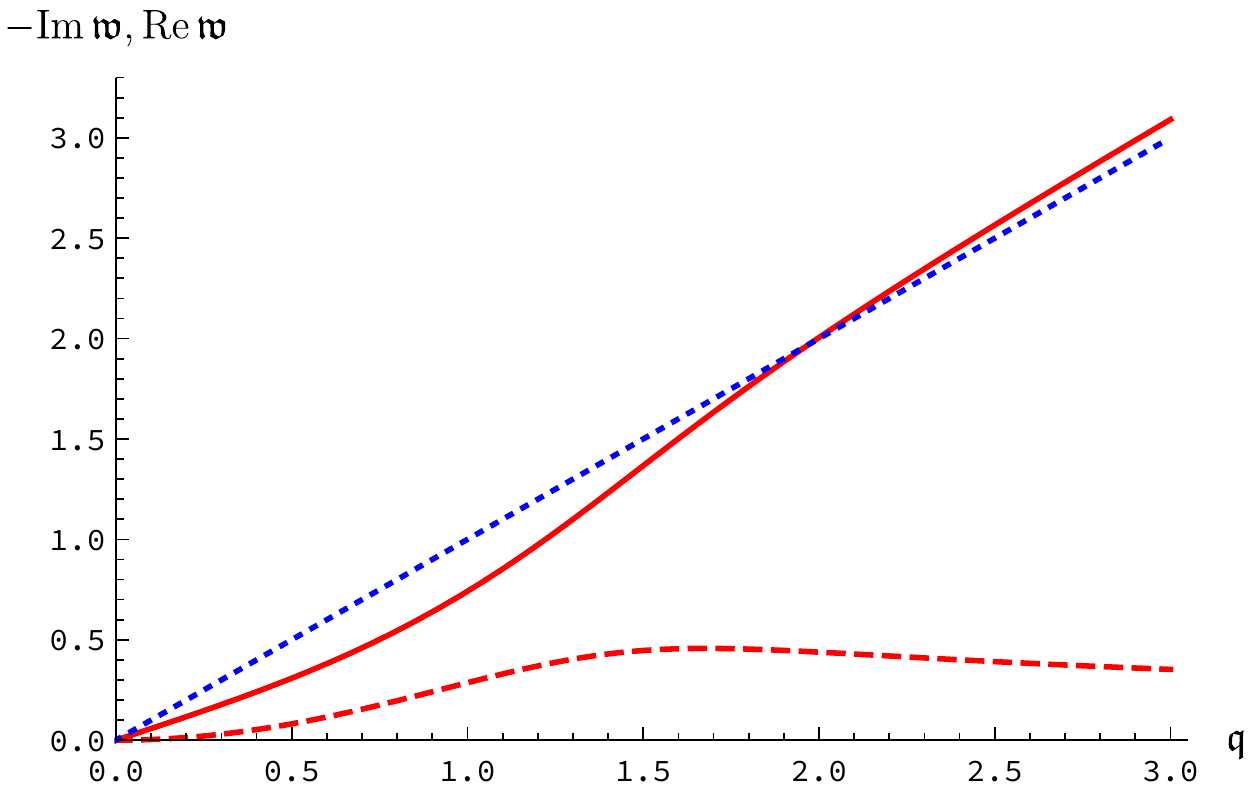}
\caption{
{\small Dispersion relations of the hydrodynamic modes in the strongly coupled ${\cal N}=4$ SYM theory, obtained using the dual holographic description. The dispersion relations are plotted in terms of dimensionless $\wfr \equiv\omega/2\pi T$  and $\qfr\equiv |\q|/2\pi T$, with complex $\wfr$ as functions of real positive $\qfr$. The left panel shows $\wfr_{\rm shear}(\qfr)$ for the shear mode, the right panel shows $\wfr_{\rm sound}^{+}(\qfr)$ for one of the two sound modes. In the left panel, the actual $-{\rm Im}\,\wfr_{\rm shear}(\qfr)$ for the shear mode is shown by the solid red curve, and the analytic hydrodynamic approximation to $O(\qfr^8)$ (computed in sec.~\ref{shear-n=4-hydro}) is shown by the dashed blue curve. The blue dot indicates the pole-skipping point at $\qfr_* =\sqrt{3/2}$, $\mathfrak{w}_* =-i$, discussed in sec.~\ref{sec:Chaos}. The right panel shows ${\rm Re}\,\wfr_{\rm sound}^{+}(\qfr)$ (solid red curve) and $-{\rm Im}\,\wfr_{\rm sound}^{+}(\qfr)$ (dashed red curve) for the ``+'' sound mode. The straight dotted line indicates the light cone $\mbox{Re}\,\mathfrak{w} =\qfr$.
}}
\label{fig:disp-rel-n=4}
\end{figure}

Our main example, considered in section \ref{sec:SYM}, is the ${\cal N}=4$ supersymmetric $SU(N_c)$ Yang-Mills theory at infinite $N_c$ and infinite 't Hooft coupling, which we will abbreviate as ``strongly coupled ${\cal N}=4$ SYM theory''. In this theory, real-time equilibrium correlation functions can be calculated by the methods of holographic duality~\cite{Son:2002sd, Policastro:2002se} (see for example refs.~\cite{Ammon:2015wua, CasalderreySolana:2011us,zaanen2015holographic,hartnoll2018holographic} for an introduction to the holographic duality and applications). The dispersion relations of the shear and sound modes in the strongly coupled ${\cal N}=4$ SYM theory obtained by holographic methods are shown in fig.~\ref{fig:disp-rel-n=4}. Using the units $\hbar=c=1$, we plot the dispersion relations in terms of dimensionless variables $\wfr\equiv \omega/2\pi T$ and $\qfr\equiv |\q|/2\pi T$. The function $\wfr(\qfr)$ for the shear mode is purely imaginary for real $\qfr$, while $\wfr(\qfr)$ for the sound mode has both real and imaginary parts for real $\qfr$. The functions $\wfr(\qfr)$ in fig.~\ref{fig:disp-rel-n=4} appear to be smooth and generally unremarkable functions of real positive~$\qfr$. Their behaviour at $\qfr \ll 1$ has a clear hydrodynamic interpretation \cite{Policastro:2002tn,Kovtun:2005ev} fully compatible with Eqs.~\eqref{hydro-modes-shear}, \eqref{hydro-modes-sound}, and their asymptotics as $\qfr\rightarrow \infty$ were studied in refs.~\cite{Festuccia:2008zx,Fuini:2016qsc}. Thus, at least in the case of the ${\cal N}=4$ SYM theory, if the series \eqref{hydro-modes-shear}, \eqref{hydro-modes-sound} have finite radii of convergence, the obstacle to convergence must lie either at negative $\qfr=\sqrt{\qfr^2}$, or more generally, at complex momenta. By studying complex $\qfr$, one indeed finds that the functions $\wfr(\qfr)$ in the ${\cal N}=4$ SYM theory have finite non-zero radii of convergence: $|\qfr_{\rm c}^{\rm sound}| = \sqrt{2}$, and $|\qfr_{\rm c}^{\rm shear}| \approx 1.49$~\cite{Grozdanov:2019kge}. In section \ref{sec:SYM}, we show in detail how the finite radius of convergence arises from the quasinormal mode level-crossing in the shear and sound channels, and demonstrate that the level-crossing phenomenon is also observed in the scalar channel of the correlators.

\begin{figure}[t!]
\centering
\includegraphics[width=0.6\textwidth]{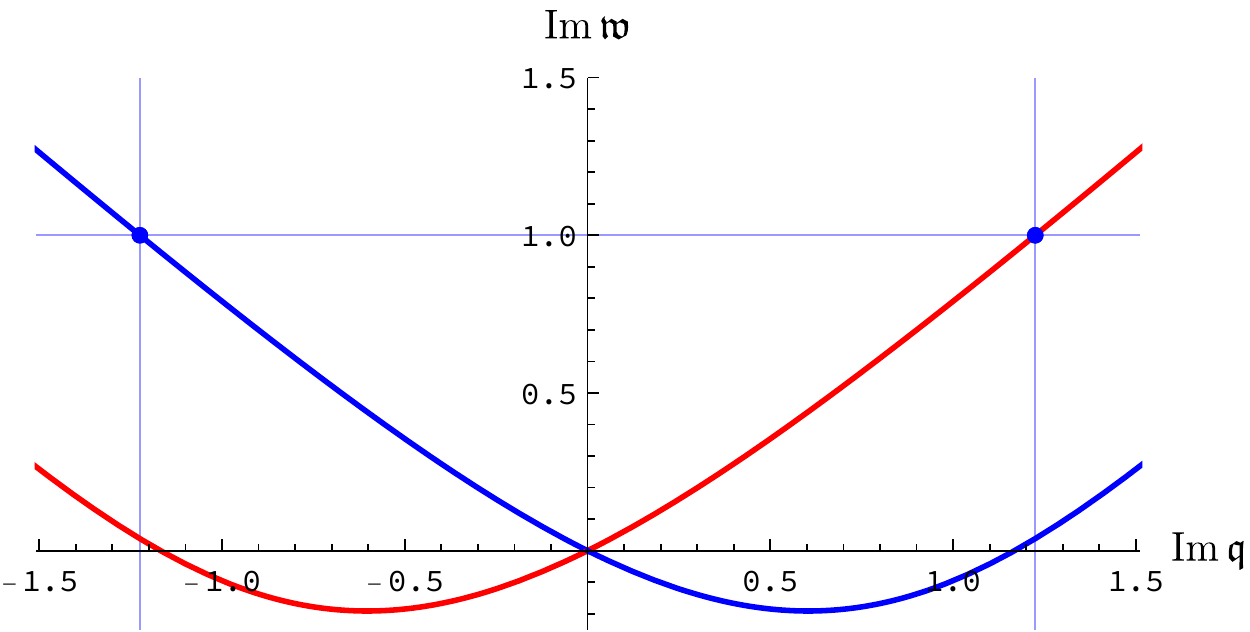}
\caption{{\small 
The analytically continued sound mode frequencies in the strongly coupled ${\cal N}=4$ SYM theory, obtained using the dual holographic description. The dimensionless frequencies $\wfr_{\rm sound}^{\pm}$ of the two sound modes are plotted for purely imaginary dimensionless spatial momentum $\qfr$, with the ``+'' branch in red and the ``$-$'' branch in blue. The frequencies $\wfr_{\rm sound}^{\pm}$ are purely imaginary at imaginary $\qfr$. At small momenta, the curves are linear with slopes $\pm v_s$, with $v_s=1/\sqrt{3}$.  The curves pass through pole-skipping points $(\qfr_*,\wfr_*) =(\pm i\sqrt{3/2}, i)$  indicated by the blue dots.
}}
\label{fig-chaos}
\end{figure}

\begin{figure}[t]
\centering
\includegraphics[width=0.6\textwidth]{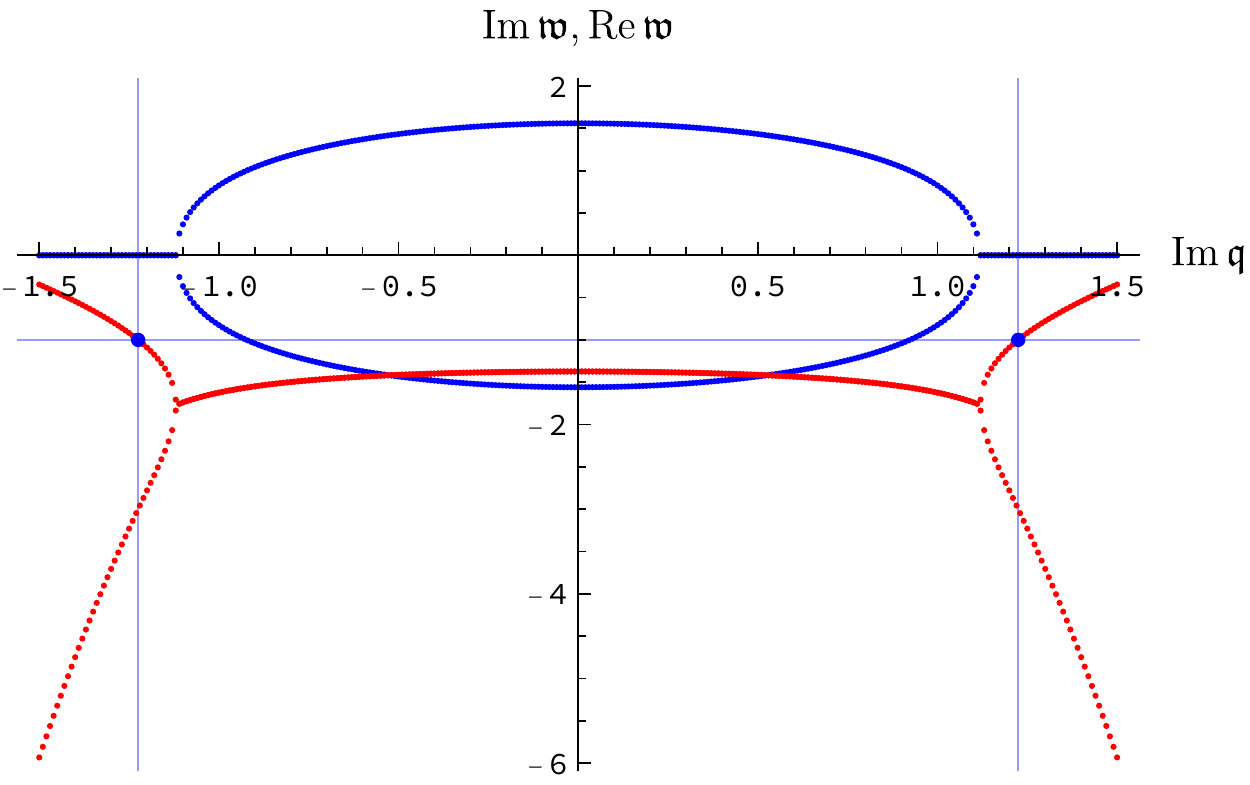}
\caption{{\small 
The first two (closest to the origin) poles of the retarded function of $T^{xy}$ in the strongly coupled ${\cal N}=4$ SYM theory, obtained using the dual holographic description. The locations of the  poles are plotted as functions of the dimensionless wave vector for $\qfr$ purely imaginary, with ${\rm Re}\, \wfr$ shown in blue, and ${\rm Im}\, \wfr$ shown in red. The dots indicate the points $(\qfr_*,\wfr_*)=(\pm i\sqrt{3/2},-i)$, where the response function of $T^{xy}$ exhibits pole-skipping.
}}
\label{fig-chaos-scalar}
\end{figure}

One of our main results concerns the response functions of the energy-momentum tensor in the strongly coupled ${\cal N}=4$ SYM theory. It was shown in refs.~\cite{Grozdanov:2017ajz, Blake:2018leo} that there is a pole-skipping singularity in the retarded two-point function of the energy density at $(\qfr^2_*,\wfr_*) = (-3/2,i)$, at which point the sound pole is ``skipped''. The sound dispersion curves pass through the point $(\qfr^2_*,\wfr_*)$, as illustrated in fig.~\ref{fig-chaos}. We observe that in close analogy with the sound mode, the shear mode dispersion relation \eqref{hydro-modes-shear} passes through the point $\qfr^2_* = 3/2$, $\wfr_* = -i$ (see fig.~\ref{fig:disp-rel-n=4}). It turns out that this is not an accident: we will show that the pole-skipping phenomenon in the strongly coupled ${\cal N}=4$ SYM theory is exhibited not only by the response functions which give rise to the sound mode (energy density correlations), but also by the response functions which give rise to the shear mode (transverse momentum density correlations). Moreover, we find that the pole-skipping at non-zero complex momentum also occurs in response functions of those components of the energy-momentum tensor that are not related to hydrodynamic modes. For example, for $\q$ along the $z$ direction, the fluctuations of $T^{xy}$ are gapped, and the response function of $T^{xy}$ has no hydrodynamic singularities. Nevertheless, the gapped singularities of the retarded function of $T^{xy}$ pass through $(\qfr^2_*,\wfr_*)=(-3/2, -i)$, at which point one of the gapped poles is ``skipped''. This is illustrated in fig.~\ref{fig-chaos-scalar}. In other words, all retarded functions of $T^{\mu\nu}$ in strongly coupled ${\cal N}=4$ SYM theory exhibit pole-skipping at $|\qfr_*|=\sqrt{3/2}$ and $|\wfr_*|=1$. The retarded functions of the energy-momentum tensor and the convergence of the derivative expansion in the ${\cal N}=4$ SYM theory are discussed in section~\ref{sec:SYM}.


A natural question to ask is whether pole-skipping happens within the domain of validity of the hydrodynamic approximation, as far as the convergence of the hydrodynamic derivative series is concerned. In other words, if the hydrodynamic dispersion relation $\wfr_i(\qfr)$ has a finite radius of convergence $|\qfr_{\rm c}^i|$ and pole-skipping in the corresponding response function happens at $|\qfr_*^i|$, how does $|\qfr_{\rm c}^i|$ compare with $|\qfr_*^i|$? In the strongly coupled ${\cal N}=4$ SYM theory we have $|\qfr_{\rm c}^{\rm sound}| = \sqrt{2}$, $|\qfr_{\rm c}^{\rm shear}| \approx 1.49$~\cite{Grozdanov:2019kge}, and $|\qfr_*|=\sqrt{3/2}$. Thus $|\qfr_*|<|\qfr_{\rm c}|$, and therefore pole-skipping in correlation functions takes place within the convergence domain of the hydrodynamic derivative expansion. On the other hand, in the model of ref.~\cite{Davison:2014lua}, we have $|\qfr_{\rm c}| =1/2$, $|\qfr_*| = \sqrt{2}$, hence $|\qfr_*| > |\qfr_{\rm c}|$, so pole-skipping occurs outside the convergence domain of the hydrodynamic derivative expansion. This indicates that the ``skipping'' of hydrodynamic poles in retarded functions of energy and momentum densities is not directly related to the convergence radius of the derivative expansion in hydrodynamics.

More generally, pole-skipping singularities in real-time response functions at non-zero momentum do not have to have any relation to hydrodynamics at all. As an example, we consider response functions of spin-zero operators in 1+1 dimensional conformal field theory (CFT). For a primary operator, the Euclidean correlation function on ${\mathbb R}^2$ is fixed by conformal symmetry. Performing a conformal transformation to the cylinder ${\mathbb R}\times {\mathbb S}^1$ gives Euclidean thermal correlations functions~\cite{DiFrancesco:1997nk}, which can be Fourier transformed and analytically continued to produce exact real-time retarded functions $G^R(\wfr,\qfr)$~\cite{Son:2002sd}. These functions have no hydrodynamic poles, yet we will see that there is an infinite number of pole-skippings at non-zero values of $(\qfr_*,\wfr_*)$. We discuss this in detail in section~\ref{sec:pskk}.  Our conclusions and discussion of the issues raised in 
the paper appear in section \ref{disc}.

\section{Hydrodynamic dispersion relations as Puiseux series}
\label{rel-hydro-det}

\subsection{Hydrodynamic spectral curves}
\label{hscx}
We start with a brief review of how the hydrodynamic dispersion relations are derived. Consider hydrodynamics of a neutral homogeneous and isotropic relativistic fluid in flat space in $d_s$ spatial dimensions. We are interested in linearised fluctuations in a homogeneous and isotropic equilibrium state, $T^{\mu\nu} = T^{\mu\nu}_{\rm eq.} + \delta T^{\mu\nu}$, where $T^{\mu\nu}$  denotes the expectation value of the symmetric energy-momentum tensor operator, and the equilibrium state is characterised by $T^{0 0}_{\rm eq.}=\epsilon$, $T^{ij}_{\rm eq.}=p \delta^{ij}$, $T^{0i}_{\rm eq.}=0$, where $\epsilon$ and $p$ are the equilibrium energy density and pressure. The equations of hydrodynamics follow from the conservation of the energy-momentum tensor, $\partial_\mu T^{\mu\nu}=0$. Translation invariance of the equilibrium state implies that we can Fourier transform the fluctuations and take all variables to be proportional to $\exp{(-i \omega t + i \q {\cdot}\x)}$. Furthermore, rotation invariance allows us to choose the direction of the $z$ axis along $\q$. We then have the following system of conservation equations for the linearised fluctuations:
\begin{subequations}
\label{eq:linearized-hydro-1}
\begin{align}
- \omega \, \delta T^{0a} + q_z \, \delta T^{za} &=0\,, \label{conservation-eqs-1} \\
- \omega \,  \delta T^{00} + q_z \, \delta T^{z0}& =0\,, \label{conservation-eqs-2} \\
- \omega \, \delta T^{0z} + q_z \, \delta T^{zz} &=0\,, 
\label{conservation-eqs-3}
\end{align}
\end{subequations}
where we use the index $a$ and subsequent Latin indices to denote any of the $d_s-1$ spatial directions orthogonal to $z$.

The above conservation equations need to be supplemented by the constitutive relations which express $\delta T^{\mu\nu}$ in terms of the hydrodynamic degrees of freedom. For linearised hydrodynamics, a convenient choice of the degrees of freedom is the energy density $\delta T^{00}$ and momentum density $\delta T^{0i}$. This choice implies that we only need the constitutive relations for the spatial stress, $\delta T^{ij} = \delta T^{ij}(\delta T^{00}, \delta T^{0k})$. The constitutive relations will contain derivatives of $\delta T^{00}$ and $\delta T^{0k}$, as is needed for example to describe the viscosity of fluids. We will organise the constitutive relations according to the number of derivatives of the hydrodynamic variables. The hydrodynamics of $k$-th order is determined by the constitutive relations in which $\delta T^{ij}$ contains up to $k$ derivatives of $\delta T^{00}$ and $\delta T^{0i}$. It is then straightforward to write down the linearised constitutive relations at any order, by noting that under the spatial $SO(d_s)$, the stress fluctuation $\delta T^{ij}$ is a rank-two tensor, momentum density $\delta T^{0i}$ is a vector, and the energy density $\delta T^{00}$ is a scalar. For example, in the first-order hydrodynamics of ref.~\cite{landau}, we have
\begin{align}
 \delta T^{ij} \,& = \delta^{ij}  {\partial p \over \partial \epsilon}\,
\delta T^{00} \nonumber \\ \,&
- {1\over \epsilon + p} \left[ \eta \left(
\partial_i \,\delta T^{0j} +
\partial_j \,\delta T^{0i} - {2\over d_s}  \delta^{ij} \partial_k \delta T^{0k}\right)
+ \zeta  \delta^{ij} \partial_k \, \delta T^{0k}\right] + \cdots\,,
\label{crel-1}
\end{align}
where $\eta$ is the shear viscosity, $\zeta$ is the bulk viscosity, and the ellipses denote terms with more than one derivative of $\delta T^{00}$, $\delta T^{0i}$. Combining the constitutive relations \eqref{crel-1} with the conservation equations \eqref{eq:linearized-hydro-1} gives a system of linear equations for the fluctuations $\delta T^{00}$ and $\delta T^{0i}$. The equations have non-trivial solutions provided the corresponding determinant vanishes:
\begin{align}
\label{eq:Pwq-1h}
  P_1(\q^2, \omega) \equiv \left( \omega + i D  \q^2 \right)^{d_s-1} \left(\omega^2 + i  \Gamma_{\!s} \omega  \q^2 - v_s^2   \q^2  \right) = 0 \, ,
\end{align}
where $v_s^2 = \partial p/\partial\epsilon$ is the speed of sound squared, and $D$, $\Gamma_{\!s}$ are defined by eqs.~\eqref{hydro-modes-coeffi-D}, \eqref{hydro-modes-coeffi-Gamma}. 

In fact, rotation invariance implies that the most general linearised constitutive relations in momentum space take the following form:
\begin{align}
\delta T^{nm} = \,& -i A \left( q^n \delta T^{0m} + q^m \delta T^{0n} \right) + 
\delta T^{00} \left( B q^n q^m + C \delta^{nm}\right) \nonumber 
\\ &\, + i q_l \delta T^{0l} \left( D q^n q^m + E \delta^{nm}\right)\,,
\label{c-rel-x}
\end{align}
where $A$, $B$, $C$, $D$, $E$ are scalar functions of $\omega$ and $\q^2$. Substituting the constitutive relations \eqref{c-rel-x} into the conservation equations \eqref{eq:linearized-hydro-1}, we find a system of $d_s{+}1$ linear equations for $d_s{+}1$ hydrodynamic variables. This system has non-trivial solutions provided the determinant of the corresponding matrix vanishes. The vanishing of the determinant is equivalent to the vanishing of
\begin{align}
  P(\q^2,\omega) \equiv F_{\rm shear} ^{d_s -1} F_{\rm sound} \,,
  \label{reduced-c}
\end{align}
where
\begin{align}
\,&  F_{\rm shear} (\q^2, \omega) \equiv \omega + i \q^2 \gamma_\eta (\q^2,\omega)=0\,, \label{disp-rel-general-shear} \\
\,&  F_{\rm sound}(\q^2, \omega) \equiv \omega^2 + i \omega \q^2 \gamma_s (\q^2,\omega) - \q^2 H(\q^2,\omega) =0\,.
\label{disp-rel-general-sound}
\end{align}
Here the coefficients are $\gamma_\eta \equiv A$, $\gamma_s \equiv 2 A - E - D \q^2$, $H = B \q^2 +C$. Thus, the shear and the sound modes decouple as a consequence  of rotation invariance.%
\footnote{
See refs.~\cite{Bu:2014sia, Bu:2014ena} for a study of ``resummed'' holographic hydrodynamics to all orders in the derivative expansion. In our language, this amounts to studying the functions $\gamma_\eta(\q^2,\omega)$, $\gamma_s(\q^2,\omega)$, $H(\q^2,\omega)$ in a holographic model. 
}

If the constitutive relations \eqref{c-rel-x} are given by a local derivative expansion, then the 
functions $\gamma_\eta (\q^2,\omega)$,  $\gamma_s (\q^2,\omega)$ and  $H(\q^2,\omega)$ are power series in $\omega $ and $\q^2$ with finite values at $\omega=0$, $\q^2=0$:
\begin{align}
\,&  \gamma_\eta (0,0) = D\,, \qquad 
 \gamma_s (0,0) = \Gamma_{\!s} \,, \qquad 
H(0,0) = v_s^2\,,
\label{local-func}
\end{align}
with $v_s$, $D$ and $\Gamma_{\!s}$ as above. Truncating the derivative expansion at order $k$ then gives a sequence of algebraic equations defined by finite polynomials in $\q^2$ and $\omega$,
\begin{align}
\,&  F_{\rm shear}^{(k)} (\q^2, \omega) =0\,, \label{disp-rel-general-shear-k} \\
\,&  F_{\rm sound}^{(k)} (\q^2, \omega) =0\,.
\label{disp-rel-general-sound-k}
\end{align}
For general complex $\omega$ and $\q^2$, eqs. \eqref{disp-rel-general-shear}, \eqref{disp-rel-general-sound}, or \eqref{disp-rel-general-shear-k}, \eqref{disp-rel-general-sound-k} define complex 
algebraic curves\footnote{ Eq.~\eqref{reduced-c}  is an example of a 
reduced  curve $f(x,y)=\prod_i g_i (x,y)$, where each $g_i$ can be considered independently \cite{wall}.}\textsuperscript{,}\footnote{ The equations  $F_{\rm shear}^{d_s-1} =0$ and $F_{\rm shear} =0$ define the same curve. To avoid any confusion, by the ``shear curve'', we shall always mean the definition $F_{\rm shear} =0$.} which we will call hydrodynamic spectral curves. The complete dispersion relations of the $i$-th mode, $\omega_i = \omega_i (\q^2)$, can be obtained by solving eqs.~\eqref{disp-rel-general-shear}, \eqref{disp-rel-general-sound} for $\omega$, while the corresponding approximate expressions arising in $k$-th order hydrodynamics are found from eqs.~\eqref{disp-rel-general-shear-k}, \eqref{disp-rel-general-sound-k}. The hydrodynamic dispersion relations are the solutions which satisfy $\omega_i(\q^2\to0) = 0$. They correspond to infinite relaxation times for infinite-wavelength perturbations of conserved densities, i.e.\ to the conservation of energy and momentum.

Note that the polynomials $F_{\rm shear}^{(k)} (\q^2, \omega)$, $F_{\rm sound}^{(k)} (\q^2, \omega)$ are not defined uniquely because of the freedom to organise the derivative expansion in hydrodynamics, such as the choice of ``frame'' and the use of on-shell conditions~\cite{Kovtun:2012rj,Kovtun:2019hdm}. As an example, the conservation equations \eqref{eq:linearized-hydro-1} imply that the factors of $\omega$ in the constitutive relations \eqref{c-rel-x} can be eliminated at each order in the derivative expansion. Thus one can organise the derivative expansion in such a way that the functions $\gamma_\eta(\q^2,\omega)$, $\gamma_s(\q^2,\omega)$ and $H(\q^2,\omega)$ are all $\omega$-independent at each given order in the expansion. Then eqs.~\eqref{disp-rel-general-shear-k}, \eqref{disp-rel-general-sound-k} give simple explicit expressions for $\omega_{\rm shear}(\q^2)$ and $\omega_{\rm sound}(\q^2)$ in terms of three scalar functions $\gamma_\eta(\q^2)$, $\gamma_s(\q^2)$ and $H(\q^2)$, whose small-$\q$ limits are given by eq.~\eqref{local-func}. In this way of implementing the derivative expansion, the hydrodynamic dispersion relations are the only solutions to \eqref{disp-rel-general-shear-k}, \eqref{disp-rel-general-sound-k}. Of course, other choices of organising the derivative expansion are possible where the $\omega$-dependence in $\gamma_\eta(\q^2,\omega)$, $\gamma_s(\q^2,\omega)$ and $H(\q^2,\omega)$ is retained, and non-hydrodynamic (gapped) modes appear in addition to the hydrodynamic modes.

The above discussion was in the context of classical hydrodynamics. A similar factorisation of the shear and sound modes happens in the full response functions of the energy-momentum tensor, without any hydrodynamic assumptions~\cite{Kovtun:2005ev}. For example, for the wave vector along $z$, the shear mode is described by fluctuations of $\delta T^{0a}$, where the direction $a$ is orthogonal to $z$. The condition that the inverse of the equilibrium response function of the $T^{0a}$ operator vanishes can be written as $P_{\rm shear}(\q^2,\omega) = 0$. In general, $P_{\rm shear}(\q^2,\omega)$ is a complicated function which describes both hydrodynamic (long-distance, long-time) and non-hydrodynamic (short-distance, short-time) physics. For small (appropriately defined) $\q^2$ and $\omega$, the exact $P_{\rm shear}(\q^2,\omega)$ will reduce to  the above $F_{\rm shear}(\q^2,\omega)$, provided hydrodynamics is a valid effective description of the system. The same applies to the sound mode: the condition that the inverse of the equilibrium response function of the $T^{00}$ operator vanishes can be written as $P_{\rm sound}(\q^2,\omega) = 0$. For small (relative to the appropriately defined scale) $\q^2$ and $\omega$, the exact $P_{\rm sound}(\q^2,\omega)$ will reduce to  the above $F_{\rm sound}(\q^2,\omega)$, provided hydrodynamics is a valid effective description of the system. In this paper, we will only be studying physical systems in which the near-equilibrium physics governed by the conserved densities can be described by classical hydrodynamics. In other words, we will assume that the functions $\gamma_\eta(\q^2,\omega)$, $\gamma_s(\q^2,\omega)$, $H(\q^2,\omega)$ are defined by the exact response functions of the $T^{\mu\nu}$ operator.

\subsection{Small-momentum expansions}
\label{hydro-spectral-sound}

For a given spectral curve, the small-$\q^2$ expansion of the hydrodynamic dispersion relation $\omega_i(\q^2)$ can be found using the theorem of Puiseux (see Appendix \ref{puiseux-app} and refs.~\cite{wall,walker,sendra}).

Starting with the shear mode, let us assume that $\gamma_\eta (\q^2,\omega)$ is analytic at $(\q^2,\omega)$=$(0,0)$ and thus eq.~\eqref{disp-rel-general-shear} defines an analytic curve for complex $(\q^2, \omega) \in \mathbb{C}^2$. 
This is of course not guaranteed {\it a priori} and should be established by independent methods, e.g.\ by finding the exact response function.\footnote{ Moreover, the analyticity fails when the statistical fluctuation effects are taken into account, see footnote \ref{fn1} and ref.~\cite{Kovtun:2011np}.
} 
The analyticity of $\gamma_\eta (\q^2,\omega)$ implies that  $F_{\rm shear}(\q^2,\omega)$ is analytic at the origin as well, and, since $\partial F_{\rm shear}/\partial \omega =1 \neq 0$ at $(0,0)$, the origin is a regular point of the analytic curve  \eqref{disp-rel-general-shear}. Then the analytic implicit function 
theorem  (see Appendix \ref{puiseux-app}) guarantees that for sufficiently small  $\q^2$ and $\omega$, there exists a unique solution of eq.~\eqref{disp-rel-general-shear}  of the form 
\begin{align}
\label{eq:shear-1x}
   \omega_{\rm shear} (\q^2) =  -i \sum_{n=1}^{\infty} c_n \q^{2 n} = -i c_1 \q^2 + O(\q^4)\, ,
\end{align}
convergent in a neighbourhood of $\q^2=0$. The radius of convergence is determined by the location of the nearest to the origin critical point of the curve  \eqref{disp-rel-general-shear}. 

\begin{figure}[h]
\centering
\includegraphics[width=0.6\textwidth]{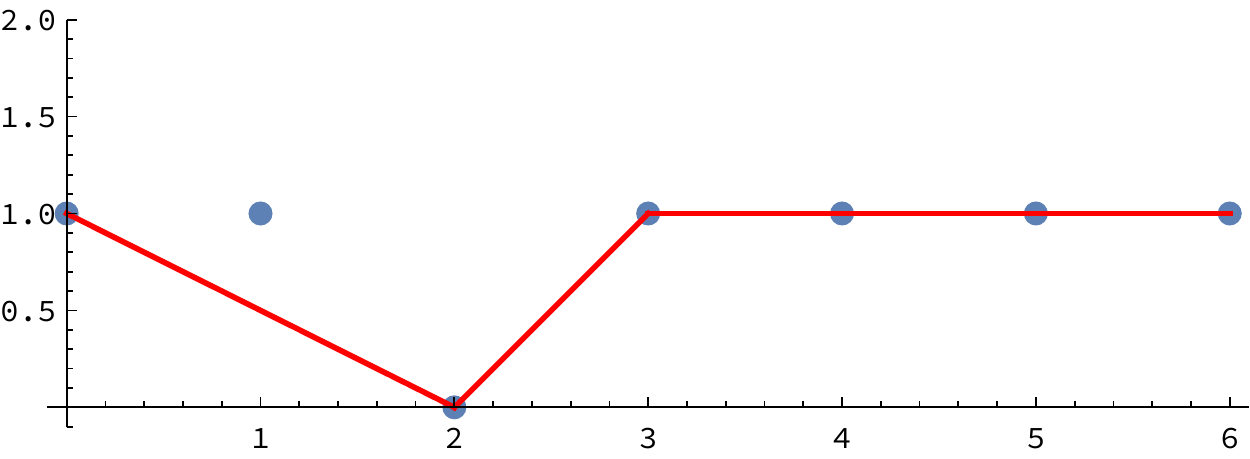}
\caption{{\small The Newton polygon for the sound mode.}}
\label{fig-np}
\end{figure}

Continuing with the sound mode, let us again assume that $\gamma_s (\q^2,\omega)$ and $H (\q^2,\omega)$ are analytic functions at $(0,0)$. The function $F_{\rm sound}(\q^2,\omega)$ is then analytic at the origin as well. Now we have $\partial F_{\rm sound}/\partial \omega = 0$ at $(0,0)$, and thus the origin is a critical point of the spectral curve. On the other hand, $\partial^2 F_{\rm sound}/\partial \omega^2 = 2\neq 0$ at $(0,0)$, thus we expect the sound dispersion relation to have $p=2$ branches. The Puiseux series expansions are then given by eqs.~\eqref{puiseux-1}, \eqref{puiseux-2}, in other words $\omega_{\rm sound}^{(j)}(\q^2)$ can be represented by series in non-negative powers of $(\q^2)^{1/m_j}$, where $m_j$ are positive integers, and $j=1,2$ labels the two branches corresponding to the two sound modes. Following the general analysis of algebraic curves, the integers $m_j$ may be found using the Newton's polygon method (see refs.~\cite{wall,walker,sendra} for details). For analytic $\gamma_s$ and $H$, we have the expansions
\begin{align}
\,& \gamma_s (\q^2,\omega) = \sum\limits_{n,m=0}^\infty \gamma^s_{nm} \omega^n \q^{2m}\,, \\
\,& H (\q^2,\omega) = \sum\limits_{n,m=0}^\infty H_{nm} \omega^n \q^{2m}\,.
\end{align}
 The coefficients in front of various powers of $\omega$ in the expression \eqref{disp-rel-general-sound} are then given by
\begin{align}
\,& \omega^0:  \qquad -  \sum\limits_{k=0}^\infty H_{0k} \q^{2k+2} \,,\\
\,& \omega^1:  \qquad  i \q^2 \gamma^s_{00} + \sum\limits_{k=1}^\infty \left( i \gamma^s_{0k} -H_{1k}\right) \q^{2k+2} \,,  \\
\,& \omega^2:  \qquad 1+ \sum\limits_{k=0}^\infty \left( i \gamma^s_{1k} -H_{2k}\right) \q^{2k+2} \,,  \\
\,& \omega^3:  \qquad \sum\limits_{k=0}^\infty \left( i \gamma^s_{2k} -H_{3k}\right) \q^{2k+2} \,,  \\
\,& \vdots \, \\
\,& \omega^n:  \qquad  \sum\limits_{k=0}^\infty \left( i \gamma^s_{n-1 k} -H_{n k}\right) \q^{2k+2} \,.  
\end{align}
The vertices of the Newton polygon are thus given by $(0,1+k_0)$, $(1,1+k_1)$, $(2,0)$, $(3,1+k_3)$, $(4,1+k_4)$, $\ldots$, where $k_0, k_1, \ldots$ are the smallest indices such that $H_{0 k_0}\neq 0$ as well as $H_{0 k_1}\neq 0$ 
or/and $\gamma^s_{0 k_1}\neq 0$, etc. The Newton polygon for the sound mode is shown in fig.~\ref{fig-np}, where it is assumed that 
$H_{00}\neq 0$ and that either  $H_{n,0}\neq 0$ or  $\gamma^s_{n-1,0}\neq 0$ (or both) are non-zero for $n=3,4, \ldots$. The exponents 
of  $x\equiv \q^2$ in the Puiseux series are given by the negative slopes of the polygon's lines, i.e.\ by $1/2$ for 
$H_{00}\neq 0$. Thus $m_j=2$, and the lowest order term in the two branches is
\begin{align}
\omega = \pm \sqrt{H_{00}}\, (\q^2)^{\frac{1}{2}} +\cdots \,.
\end{align}
From the Newton polygon, it is clear that $H_{00}\neq 0$ is the necessary and sufficient condition for the fractional powers of $\q^2$ to appear in the dispersion relation. In the hydrodynamic derivative expansion, $H_{00} = \partial p/\partial \epsilon = v_s^2$ is the speed of sound squared. One may have $H_{00}=0$ at the point of a 
phase transition (see e.g. ref.~\cite{Parnachev:2005hh}) in which case the dispersion relation contains only positive powers of $\q^2$.

Generically, for $H_{00}=v_s^2\neq 0$, the sound mode dispersion relation will be given by the two branches of 
Puiseux series in $(\q^2)^{1/2}$ converging in some neighbourhood of the point $\q^2=0$,
\begin{align}
\label{eq:sound-1x}
  \omega_{\rm sound}^\pm (\q^2) = -i  \sum_{n=1}^{\infty} a_n e^{\pm \frac{i \pi n}{2}} (\q^2)^{\frac{n}{2}} = 
\pm a_1 (\q^2)^{\frac{1}{2}} + i a_2 \q^2 \mp a_3 (\q^2)^{\frac{3}{2}} +  O(\q^4)\, ,
\end{align}
where $a_n \in \mathbb{R}$ and $a_1=c_s$ is the speed of sound. In particular, for $\q^2 = e^{\pm i \pi} |\q^2|$, the functions $\omega_{\rm sound}^\pm (\q^2)$ are purely imaginary as anticipated in ref.~\cite{Grozdanov:2017ajz}.

\subsection{Convergence of the hydrodynamic series}
\label{hydro-series-convergence}
The shear and sound dispersion relation series \eqref{eq:shear-1x} and \eqref{eq:sound-1x} have non-zero radii of convergence as long as the corresponding spectral curves \eqref{disp-rel-general-shear} and \eqref{disp-rel-general-sound} are given by the functions  of $\q^2$ and $\omega$ analytic at the origin $(0,0)$. One way to find the radius of convergence of the series is to analyse the behaviour of the coefficients $a_n$, $c_n$ at large $n$. This behaviour will of course depend on the microscopic details of the particular physical system, and may be difficult to study in practice (see, however, ref.~\cite{Withers:2018srf}). Instead, here we use the spectral curves to determine the radii of convergence. 

The Puiseux analysis implies that the domain of convergence of Puiseux series centred e.g. at the origin is the circle whose radius is set by the distance from the origin to the nearest critical point of the associated spectral curve.  Critical points of the spectral curve $F(\q^2,\omega)=0$, where $\q^2, \omega \in \mathbb{C}$, are determined by the conditions
\begin{align}
F(\q^2_{\rm c}, \omega_{\rm c} ) =0\,, \qquad \frac{\partial F}{\partial \omega} (\q^2_{\rm c}, \omega_{\rm c} ) =0 \, .
\label{c-curve-critical-x}
\end{align}
There are $p>1$ branches of the solution $\omega=\omega(\q^2)$ in the vicinity of the critical point, provided that 
\begin{align}
F(\q^2_{\rm c}, \omega_{\rm c} ) =0\,, \qquad \frac{\partial F}{\partial \omega} (\q^2_{\rm c}, \omega_{\rm c} ) =0 \,, \, \ldots, \,
\frac{\partial^{p} F}{\partial \omega^{p}} (\q^2_{\rm c}, \omega_{\rm c} ) \neq 0 \,.
\label{c-curve-critical-xp}
\end{align}
For example, the origin $(0,0)$ is the critical point (with $p=2$) of the sound hydrodynamic spectral curve \eqref{disp-rel-general-sound}, as discussed in section \ref{hydro-spectral-sound}. If the spectral curves happen to be known exactly or approximately (as will be the case in the holographic models we study below), eqs.~\eqref{c-curve-critical-x} provide an efficient method to find the radii of convergence, without performing the large-$n$ analysis of the expansion coefficients.

When the function $F(\q^2,\omega)$ is a polynomial, the condition \eqref{c-curve-critical-x} means that the equation $F(\q^2,\omega)=0$, where $F(\q^2,\omega)$ is considered as a polynomial in $\omega$ with $\q^2$-dependent coefficients, has multiple roots at $\omega = \omega_{\rm c}$. This is equivalent to the condition that the discriminant of the polynomial $F(\q^2,\omega)$ vanishes. As an example, consider the first-order hydrodynamics of~\cite{landau}, where the spectral curves following from eq.~\eqref{eq:Pwq-1h} are\footnote{ As mentioned in section \ref{hscx}, the expressions for the truncated spectral curves $F^{(k)}$ at each order $k$ of the hydrodynamic derivative expansion are not unique due to the freedom allowed by the ``frame'' choice. Correspondingly, the critical points determined by the approximate spectral curves $F^{(k)}$ depend on the ``frame'' choice. This dependence becomes less and less pronounced with $k$ increasing and disappears in the limit $k\rightarrow \infty$. Thus, although the critical points of the exact spectral curve are ``frame''-independent, the rate of convergence of the approximate location of the critical points to the exact values can be affected by the ``frame'' choice.}
\begin{align}
\,&  F_{\rm shear}^{(1)} (\q^2, \omega) = \omega + i  D  \q^2 =0\,, \label{disp-rel-general-shear-1} \\
\,&  F_{\rm sound}^{(1)} (\q^2, \omega) = \omega^2 + i  \Gamma_{\!s}  \omega  \q^2  - v_s^2  \q^2  =0\,.
\label{disp-rel-general-sound-1}
\end{align}
These equations are simple enough to be solved explicitly: eq.~\eqref{disp-rel-general-shear-1} is solved by $\omega = -i D \q^2$, whereas the solutions of eq.~\eqref{disp-rel-general-sound-1} are
\begin{align}
 \omega^\pm_{\rm sound}(\q^2) = - \frac{i \Gamma_{\!s}}{2}\, \q^2 \pm \sqrt{v_s^2 \q^2 \left( 1 - \frac{\Gamma_{\!s}^2 \q^2}{4 v_s^2}\right)} = 
\pm v_s q - \frac{i \Gamma_{\!s}}{2}\, q^2 + \cdots\,,
\label{sol-sound-1}
\end{align}
where in the expansion we only kept terms quadratic in $q\equiv \sqrt{\q^2}$, since we expect the 
coefficients in front of the higher powers of $q$ to be modified by higher-derivative terms in the hydrodynamic expansion. The series in $q$ on the right-hand side of eq.~\eqref{sol-sound-1} has the radius of convergence
\begin{align}
R^{(1)}_{\rm sound} = \frac{2 v_s}{\Gamma_{\!s}}\,,
\label{rad-sound-1}
\end{align}
determined by the branch points of the square root in eq.~\eqref{sol-sound-1} or, equivalently, by the zeros of the discriminant%
\footnote{
Similar methods have been used in spectroscopy \cite{Bhattacharya2006DetectingLC}.
}  
$(-\Gamma_{\!s}^2 \q^4 + 4 v_s^2 \q^2)$ of the polynomial \eqref{disp-rel-general-sound-1}. Since first-order 
hydrodynamics  can only be trusted for $|q|\ll 2 v_s/\Gamma_{\!s} = R^{(1)}_{\rm sound}$, 
the result \eqref{rad-sound-1} is only an approximation.  Alternatively, applying the condition \eqref{c-curve-critical-x} to the spectral curve 
\eqref{disp-rel-general-sound-1}, we find 
\begin{align}
\q^2_{\rm c} = \frac{4 v_s^2}{\Gamma_{\!s}^2}\,, \qquad \omega_{\rm c}^{\rm (sound)} = - i \frac{2 v_s}{\Gamma_{\!s}}\,,
\label{rad-sound-1a}
\end{align}
which  coincides with \eqref{rad-sound-1}. In what follows, we will be studying models where the convergence radii of the small-$\q$ expansions can be determined from the exact response functions of the theory, without resorting to the derivative expansion of hydrodynamics.


\section{A holographic model with translation symmetry breaking}
\label{DG-model}
To illustrate the methods discussed in section \ref{hydro-series-convergence},  we consider the
 holographic model with translation symmetry breaking  \cite{Andrade:2013gsa}, studied, in 
particular, in ref.~\cite{Davison:2014lua}. The model is a bottom-up gravity construction in $4d$ describing a hypothetical dual $2+1$-dimensional QFT with broken translational invariance. In the context of the present paper, the significance of the construction discussed in ref.~\cite{Davison:2014lua} lies in the fact that it provides  exact analytic formulae for the current and energy-momentum correlator two-point functions at a special, self-dual symmetry point (see section 4 of ref.~\cite{Davison:2014lua} for details). In particular, among the poles of the correlation function, one finds a gapless excitation whose dispersion relation is known analytically, possibly a unique example in holography.

The bulk action of the model is given by \cite{Andrade:2013gsa,Davison:2014lua} (see also ref. \cite{Blake:2018leo})
\begin{align}
S = \int \, d^4x \sqrt{-g} \left( R - 2\Lambda - \frac{1}{2} \sum_{i=1}^2 \partial^\mu\phi_i \partial_\mu \phi_i - \frac{1}{4} F_{\mu\nu} F^{\mu\nu}\right),
\label{rb-action}
\end{align}
where $\Lambda = -3/L^2$ (we  set $L=1$ in the following). The background solution of interest involves the AdS-Schwarzschild black brane with translational invariance broken by the linear dilaton fields, and a vanishing Maxwell field.

First, we recast the setup of ref.~\cite{Davison:2014lua} in the form more convenient for our purposes, having in mind the variables used in  ref.~\cite{Kovtun:2005ev}. The metric at the special symmetry point has the form (see section 4 of  ref.~\cite{Davison:2014lua})
\begin{align}
ds^2 = -\frac{r_0^2}{u} f(u)\, dt^2 + \frac{r_0^2}{u}\left( dx^2 +dy^2\right) + \frac{du^2}{4 f u^2}\,,
\label{rb-metric-x}
\end{align}
where $f=1-u$, and we have used the coordinate $u=r_0^2/r^2$. The horizon is located at $u=1$ and the boundary at $u=0$. The Hawking temperature of the background is $T=r_0/2\pi$. The dilaton fields are given by $\phi_1 = \sqrt{2} r_0 x$ and $\phi_2 = \sqrt{2} r_0 y$. They will not play any role in the following.

Considering the Maxwell field fluctuations in the background \eqref{rb-metric-x}, coupled to the current operator on the boundary, one finds the equations of 
motion 
\begin{align}
a_t'' + \frac{1}{2u} a_t' - \frac{1}{4 u f}\left( \qfr^2 a_t + \wfr  \qfr \, a_x\right)&=0\,, \\
a_x'' + \frac{1-3u}{2uf} a_x' + \frac{1}{4 u f^2}\left( \qfr \wfr  a_t + \wfr^2  a_x\right)&=0\,, \\
\wfr a_t' +\qfr f a_x' &=0\,,
\end{align}
where $\wfr = \omega/2\pi T = \omega/r_0$, $\qfr = k/2\pi T = k/r_0$, and the momentum $k$ is directed along $x$. For the gauge-invariant variable (the longitudinal component of the electric field) $E_x = \qfr a_t + \wfr a_x$ \cite{Kovtun:2005ev}, the equation of motion reads
\begin{align}
E_x'' + \frac{\wfr^2(1-3u)-\qfr^2 f^2}{2 u f (\wfr^2 -\qfr^2 f)}\, E_x' + \frac{\wfr^2 -\qfr^2 f}{4 u f^2}\, E_x =0\,.
\label{rb-eom-ex-x}
\end{align}
The equation \eqref{rb-eom-ex-x} has 4 regular singular points (located at $u=0,1,1-\wfr^2/\qfr^2,\infty)$ and thus is of the Heun type. The indices of this equation are, respectively,
\begin{align}
\,u&=0:   &&0,1/2\,\\
\,u&=1:  &&\pm i\wfr/2 \, \\
\,u&=1 - \frac{\wfr^2}{\qfr^2}:& &0,2\, \\
\,u&=\infty: && \frac{1}{4} \left(  1 \pm \sqrt{1-4\qfr^2}\right)\,.
\end{align}
Note that the local 
solution at $u=0$ does not contain logarithms. The exact solution to eq.~\eqref{rb-eom-ex-x} can be written as
\begin{align}
E_x (u) = \frac{4 u f}{\qfr} G'(u) + \frac{2 f}{\qfr} G(u)\,,
\end{align}
where $G\equiv a_t'$ is the solution of the hypergeometric equation\footnote{ The very fact that an exact solution to the Heun equation can be found via the supplementary hypergeometric equation is rather curious and may warrant further reflection.}
\begin{align}
G'' - \frac{5 u-3}{2 u f} G' + \frac{\wfr^2 -\qfr^2 f - 2 f}{4 u f^2}\, G=0\,,
\label{rb-eom-G}
\end{align}
obeying the incoming wave boundary condition at $u=1$:
\begin{align}
G = \frac{\qfr}{2 i\wfr} x^{-i \wfr/2} \,_2 F_1 \left( \frac{ 3 - 2 i \wfr -\sqrt{1-4 \qfr^2}}{4}, \frac{
  3 - 
2 i \wfr +\sqrt{1-4 \qfr^2}}{4}; 1-i\wfr; x\right)\,,
\end{align}
where $x=1-u$ and the normalisation is chosen to ensure $E_x \rightarrow (1-u)^{-i \wfr/2}\left( 1+\cdots\right)$ 
at $u\rightarrow 1$.

\subsection{The exact spectral curve}

For the boundary value of the electric field, we find 
\begin{align}
E_x (u=0; \qfr^2, \wfr) \equiv F(\qfr^2, \wfr) = \frac{2 \sqrt{\pi} \ \Gamma (-i\wfr) }{ \Gamma \left[ A_+ (\qfr^2, \wfr )\right]  
\Gamma \left[ A_- (\qfr^2, \wfr )\right]  }\,,
\label{rb-sol-0-x}
\end{align}
where 
\begin{align}
A_\pm = \frac{1}{4}\left( 1 \pm \sqrt{1-4 \qfr^2} - 2 i \wfr \right)\,.
\end{align}
The condition $E_x(u=0)=0$ determines the quasinormal modes and thus the poles of the corresponding current-current correlators \cite{Son:2002sd,Kovtun:2005ev}. One can also write down the explicit analytic expressions for the correlators themselves \cite{Davison:2014lua}, but this will not be necessary: the expression $F(\qfr^2, \wfr) =0$, where $F(\qfr^2, \wfr)$ is given by eq.~\eqref{rb-sol-0-x}, is the exact spectral curve containing all information about the poles of the two-point function. There are two sequences of quasinormal frequencies
\begin{align}
\wfr_n^{\pm} (\qfr^2) = - i \left( 2 n_\pm +\frac{1}{2} \right) \pm \frac{i}{2} \sqrt{1-4 \qfr^2}\,, \qquad n_\pm=0,1,2\ldots\,.
\label{qnm-modes-w-x}
\end{align}
The solutions $E_{x,n_\pm}^{\pm}(u)$ themselves (the quasinormal modes) have the form
\begin{align}
E^\pm_{x,n_\pm} (u) = \sqrt{u}\, \left(1-u \right)^{-n_\pm-\frac{1}{4} \pm \frac{1}{4}\sqrt{1-4 \qfr^2}}\, P^{\pm}_{n_\pm}(u,\qfr^2)\,,
\label{qnm-modes-solutions-x}
\end{align}
where $P^{\pm}_{n_\pm}(u,\qfr^2)$ are polynomials of degree $n_\pm$ in $u$ with $\qfr^2-$dependent coefficients. In particular,
\begin{align}
E^\pm_{x,0} (u) = \sqrt{u}\, \left(1-u \right)^{-\frac{1}{4} \pm \frac{1}{4}\sqrt{1-4 \qfr^2}}\,.
\label{qnm-modes-x}
\end{align}
Note that apart from the prefactors determined by the indices, the solutions \eqref{qnm-modes-solutions-x} are polynomials, with $P_0^\pm=1$. We are especially interested in the gapless mode 
\begin{align}
\wfr = \wfr_0^+ \,&= -\frac{i}{2}\left( 1 - \sqrt{1-4 \qfr^2}\right) =\frac{i}{2} \sum_{n=1}^\infty (-1)^n {1/2\choose n} \left( 4 \qfr^2 \right)^n  \nonumber \\
\,&=  -i \qfr^2 - i \qfr^4 - 2 i  \qfr^6  - 5 i \qfr^8 - 14 i \qfr^{10} + \cdots\,.
\label{gapless-mode-x}
\end{align}
The  power series in the second line of eq.~\eqref{gapless-mode-x} converges in the circle $|\qfr^2|<1/4$, due to the branch point singularity of the function at $\qfr^2=1/4$ evident from eq.~\eqref{gapless-mode-x}. The same conclusion can be obtained by analysing critical points of the spectral curve  \eqref{rb-sol-0-x}. Indeed, the critical points are determined by the equations \eqref{c-curve-critical-x} whose solutions are $\wfr_{n_+}^+(\qfr^2) = \wfr_{n_-}^-(\qfr^2)$, i.e.
\begin{subequations}
\begin{align}
\,& \qfr^2_{\rm c} = \frac{1}{4} - \left( n_+-n_-\right)^2,\\
\,& \wfr_{\rm c} = -\frac{i}{2} \left[ 1 + 2 \left(n_++n_-\right)\right].
\label{critical-pts-dgm}
\end{align}
\end{subequations}
It is also clear that $\partial^2_\wfr F(\qfr^2_{\rm c}, \wfr_{\rm c}) \neq 0$ and thus there are two branches of the spectral curve emerging at each critical point. Put simply, the critical points occur when the two quasinormal frequencies in eq.~\eqref{qnm-modes-w-x} collide. This happens for real $\qfr$ in case of  $n_+=n_-$, and for purely imaginary $\qfr$ for  $n_+ \neq n_-$. The critical point closest to the origin $\qfr^2=0$ is at $\qfr^2=1/4$ (it corresponds to the collision of the modes $\wfr_0^+$ and $\wfr_0^-$). This value determines the radius of convergence of the series in eq.~\eqref{gapless-mode-x}.

Exact spectral curves are rare: in addition to \eqref{rb-sol-0-x}, we are aware of only one example 
(involving the exact R-current two-point correlators in ${\cal N}=4$ SYM theory in appropriate limit \cite{Myers:2007we}) for a QFT in dimension higher than $1+1$, and even in that case it is only known exactly for $\qfr^2 = 0$:
\begin{align}
F_R (\qfr^2 =0,\wfr) = 2^{-\frac{(1+i)\wfr}{2}}\,
 \frac{\Gamma \left[1-i\wfr\right] }{\Gamma \left[ 1 
- \frac{(1+i)\wfr}{2}\right] \Gamma \left[ 1 + \frac{(1-i)\wfr}{2}\right]}\,.
\label{rob}
\end{align}
One can use the Weierstrass decomposition 
\begin{align}
\frac{1}{\Gamma (z)} = z e^{\gamma z} \prod_{n=1}^\infty \left( 1+\frac{z}{n}\right) e^{-\frac{z}{n}}\,,
\end{align}
where $\gamma$ is the Euler–Mascheroni constant,
to write eqs.~\eqref{rb-sol-0-x}, \eqref{rob} as  infinite products: this ilustrates explicitly why the critical points 
determined by the condition \eqref{c-curve-critical-x} correspond to multiple roots of (infinite order) polynomials.

\subsection{The hydrodynamic approximation to the spectral curve}
\begin{figure}[h!]
\centering
\includegraphics[width=0.7\textwidth]{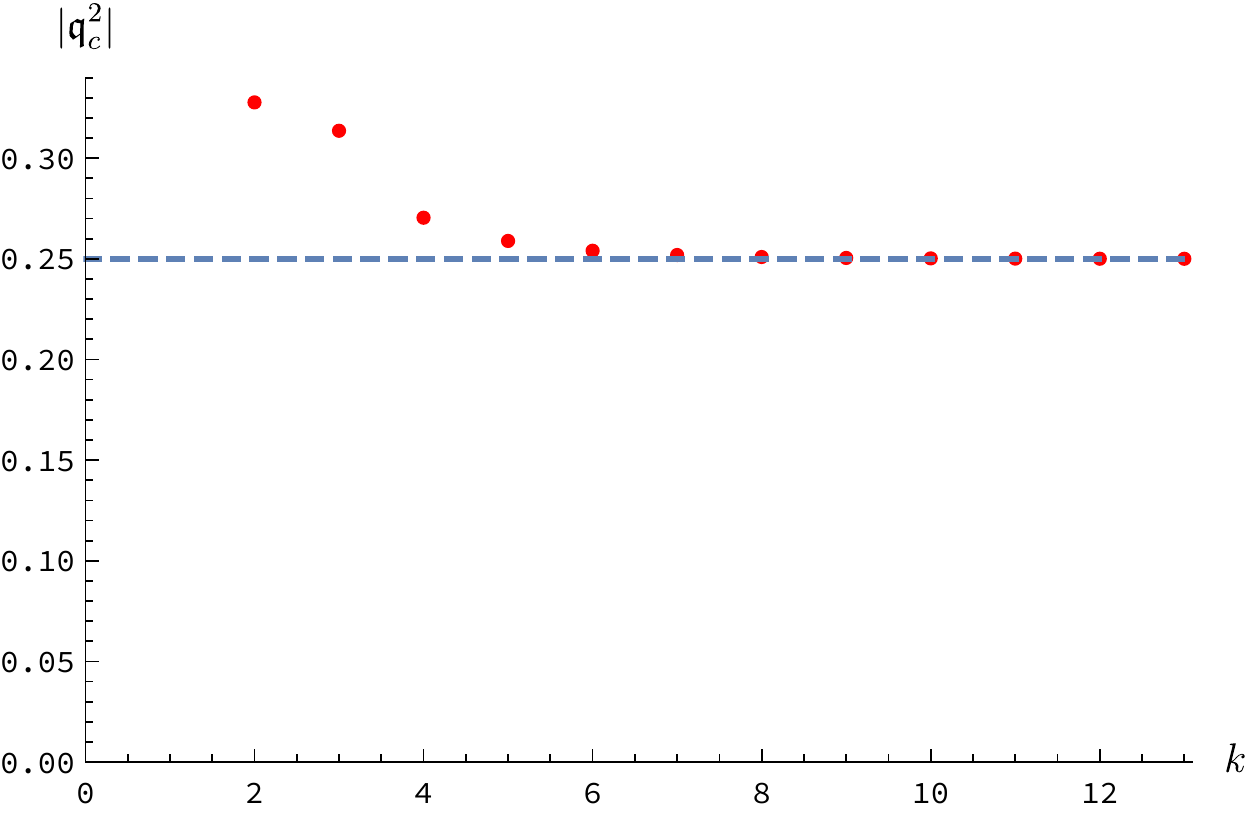}
\caption{ {\small The approximations to the exact position of the critical point $|\qfr_{\rm c}^2|=1/4$ in the holographic model with broken translation symmetry determined from the hydrodynamic algebraic curves $F_k(\qfr^2,\wfr)=0$ as a function of $k$. }}
\label{fig-cpts-b}
\end{figure}

The expansion of the expression $\wfr F(\qfr^2, \wfr)$  for small $\wfr$, $\qfr$ (assuming the scaling $\wfr\rightarrow \epsilon\wfr$, 
$\qfr^2\rightarrow \epsilon\qfr^2$ with $\epsilon \to 0$) truncated at order $\wfr^k$, $\qfr^{2k}$ is a 
polynomial $F_k(\qfr^2,\wfr)$, with
\begin{align}
\,& F_1(\qfr^2,\wfr) = \wfr + i\qfr^2\,, \\
\,& F_2(\qfr^2,\wfr) = \wfr + i\qfr^2 - i\left[ \wfr^2 \ln{2} - \qfr^4 \left( 1 - \ln{2}\right)\right] \,, \\
\,& F_3(\qfr^2,\wfr) = \wfr + i\qfr^2 - i\left[ \wfr^2 \ln{2} - \qfr^4 \left( 1 - \ln{2}\right)\right] \nonumber \\
\,& -\frac{i}{12}\left[
\qfr^6 \left( \pi^2  - 6 ( \ln{2} -2)^2\right) - 6 i \qfr^4 \wfr \ln^2 2 - 
   6 i \wfr^3 \ln^2 2 + \qfr^2 \wfr^2 (\pi^2 - 6 \ln^2 2)\right]\,,
\label{gapless-mode-exp-x-x}
\end{align}
and so on. Using this expansion to solve the equation $\wfr  F(\qfr^2, \wfr) =0$ for $\wfr$ {\it perturbatively} in $\qfr^2$, one reproduces the series in eq.~\eqref{gapless-mode-x}. The equation $F_k(\qfr^2,\wfr)=0$ defines the hydrodynamic spectral curve of order $k$ as discussed in section \ref{rel-hydro-det}. At each order, the critical points are determined by  the condition \eqref{c-curve-critical-x}. In fig.~\ref{fig-cpts-b}, we plot the corresponding value $|\qfr_{\rm c}^2|$ for each of the spectral curves $F_k(\qfr^2,\wfr)=0$ for $k=2,3,\ldots,13$. The resulting points converge  rapidly to the exact value $|\qfr_{\rm c}^2|=1/4$.

\subsection{Pole-skipping in the full response functions}
Finally, we comment on the relationship between the critical point defining the hydrodynamic series radius of convergence and the  pole-skipping point in the holographic model with broken translation symmetry, which occurs at $|\qfr_*^2|=2>|\qfr_{\rm c}^2|$  \cite{Blake:2018leo}. Because gapless excitations in the current and the energy density correlators have the same dispersion relation, we can directly use $\wfr(\qfr)$ from eq.~\eqref{gapless-mode-x} to discuss both the charge and the energy sectors. At the pole-skipping 
point, the hydrodynamic series stated in the second line of eq.~\eqref{gapless-mode-x} diverges 
but  can be resummed using the 
Borel transform\footnote{ An alternative analytic continuation can be provided e.g. by the 
Mittag-Leffler summation. See e.g. ref. \cite{hardy1967divergent}.} 
\begin{align}
\CB \wfr (\qfr ) &= \frac{i}{2} \sum_{n=1}^\infty \frac{(-1)^n}{n!} {1/2\choose n}  \left( 4\qfr^2 \right)^n  \\
&= - \frac{i}{2} \left[ 1 - e^{2\qfr^2 }  \left( (1-4\qfr^2) I_0 (2\qfr^2) + 4\qfr^2 I_1(2\qfr^2)    \right)   \right]\, ,
\end{align} 
where $I_n(x)$ is the modified Bessel function. Since this is a series in $\qfr^2$ and not $\qfr$, the corresponding 
Borel integral has the form 
\begin{align}\label{DG-Borel-Integral-x}
\mathbf{\Omega}(\qfr ) = \int_0^\infty d t \, e^{-t} \CB \wfr (\qfr \sqrt{t}) = \CI_1 + \CI_2 + \CI_3 + \CI_4 \,,
\end{align}
where
\begin{align}
\CI_1 &= - \frac{i}{2} \int_0^\infty d t \, e^{-t} = - \frac{i}{2} \, , \\
\CI_2 &= \frac{i}{2} \int_0^\infty dt \, e^{-t}e^{2\qfr^2  t} I_0 (2 \qfr^2 t) = \frac{i}{2} \frac{1}{\sqrt{1-4\qfr^2}} \,,\\
\CI_3 &= - i  \int_0^\infty dt \, e^{-t}  ( 2\qfr^2  t ) e^{2\qfr^2 t} I_0 (2\qfr^2 t) = - 2 i \qfr^2 \frac{(1-2\qfr^2)}{(1-4\qfr^2)^{3/2}} \, , \\
\CI_4 &= i  \int_0^\infty dt \,e^{-t}  (2\qfr^2 t ) e^{2\qfr^2 t} I_1 (2\qfr^2 t) = 4 i \frac{ \qfr^2 \sqrt{\qfr^4} }{(1-4\qfr^2)^{3/2}} \,,
\end{align}
each with their respective region of $\qfr$ for which the integral converges. Together, we find that the Borel integral representation \eqref{DG-Borel-Integral-x} of the series converges for $\qfr \in \mathbb{C}$ in the region defined by the function $\CC(\qfr)$:
\begin{align}
\CC(\qfr) \equiv \re[\qfr^2]  < 1/4 \, .
\end{align}
This is a significant improvement over the convergence region of the hydrodynamic power series, 
$| \qfr^2 | < |\qfr_{\rm c}^2|=1/4$. In particular, the Borel series is well suited for studying purely imaginary $\qfr$. By writing $\qfr = i \kfr$, with $\kfr \in \mathbb{R}$, we see that   
\begin{align}
\CC(i \kfr) = \re[- \kfr^2] = - \re[\kfr^2] \leq 0 <1/4 \, .
\end{align}
Hence, the Borel representation of the series converges for all values of imaginary $\qfr$, including $\wfr(\qfr)$ at the chaos point $\wfr_*(\qfr_* = \sqrt{2} i) = i$. Finally, for $\qfr$ such that $\CC(\qfr)  < 1/4$, it is easy to check that the sum of four terms in \eqref{DG-Borel-Integral-x} indeed reproduces the full dispersion relation \eqref{gapless-mode-x}.


\section{Response functions in strongly coupled ${\cal N}=4$ SYM theory}
\label{sec:SYM}
In this section, we use holographic duality to find the spectral curves, determine the radii of convergence of hydrodynamic series and analyse the pole-skipping phenomenon in the three channels of the response function of the energy-momentum tensor in the ${\cal N}=4$ $SU(N_c)$ SYM theory at infinite 't Hooft coupling and infinite $N_c$. The details of the duality are well known, and the necessary ingredients 
can be found e.g.~in refs.~\cite{Son:2007vk,Policastro:2002se,Kovtun:2005ev}. In short, holography reduces the study of the response functions to the analysis of the fluctuations of the dual gravitational background involving a black hole with  translationally invariant horizon---the AdS-Schwarzschild black brane.

The equations of motion describing fluctuations of the gravitational background dual to finite-temperature ${\cal N}=4$ SYM theory are of the Heun type \cite{Starinets:2002br}, and the exact analytic solution for the spectral curve similar to eq.~\eqref{rb-sol-0-x} is not available. The equations can be solved perturbatively and analytically  in $\wfr\ll 1$, $\qfr \ll 1$, as was done in refs.~\cite{Policastro:2002se,Policastro:2002tn}, thereby giving a hydrodynamic approximation to the spectral curve, or numerically, for arbitrary $\wfr$ and $\qfr$, along the lines of ref.~\cite{Kovtun:2005ev}. In this section, we consider and compare these two approaches.

\subsection{Shear mode: hydrodynamic approximation to the spectral curve}
\label{shear-n=4-hydro}
We start with the analysis of the shear channel of the energy-momentum tensor response function \cite{Kovtun:2005ev}. In the hydrodynamic approximation, the spectral curve can be found analytically from the boundary value of the solution to the ODE obeyed by one of the components of the shear perturbation in dual gravity (see section 6.2 of ref.~\cite{Policastro:2002se} for details):
\begin{align}
 G''(u) - \left(\frac{2u}{f} - \frac{i \wfr}{1-u}\right)\, G'(u) + \frac{1}{f} \left( 2 + \frac{i \wfr}{2} - \frac{\qfr^2}{u} + \frac{\wfr^2 [4-u(1+u)^2]}{4 u f}\right) G(u) =0\,,
\label{shear-old-r}
\end{align}
where $G$ is regular at $u=1$. Rescaling $\wfr\rightarrow \lambda^2 \wfr$ and $\qfr^2 \rightarrow \lambda^2 \qfr^2$, assuming $\lambda \ll 1$,
and looking for a perturbative solution of the form
\begin{align}
 G (u) = \sum_{n=0}^\infty \lambda^{2 n} G_n (u)\,,
\label{Shear_old_perturb_r}
\end{align}
we find the following equation for coupled components $G_n$, $G_{n-1}$, $G_{n-2}$:
\begin{align}
  G_n '' & - \frac{2 u}{f} G_n ' + \frac{2}{f} G_n  \nonumber \\
 & + \frac{i \wfr}{1-u} G_{n-1}' + \frac{i \wfr}{2 f} G_{n-1}  - \frac{\qfr^2}{u f}  G_{n-1}\nonumber \\
 &
 +   \frac{\wfr^2 [4-u(1+u)^2]}{4 u f^2}  G_{n-2} = 0\,,
\label{Shear_old_reccur-r}
\end{align}
where $G_n =0$ for $n<0$, and we can set $G_0(1)=1$, $G_i (1)=0$, $i\geq 1$, without loss of generality.\footnote{ See Appendix 
C of ref.~\cite{Grozdanov:2016fkt}.} The explicit formulae for $G_0(u)$, $G_1(u)$ and $G_2(u)$ obeying the boundary conditions at $u=1$ are written in Appendix \ref{app-cc}. The solution of the homogeneous equation 
\begin{align}
g '' - \frac{2 u}{f}\,  g ' + \frac{2}{f}\,  g  =0
 \label{Shear_old_homog}
\end{align}
is given by $g = C_1\, g_1 (u) + C_2\,  g_2 (u)$, where
\begin{align}
  g_1 &= u\,, \\
g_2 &= \frac{u}{2} \ln{\frac{1+u}{1-u}} -1\,.
 \label{Shear_old_sol}
\end{align}
Note that the Wronskian is $W(g_1,g_2)=1/f$. Then one can write the following expression for $G_n$, $n\geq 2$:
\begin{align}
G_n(u) & = g_1(u) \int_{u}^1 g_2(t) f(t) {\cal F}_n(t) dt -  g_2(u) \int_{u}^1 g_1(t) f(t) {\cal F}_n(t) dt \nonumber \\
& + C_1\, g_1 (u) + C_2\,  g_2 (u)\,,
\end{align}
where
\begin{align}
 {\cal F}_n(u) = -  \frac{i \wfr}{1-u} G_{n-1}' - \frac{i \wfr}{2 f} G_{n-1} + \frac{\qfr^2}{u f}  G_{n-1}
 -    \frac{\wfr^2 [4-u(1+u)^2]}{4 u f^2}  G_{n-2}\,.
 \label{curlyf-r}
\end{align}
Boundary conditions at $u=1$ (regularity of $G_n(u)$ at $u=1$ and $G_n(1)=0$ for $n>0$) require $C_1=0$, $C_2=0$. Thus, we have the equation
determining $G_n$ from $G_{n-1}$ and $G_{n-2}$,
\begin{align}
G_n(u) & =  u  \int_{u}^1 (1-t^2)  \left( \frac{t}{2} \ln{\frac{1+t}{1-t}} -1 \right)  {\cal F}_n(t) dt \nonumber \\
& -  \left( \frac{u}{2} \ln{\frac{1+u}{1-u}} -1 \right) \int_{u}^1 t (1-t^2)  {\cal F}_n(t) dt \,,
\end{align}
where ${\cal F}_n$ is given by eq.~\eqref{curlyf-r}. This can be written as
\begin{align}
G_n(u) &=  \frac{u}{2}  \int_{u}^1   t (1-t^2)   \ln{\left[ \frac{1+t -u -u t}{1-t+u-u t}\right]}  {\cal F}_n(t)\, dt \nonumber \\
& +  \int_{u}^1 (t-u) (1-t^2)  {\cal F}_n(t) dt \,,
\end{align}
or
\begin{align}
G_n(u) = \,&  \int_{u}^1  \left\{  \frac{u t}{2}  \ln{\left[ \frac{1+t -u -u t}{1-t+u-u t}\right]}  + t -u \right\} (1-t^2) \, {\cal F}_n(t)\, dt \,.
\end{align}
Note that ${\cal F}_n(t)\sim 1/(1-t)$ at $t\rightarrow 1$, and so all integrals converge. In particular,
\begin{align}
G_n(0) =  \int_{0}^1 t (1-t^2)  {\cal F}_n(t) dt \,.
\end{align}
The explicit expressions for $G_i(0)$, with $i=1,2,3,4$, are written in Appendix \ref{app-cc}. The results for $G_0(0)$, $G_1(0)$ and  $G_2(0)$ coincide with those in ref.~\cite{Baier:2007ix}. The results for $G_3(0)$ and  $G_4(0)$ are new. The condition $G(0)=0$ at this order in the hydrodynamic expansion defines the algebraic curve
\begin{align}
\,& F(\qfr^2,\wfr) = - i \wfr +\frac{\qfr^2}{2} \nonumber  \\
 \,& + \frac{\qfr^4}{4} - \frac{i \wfr \qfr^2 \ln{2}}{4} + \frac{\wfr^2 \ln{2}}{2} \nonumber \\
 \,& + i \wfr^3 \left( \frac{\pi^2}{24} +\ln{2} - \frac{3}{8} \ln^2 2\right) +\qfr^6 \left( \frac{\ln{2}}{4} -\frac{1}{8}\right)  + i \wfr \qfr^4 
\left( \frac{1}{4}  - \frac{\ln 2}{8}\right) \nonumber \\
\,& +   \qfr^2 \wfr^2 \left( \frac{\pi^2}{48} -\frac{\ln{2}}{2}  - \frac{\ln^2 2}{16} \right)+ \qfr ^8 \left(-\frac{1}{16}+\frac{\pi^2}{64}-\frac{\ln{2}}{8}\right)-\qfr^4 \wfr^2 \left(\frac{\pi^2}{96} + \left(12 -7\ln{2}\right)\frac{\ln{8}}{96}\right)\nonumber \\ \,& -i \qfr^6 \wfr\left(\frac{\pi^2}{96} +\left(-5+\ln{4}\right)\frac{\ln{64}}{96}\right) + \wfr^{4}\left(\left(24-5\ln{2}\right)\frac{\ln^{2}{2}}{48}+\frac{\pi^2}{48}\left(-4+\ln{8}\right)-\frac{1}{2}\zeta(3)\right) \nonumber \\ \,& +i \qfr^{2}\wfr^{3}\left(-\frac{\pi^{2}\ln{2}}{96}+\left(-24+\ln{2}\right)\frac{\ln^{2} {2}}{96}+\frac{3}{16}\zeta(3)\right) =0\,.
\label{shear-al-curve}
\end{align}
Here, $\zeta(z)$ is the Riemann zeta function. Solving eq.~\eqref{shear-al-curve} perturbatively in $\qfr^2$, one finds the dispersion relation for the shear mode
\begin{align}
\wfr  \,&= - \frac{i}{2}\,  \qfr^2 - \frac{i (1-\ln{2})}{4}\, \qfr^4 - \frac{i (24 \ln^2 2 - \pi^2)}{96}\, \qfr^6 \nonumber \\ \,& - \frac{i(2\pi^2 (5\ln{2}-1)-21\zeta(3)-24\ln{2}[1+\ln{2}(5\ln{2}-3)]    )}{384}  \qfr^{8}       \,+ O(\qfr^{10})\,.
\label{mkm}
\end{align}
The first two terms in \eqref{mkm} agree with the ones obtained in refs.~\cite{Policastro:2002se} and \cite{Baier:2007ix}, respectively.

\begin{figure}[h!]
\centering
\includegraphics[width=0.7\textwidth]{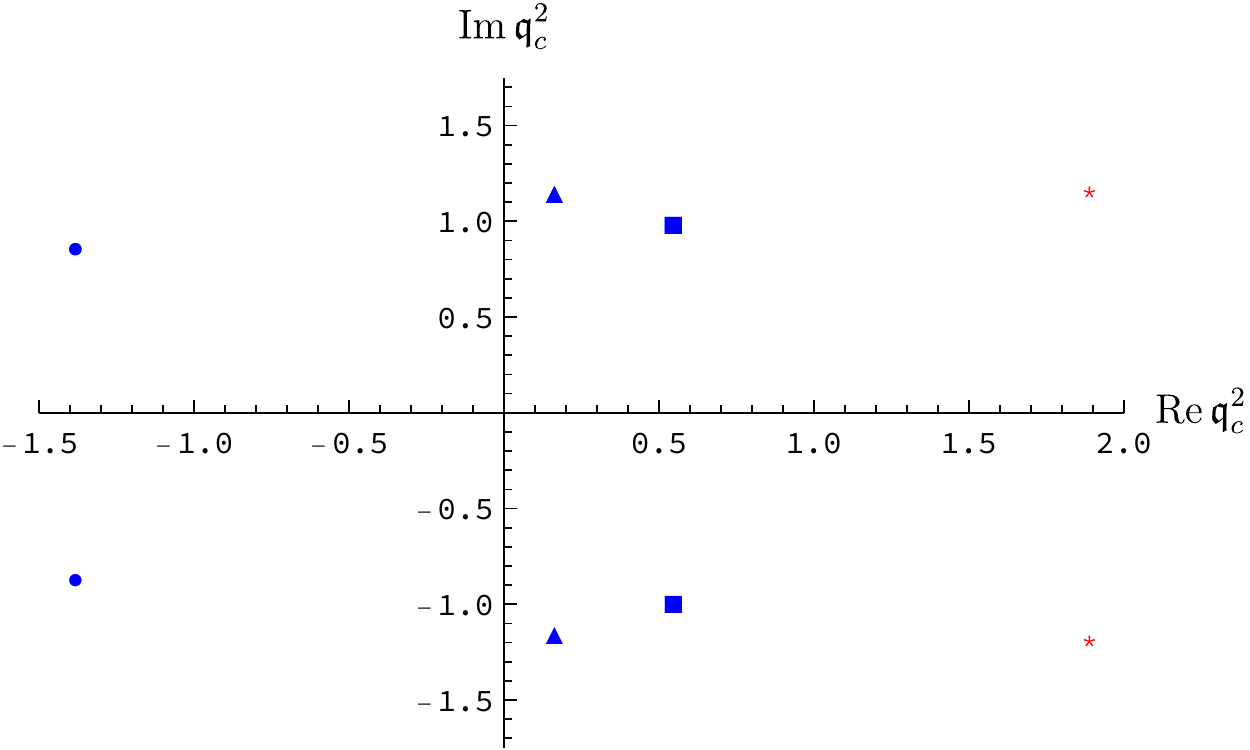}
\caption{{\small Approximations to the position of the critical point in the complex $\qfr_{\rm c}^2$  plane  in the shear channel of the ${\cal N}=4$ SYM theory arising from the order $k$ hydrodynamic polynomials   $F_k(\qfr^2,\wfr)=0$ (circles for $k=2$, triangles for $k=3$, squares for $k=4$). The ``exact'' position $\qfr_{\rm c}^2 \approx 1.8906469 \pm 1.1711505 i$ (see section \ref{shear-lc}) is marked by red asterisks.}}
\label{fig-cpts-b-shear}
\end{figure}

Encouraged by the success of finding the critical point in  the holographic model with broken translation symmetry in section \ref{DG-model}, we can apply the condition \eqref{c-curve-critical-x} to the polynomial \eqref{shear-al-curve}. Truncating \eqref{shear-al-curve} at linear (in $\wfr$ and $\qfr^2$), quadratic, cubic and quartic order, correspondingly, we obtain the algebraic curves $F_k(\qfr^2,\wfr)=0$ for $k=1,2,3,4$. Using the condition \eqref{c-curve-critical-x}, we find that there are no solutions at $k=1$, whereas for $k=2,3,4$, we have
\begin{align}
& k=2: & \qfr_{\rm c}^2 &\approx -1.380398 \pm 0.865925 i \,, & \wfr_{\rm c} &\approx   \pm
 0.216481+1.097596 i\, , \\
& k=3:  &\qfr_{\rm c}^2 &\approx 0.164953 \pm 1.151910 i \,, & \wfr_{\rm c} &\approx  \pm 0.735771 +0.164407 i  \,,\\
& k=4:  & \qfr_{\rm c}^2 &\approx 0.548173 \pm 0.988705 i \,, & \wfr_{\rm c} &\approx \pm 0.706672   -0.140043  i \,.
\end{align}
In fig.~\ref{fig-cpts-b-shear}, the $k=2,3,4$  approximations are shown in the complex plane of $\qfr_{\rm c}^2$ together with the ``exact'' value $\qfr_{\rm c}^2 \approx 1.8906469 \pm 1.1711505 i$ obtained via the quasinormal level-crossing method 
 (see section \ref{shear-lc}). In contrast with the  holographic model with broken translation symmetry (see fig.~\ref{fig-cpts-b}), the convergence is slow (admittedly, we only have 3 points in the present case). However, we learn an important lesson, namely, that the critical point can be located at a generic complex value of $\qfr^2$. In the next section, we use a more efficient method of the quasinormal modes level-crossing to determine the critical points.

\subsection{Shear mode: full spectral curve}
\label{shear-lc}
To compute the spectral curve numerically for all values of $\qfr^2$ and $\wfr$, it is more convenient to use the ODE obeyed by the gauge-invariant perturbations. The gauge-invariant shear mode gravitational perturbations of the AdS-Schwarzschild black brane are described by the  function $Z_1(u)$, where $u$ is the radial coordinate ranging from $u=0$ (asymptotic boundary) to $u=1$ (event horizon) 
\cite{Kovtun:2005ev}. The function $Z_1(u)$ obeys the equation 
\be
    Z_1'' - \frac{(\wfr^2 - \qfr^2 f)f - u\wfr^2 f'}{uf(\wfr^2-\qfr^2 f)} Z_1'
    + \frac{\wfr^2 - \qfr^2 f}{u f^2} Z_1 = 0\,,
\label{eq:Z1eqn}
\ee
where $f(u)=1-u^2$, and  the solution must obey the incoming wave boundary condition at the horizon, $Z_1(u)\sim(1-u)^{-i\wfr/2}$ as $u\to1$ \cite{Son:2002sd,Kovtun:2005ev}. The full shear spectral curve is then given by 
\begin{align}
\,& F(\qfr^2,\wfr) =Z_1 (u=0; \qfr^2,\wfr) =0\,.
\label{full-shear-spectral-curve}
\end{align}
The spectral curve \eqref{full-shear-spectral-curve} can be determined numerically by e.g. using $N$ terms of the Frobenius series solution at $u=1$ \cite{Kovtun:2005ev}.
\begin{figure*}[t]
\centering
\includegraphics[width=0.45\textwidth]{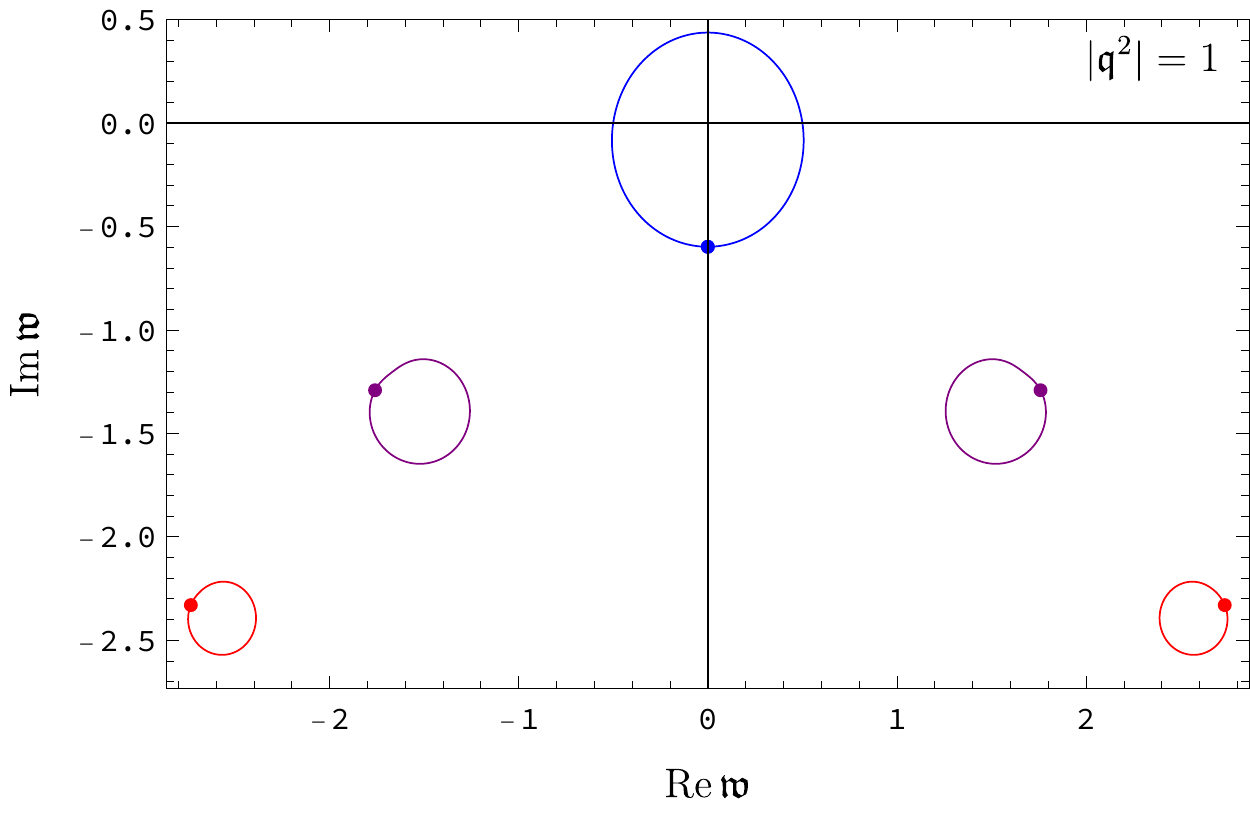}
\hspace{0.05\textwidth}
\includegraphics[width=0.45\textwidth]{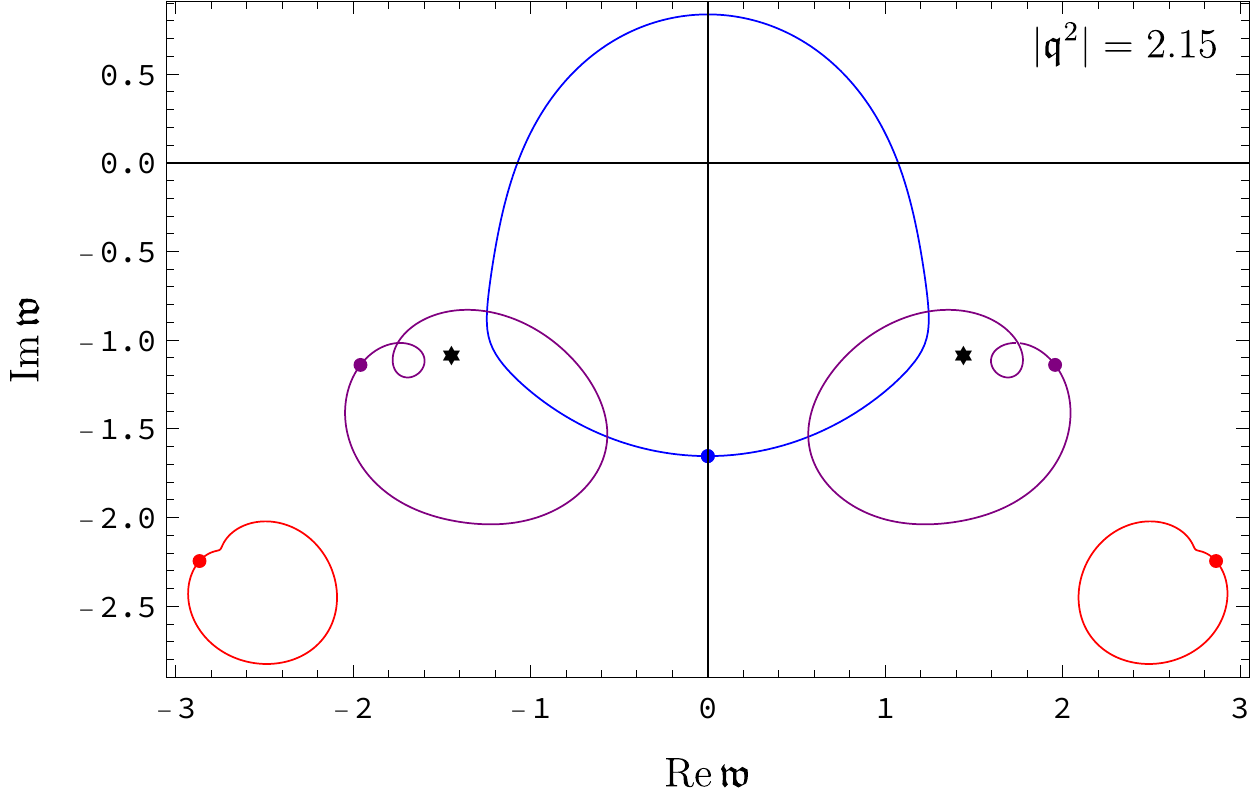}
\\
\includegraphics[width=0.45\textwidth]{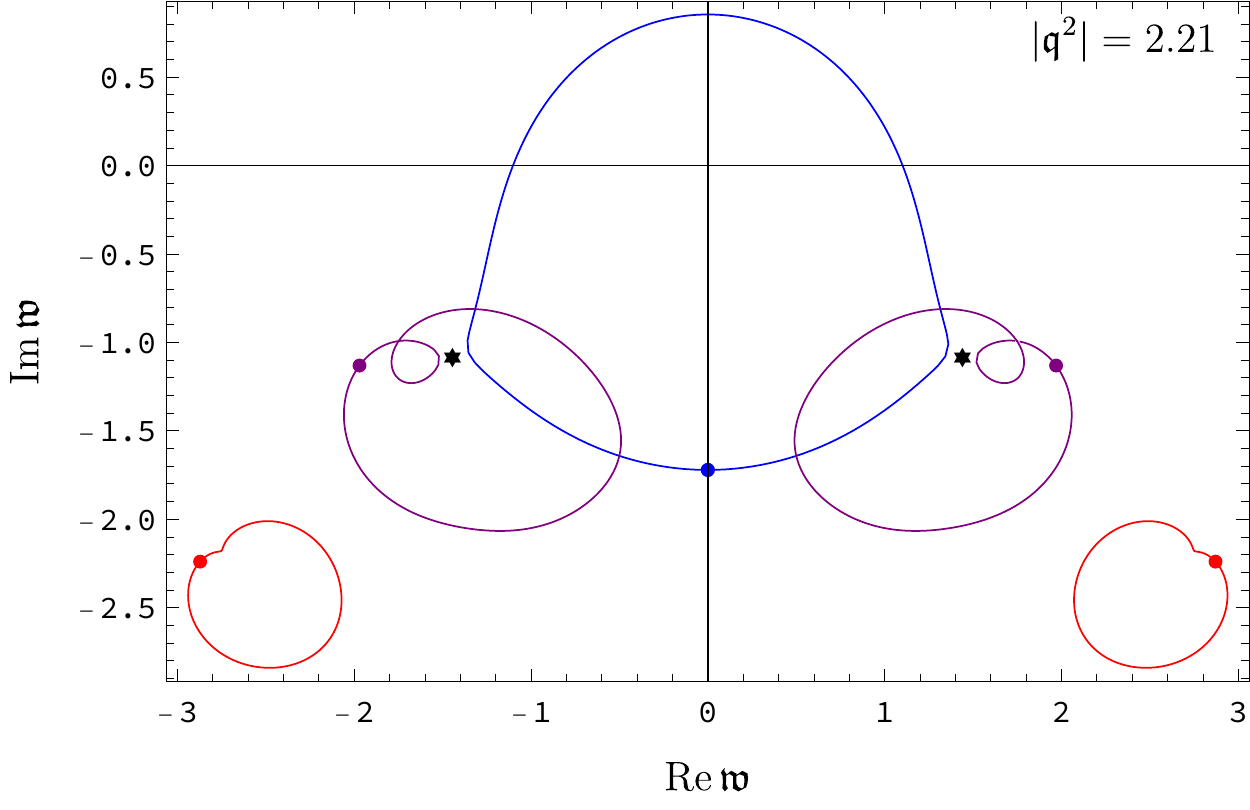}
\hspace{0.05\textwidth}
\includegraphics[width=0.45\textwidth]{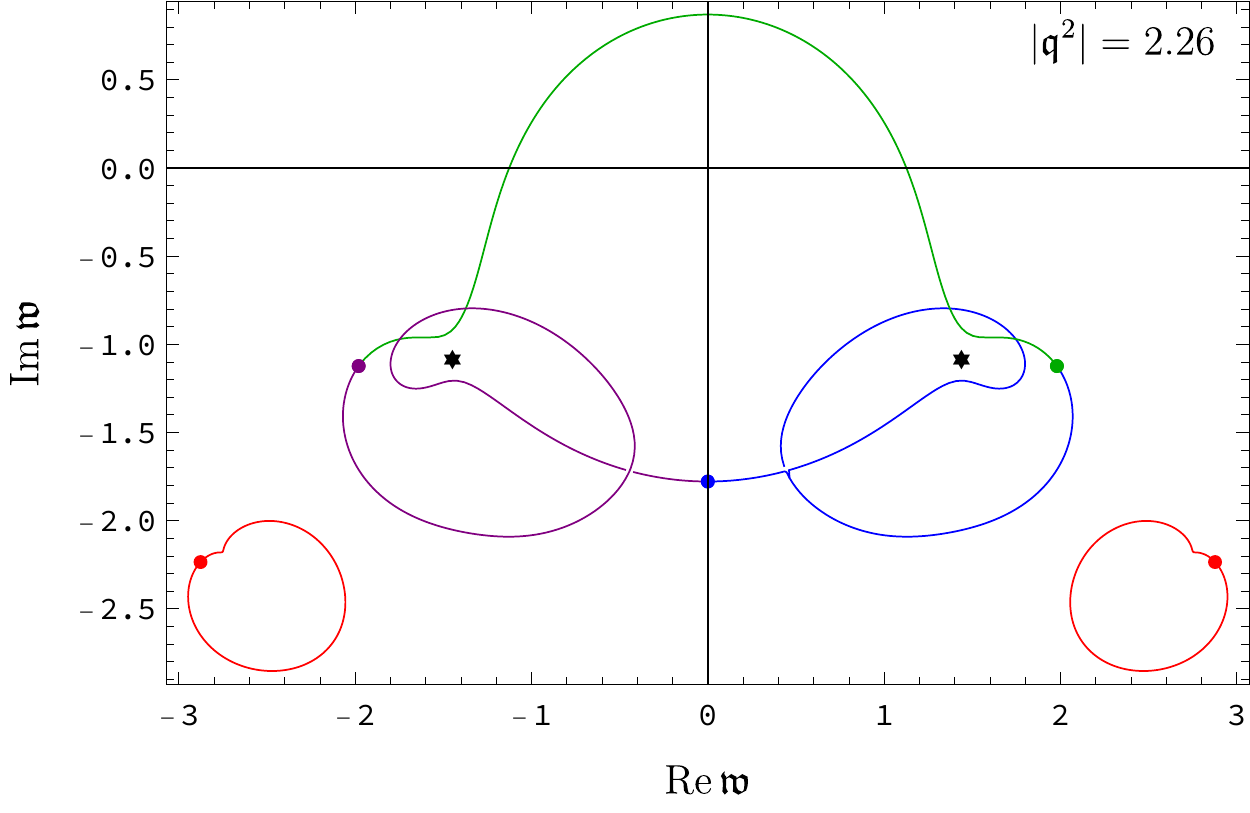}
\caption{
\label{fig:pole-collisions-complex-q-shear}
{\small  Quasinormal spectrum 
(poles of the retarded energy-momentum tensor two-point function in  the ${\cal N}=4$ SYM theory)  in the  shear channel   plotted in the complex $\wfr$-plane for different values of the complex momentum $\qfr^2 = |\qfr^2|e^{i \theta}$. Large dots in all four plots correspond to the location of the poles for purely real momentum, $\qfr^2$ (i.e. at $\theta=0$) \cite{Kovtun:2005ev}. The hydrodynamic shear pole is the blue pole closest to the real axis in the top left panel. As $\theta$ increases from $0$ to $2\pi$, each pole moves in a counter-clockwise direction, following the trajectory of its colour. At $|\qfr^2|=1$, all poles follow a closed orbit (top left panel). At $|\qfr^2|=2.15$ (top right panel), the trajectory of the hydrodynamic pole intersects the trajectories  of  the two nearest gapped poles. With $|\qfr^2|$ further increasing to $|\qfr^2|=2.21$, the poles nearly collide at the positions marked by  asterisks (bottom left panel). The actual collision occurs  at the critical point with the momentum (\ref{eq:qw-crit-shear-x}), $|\qfr_{\rm c}^2| \approx 2.224$.  At $|\qfr^2|=2.26$ (bottom right panel), the orbits of the three poles closest to the origin $(\wfr = 0)$ are no longer closed: the hydrodynamic pole and the two gapped poles exchange their positions cyclically as the phase $\theta$ increases from $0$ to $2\pi$. This is a manifestation of the quasinormal mode level-crossing. The dispersion relation $\wfr(\qfr^2)$ therefore has branch point singularities  in the complex momentum squared plane  at $\qfr_{\rm c}^2$. 
}}
\end{figure*}
The stability of the numerical procedure is ensured by checking that adding several more terms to the series does not appreciably change the result. Applying (numerically) the criterion \eqref{c-curve-critical-x} to \eqref{full-shear-spectral-curve}, we find the critical point closest to the origin,
\begin{align}\label{eq:qw-crit-shear-x}
   \qfr_{\rm c}^2 &\approx 1.8906469 \pm 1.1711505 i\,, & \wfr_{\rm c} &\approx \pm 1.4436414 - 1.0692250 i\,,
\end{align}
implying the convergence radius of the shear mode dispersion relation $|\qfr_{\rm shear}^{\rm c}| \approx 1.49131$. There are other 
critical points with $|\qfr^2|>|\qfr_{\rm shear}^{\rm c}|$. The first three of them are located at
\begin{align}
   \qfr_{\rm c,1}^2 & \approx -2.37737\,, &  \wfr_{\rm c,1} & \approx -1.64659 i\,, \label{eq:qw-crit-shear-x1} \\
   \qfr_{\rm c,2}^2 & \approx  -3.11051 \pm 0.8105 i\,, &  \wfr_{\rm c,2} & \approx \pm  1.41043 - 2.87086 i\,, \label{eq:qw-crit-shear-x2} \\ 
   \qfr_{\rm c,3}^2 & \approx  2.90692 \pm  1.66606 i \,, &  \wfr_{\rm c,3} & \approx  \pm 2.38819 - 2.13154 i \,. \label{eq:qw-crit-shear-x3}
\end{align}

\begin{figure*}[t]
\centering
\includegraphics[width=0.45\textwidth]{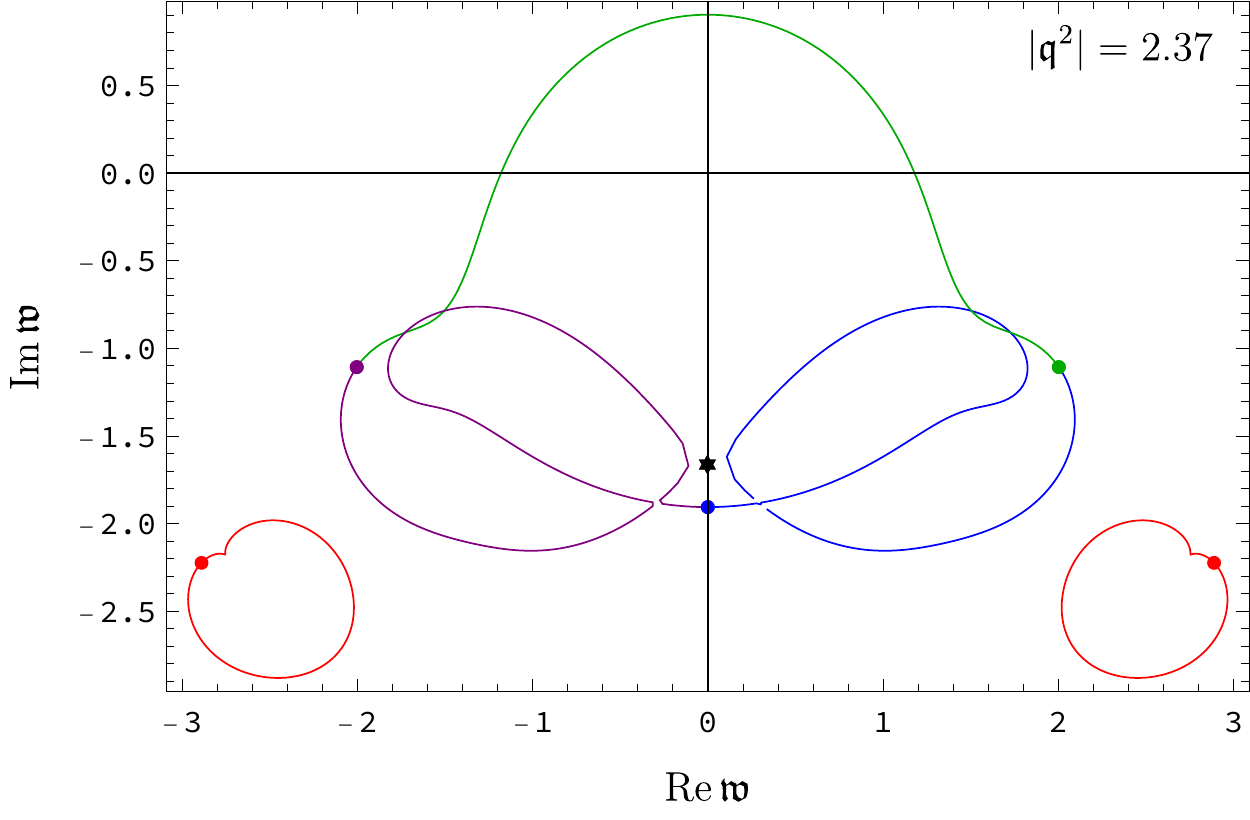}
\hspace{0.05\textwidth}
\includegraphics[width=0.45\textwidth]{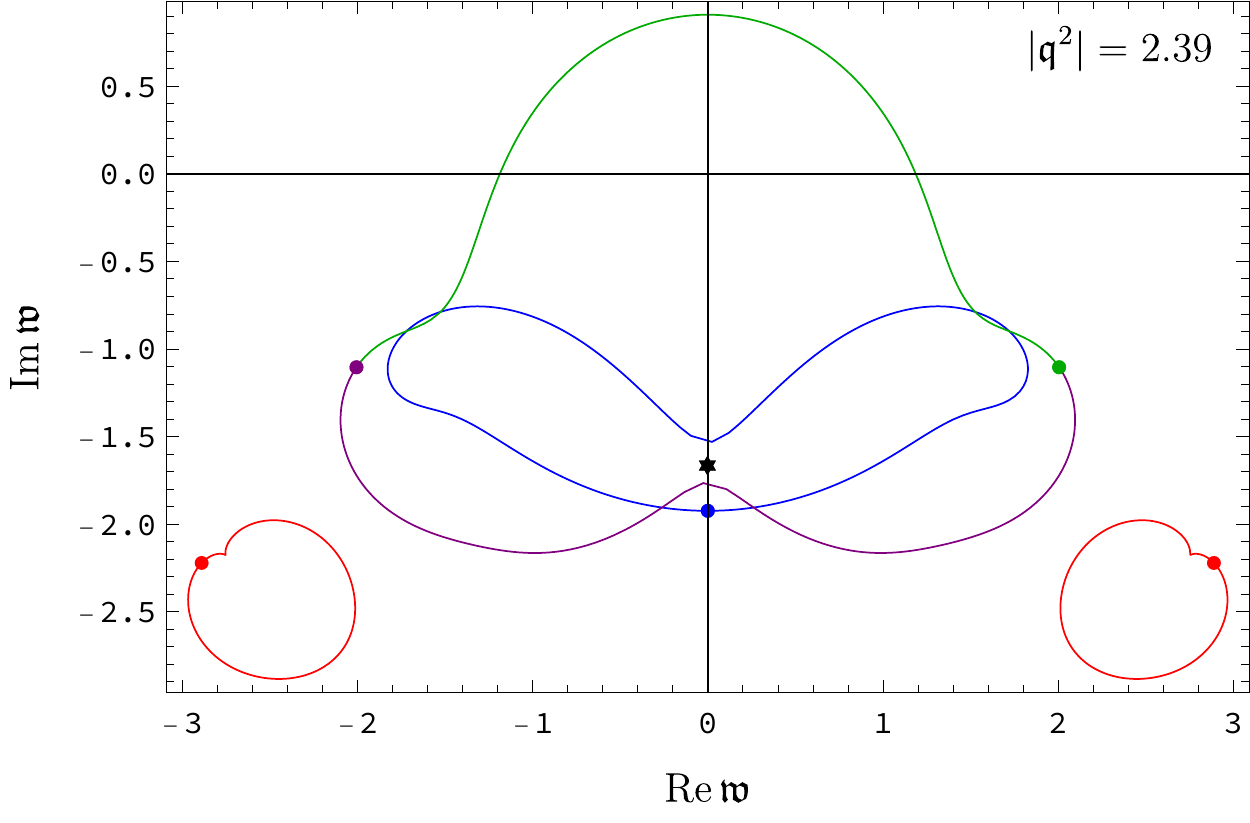}
\\
\includegraphics[width=0.45\textwidth]{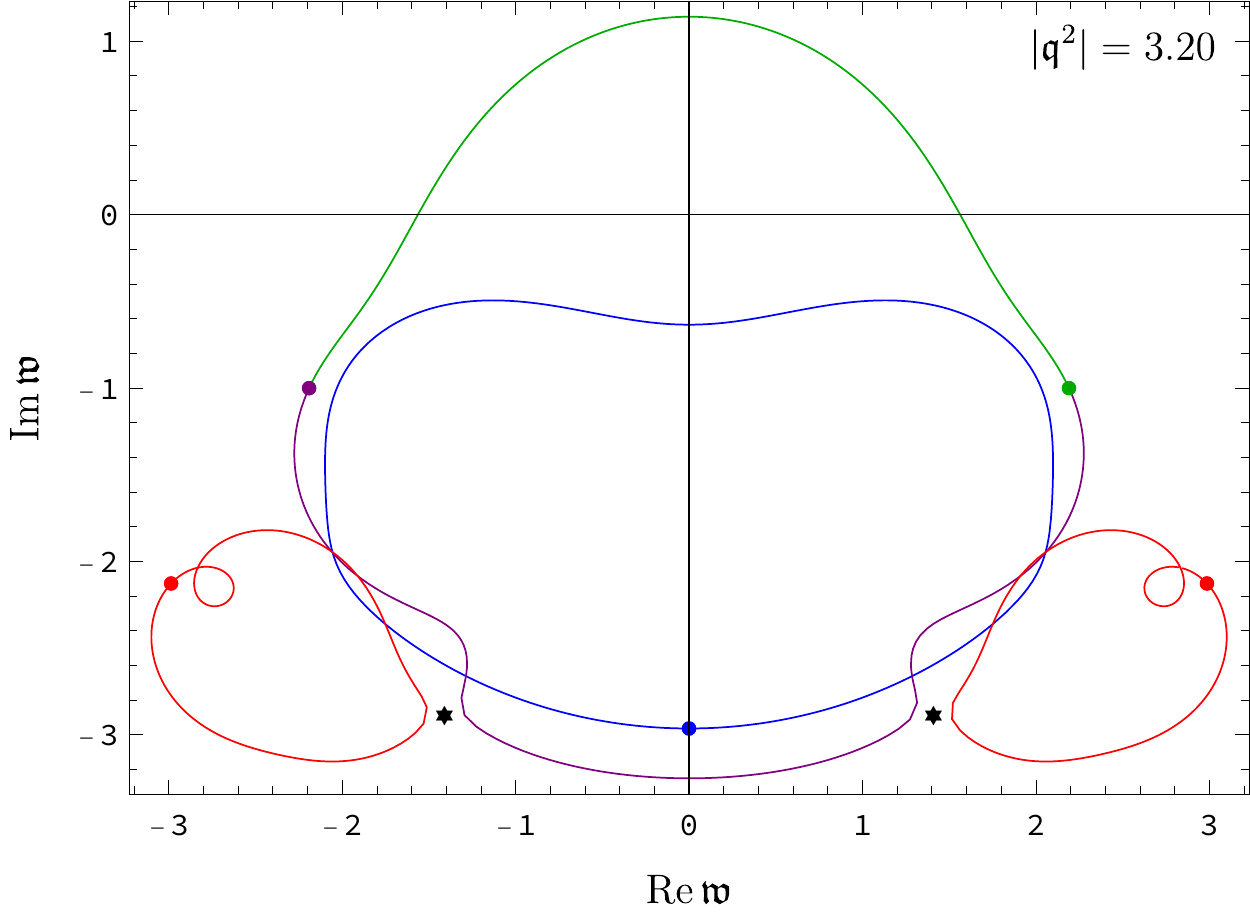}
\hspace{0.05\textwidth}
\includegraphics[width=0.45\textwidth]{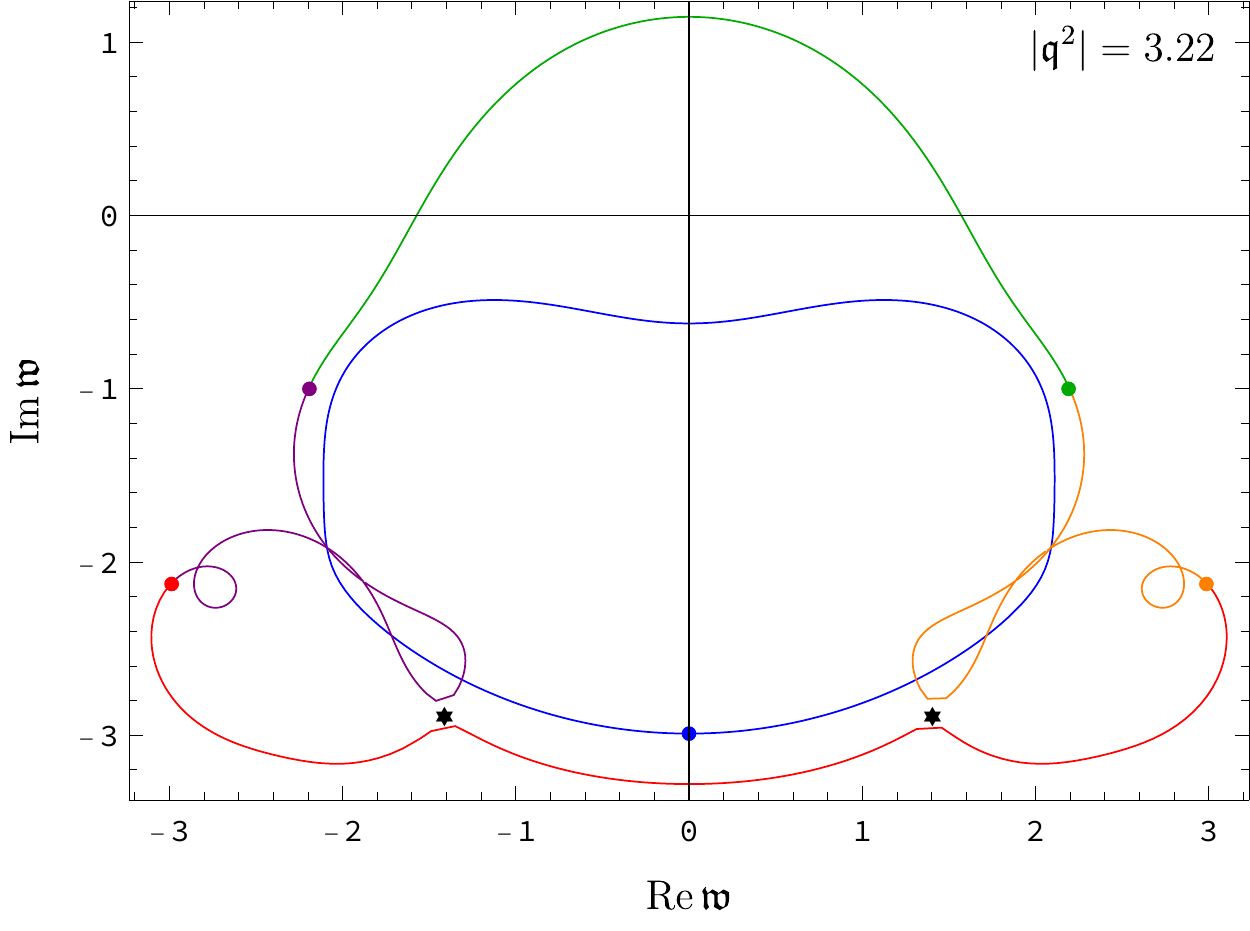}
\caption{
\label{fig:pole-collisions-complex-q-shear-1}
 {\small Quasinormal spectrum 
(poles of the retarded energy-momentum tensor two-point function in  the ${\cal N}=4$ SYM theory)  in the  shear channel plotted in the complex $\wfr$-plane for different values of the complex momentum $\qfr^2 = |\qfr^2|e^{i \theta}$. Large dots in all four plots correspond to the location of the poles for purely real momentum, $\qfr^2$ (i.e. at $\theta=0$) \cite{Kovtun:2005ev}.  The second critical point (occuring at $|\qfr_{\rm c,1}^2| \approx 2.377$) is marked
 by  asterisks in the figures showing the trajectories just before and just after the pole collision (top panels). With $|\qfr^2|$ further increasing, the second pair of gapped modes becomes involved in the collisions leading to the third critical point at $|\qfr_{\rm c,2}^2| \approx 3.214$ (bottom panels).
}}
\end{figure*}
In figs.~\ref{fig:pole-collisions-complex-q-shear} and \ref{fig:pole-collisions-complex-q-shear-1}, we show the corresponding quasinormal spectrum (solutions $\wfr = \wfr (\qfr^2)$ of eq.~\eqref{full-shear-spectral-curve}) in the complex plane of frequency $\wfr$, parametrised by $\qfr^2$. The difference with previous works is that now, we treat $\qfr^2$ as a generic complex variable, $\qfr^2 = |\qfr^2|e^{i \theta}$, and vary its phase, $\theta \in [0,2\pi]$, at specific fixed values of the magnitude $|\qfr^2|$. From eq.~\eqref{eq:Z1eqn}, it is clear that if $Z_1 (u; \qfr,\wfr)$ is a 
solution satisfying the incoming wave boundary condition at the horizon, then  $Z_1 (u; \qfr,-\wfr)$ is also a 
solution. The spectrum in figs.~\ref{fig:pole-collisions-complex-q-shear} and \ref{fig:pole-collisions-complex-q-shear-1} therefore appears to be symmetric with respect to the imaginary axis. For real $\qfr^2$, the spectrum coincides with the one originally found in ref.~\cite{Nunez:2003eq}. 

At small $|\qfr^2|$, the poles follow  closed orbits as the phase $\theta$ varies from $0$ to $2\pi$ (fig.~\ref{fig:pole-collisions-complex-q-shear}, top panels). With the parameter 
$|\qfr^2|$  increasing, the poles start feeling the presence of each other, and their orbits become more complicated. Finally, at $|\qfr_{\rm c}^2| \approx 2.224$, the hydrodynamic shear pole collides with one of the two closest non-hydrodynamic poles (see fig.~\ref{fig:pole-collisions-complex-q-shear}, bottom panels). For $|\qfr^2|>|\qfr_{\rm c}^2|$, those three poles no longer have closed orbits: as the phase $\theta$ varies from $0$ to $2\pi$, they interchange their positions cyclically  (fig.~\ref{fig:pole-collisions-complex-q-shear}, bottom right panel). For even larger $|\qfr^2|$, other gapped poles become involved in this collective motion of quasinormal modes in a similar manner (fig.~\ref{fig:pole-collisions-complex-q-shear-1}).

\begin{figure}[h!]
\centering
\includegraphics[width=0.7\textwidth]{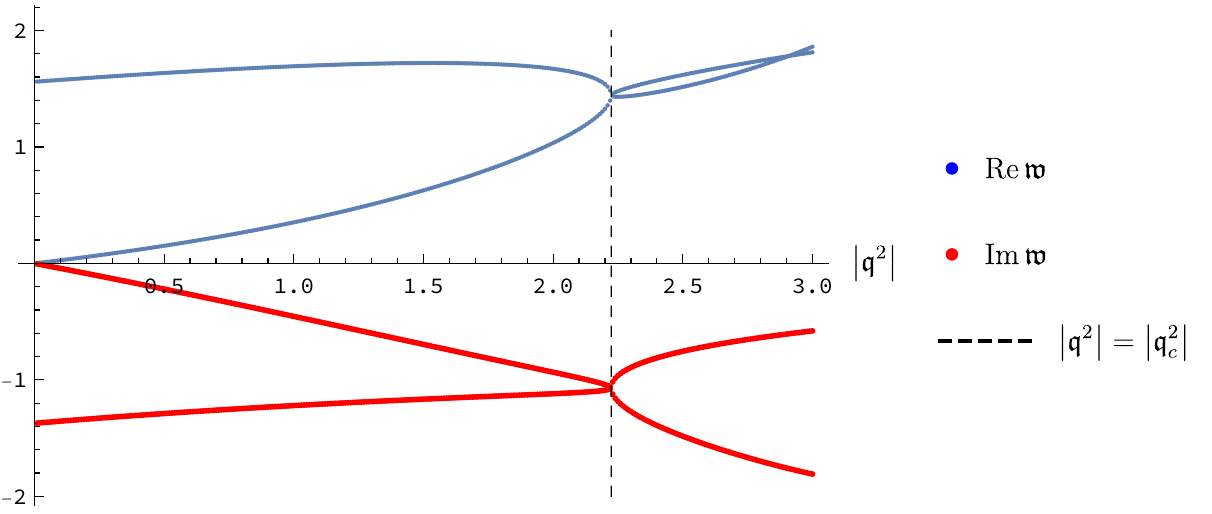}
\caption{{\small Quasinormal spectrum level-crossing: the real (blue curves) and the imaginary (red curves) parts of the hydrodynamic shear mode and the closest gapped quasinormal mode dispersion relations plotted as functions of $|\qfr^2|$ at the fixed phase $\theta \approx 0.338858 \pi$ of the complex momentum $\qfr^2 = |\qfr^2|e^{i \theta}$. At $|\qfr^2|=|\qfr_{\rm c}^2| \approx 2.224$, the level-crossing occurs.}}
\label{qnm-lc-shear}
\end{figure}

The phenomenon observed in figs.~\ref{fig:pole-collisions-complex-q-shear}, \ref{fig:pole-collisions-complex-q-shear-1} is the quasinormal spectrum level-crossing, reminiscent of the well-known effect in quantum mechanics. Indeed, for real $\qfr^2$, the real and imaginary parts of the quasinormal frequencies do not exhibit level-crossing (see e.g. figs.~13 and 14 in ref.~~\cite{Nunez:2003eq}). For complex momentum $\qfr^2 = |\qfr^2|e^{i \theta}$, the real and imaginary parts of the shear hydrodynamic mode and the nearest gapped mode do cross at  the fixed phase $\theta \approx 0.338858 \pi$ and $|\qfr^2|=|\qfr_{\rm c}^2| \approx 2.224$, as shown in fig.~\ref{qnm-lc-shear}. A similar effect at a purely imaginary momentum was observed by Withers (see fig.~3 in ref.~\cite{Withers:2018srf}), and numerous instances of pole collisions 
(i.e. quasinormal level-crossings) at real momenta were reported earlier (see e.g. \cite{Davison:2011ek,
Davison:2014lua,Grozdanov:2016vgg,Grozdanov:2016fkt,Grozdanov:2018ewh,Baggioli:2018nnp,Baggioli:2018vfc,Grozdanov:2018fic,Grozdanov:2018gfx,Gushterov:2018spg,Davison:2018ofp,Baggioli:2019jcm}).

\subsection{Sound mode: hydrodynamic approximation to the spectral curve}
\label{sound-hydro-curve}
The gauge-invariant perturbation corresponding to the sound mode  obeys the equation \cite{Kovtun:2005ev}
\begin{align}
  \,&  Z_2'' - 
    \frac{3\wfr^2 (1+u^2) + \qfr^2 ( 2u^2 - 3 u^4 -3)}
         {u f (3 \wfr^2 +\qfr^2 (u^2-3))} \, Z_2' \nonumber \\
\,&
    + \frac{3 \wfr^4 +\qfr^4 ( 3-4 u^2 + u^4) +
    \qfr^2 ( 4 u^5 - 4 u^3 + 4 u^2 \wfr^2 - 6 \wfr^2)}
    {u f^2 ( 3 \wfr^2 + \qfr^2 (u^2 -3))}\, Z_2
    = 0 \, .
\label{eq:Z2eqn-x}
\end{align}
The full sound spectral curve is constructed from the solution $Z_2 (u; \qfr^2,\wfr)$ obeying the incoming wave boundary conditions at the horizon and is given by 
\begin{align}
 F(\qfr^2,\wfr) =Z_2 (u=0; \qfr^2,\wfr) =0\,.
\label{full-sound-spectral-curve}
\end{align}
Similarly to the shear mode case, one can first find a hydrodynamic approximation to eq.~\eqref{full-sound-spectral-curve} analytically. For $\wfr\ll 1$ and $\qfr \ll 1$, eq.~\eqref{eq:Z2eqn-x} can be solved perturbatively. Writing 
\begin{align}
Z_2(u) = \left(1-u^2\right)^{-\frac{i\wfr}{2}} \sum_{n=0}^\infty\lambda^n z_n(u)
\end{align}
to enforce the boundary condition at the horizon, rescaling $\wfr \to \lambda \wfr$ and $\qfr \to \lambda\qfr$,
 and assuming $\lambda \ll 1$, we find the following recurrence relation for the functions $z_n(u)$:
\begin{align}
z_n'' -& \frac{3\left(1+u^2\right)\wfr^2 - \left(3-2u^2+3u^4\right) \qfr^2}{u\left(1-u^2\right) \left[3\wfr^2 - \left(3-u^2\right) \qfr^2 \right]} z_n' - \frac{4u^2\qfr^2}{\left(1-u^2\right) \left[ 3\wfr^2 - \left(3-u^2\right)\qfr^2  \right] } z_n \nn
&+ \frac{2iu\wfr}{1-u^2} z_{n-1}' - \frac{4 i u^2 \wfr \qfr^2}{\left(1-u^2\right) \left[ 3\wfr^2 - \left(3-u^2\right)\qfr^2 \right] } z_{n-1} \nonumber  \\    
&+ \frac{ \left(1+u+u^2\right)\wfr^2 - \left(1+u\right)\qfr^2 }{u \left(1-u\right) \left(1+u\right)^2   } z_{n-2 } = 0 \, , 
\end{align}
with $z_{-1} = z_{-2} = 0$. To fourth order in $\lambda$, we find the hydrodynamic algebraic curve to be (see also eq. (4.18) in ref.~\cite{Baier:2007ix})
\begin{align}
F(\qfr^2,\wfr) &=  \frac{\qfr^2}{2} - \frac{3 \wfr^2}{2} -i \wfr \qfr^2 + 
\frac{\wfr^4}{16}\left( \pi^2 - 12 \ln^22 + 24\ln{2}\right)
-\frac{\qfr^4}{12} \left( 2 \ln{2} -8 \right) \nonumber 
\\ &- \frac{\wfr^2\qfr^2}{48} \left(\pi^2 - 12 \ln^2 2+ 48 \ln2\right) =0\,.
\label{hsac}
\end{align}
The form of $F(\qfr^2,\wfr)$ for general $\wfr$ and $\qfr$ is at present not known to $O(\lambda^5)$. However, assuming that $\wfr$ can be expanded in a series in powers of $\qfr$, in ref.~\cite{Grozdanov:2015kqa}, $F(\qfr^2)$ was computed  to order $O(\lambda^5) = O(\qfr^5)$, which was sufficient to find the coefficient in the sound mode dispersion relation multiplying $\qfr^4$ (i.e. to third order in the gradient expansion):
\begin{align}
\wfr  \,&= \pm \frac{1}{\sqrt{3}} \qfr - \frac{i}{3} \qfr^2 \pm \frac{3-2\ln 2}{6 \sqrt{3} } \qfr^3 - \frac{ i \left(\pi^2 - 24 + 24\ln 2 - 12 \ln^2 2\right) }{108} \qfr^4  \,+ O(\qfr^{5})\,.
\label{rkr}
\end{align}
The first two terms in the dispersion relation \eqref{rkr} coincide with those obtained in ref.~\cite{Policastro:2002tn}. The third term was found in ref.~\cite{Baier:2007ix} (see also ref.~\cite{Bhattacharyya:2008jc}). The fourth term was computed in ref.~\cite{Grozdanov:2015kqa}. It appeared earlier in ref.~\cite{Bu:2014ena} (with an incorrect coefficient in front of $\ln{2}$) and (correctly) in ref.~\cite{Bu:2015bwa} in the context of the fluid-gravity correspondence.

Here, we use the form of the algebraic curve \eqref{hsac} and apply the condition \eqref{c-curve-critical-x}  to show that in the sound channel, the small $\wfr$ and $\qfr$ expansion of the spectral curve $F(\qfr^2,\wfr)$ qualitatively correctly accounts for two sets of critical points:
\begin{align}
 \qfr_{\rm c}^2 &=0\,, & \wfr_{\rm c} &= 0\,, \label{cps1} \\
\qfr_{\rm c}^2 &\approx 0.34739  \pm 0.56763i \,, & \wfr_{\rm c} &\approx   \pm
 0.71165+0.39882 i\, . \label{cps2}
\end{align}
The critical point \eqref{cps1} is expected from hydrodynamics, as discussed in section \ref{hydro-spectral-sound}. Having in mind 
the results of section \ref{shear-n=4-hydro}, we may expect that the hydrodynamic approximation \eqref{cps2} to the position of the non-trivial critical point is not accurate.  In section 
\ref{sound-lc} we confirm this by finding the critical point exactly.
\begin{figure*}[t]
\centering
\includegraphics[width=0.45\textwidth]{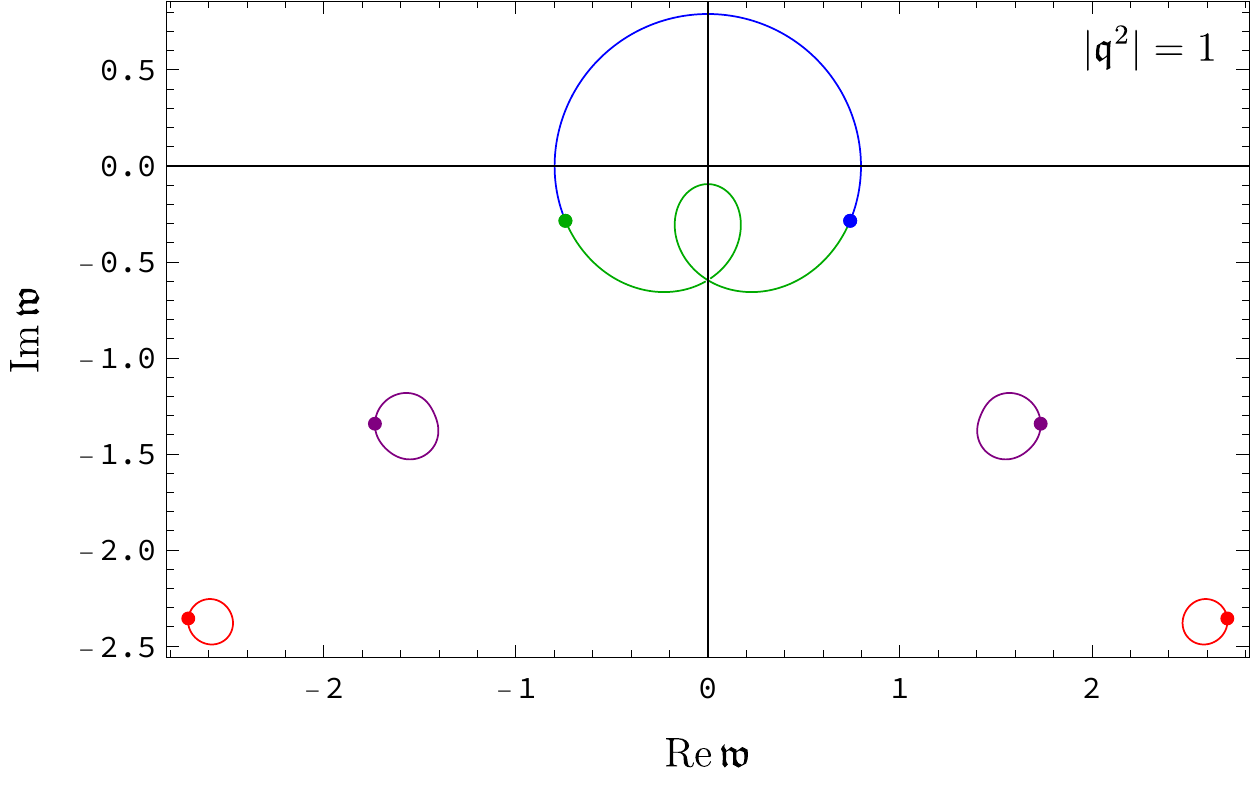}
\hspace{0.05\textwidth}
\includegraphics[width=0.45\textwidth]{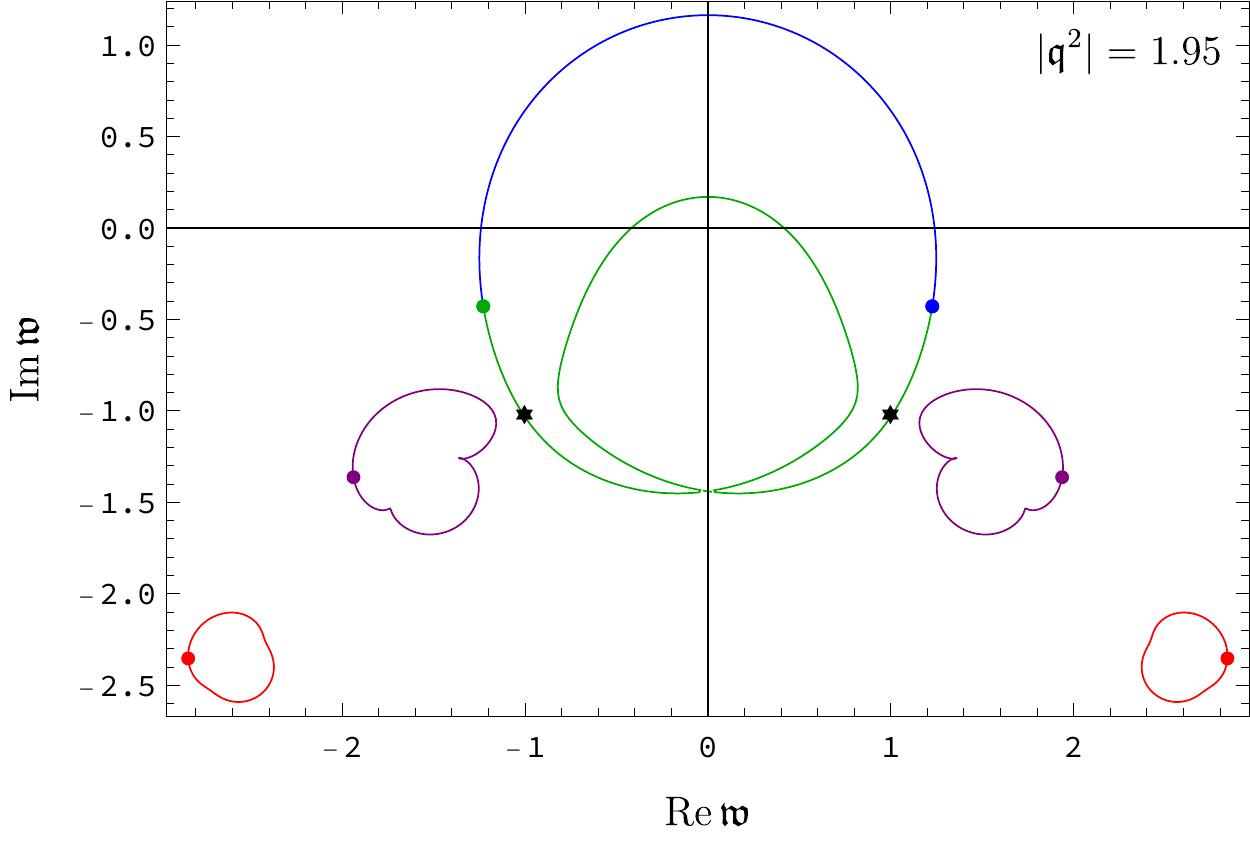}
\\
\includegraphics[width=0.45\textwidth]{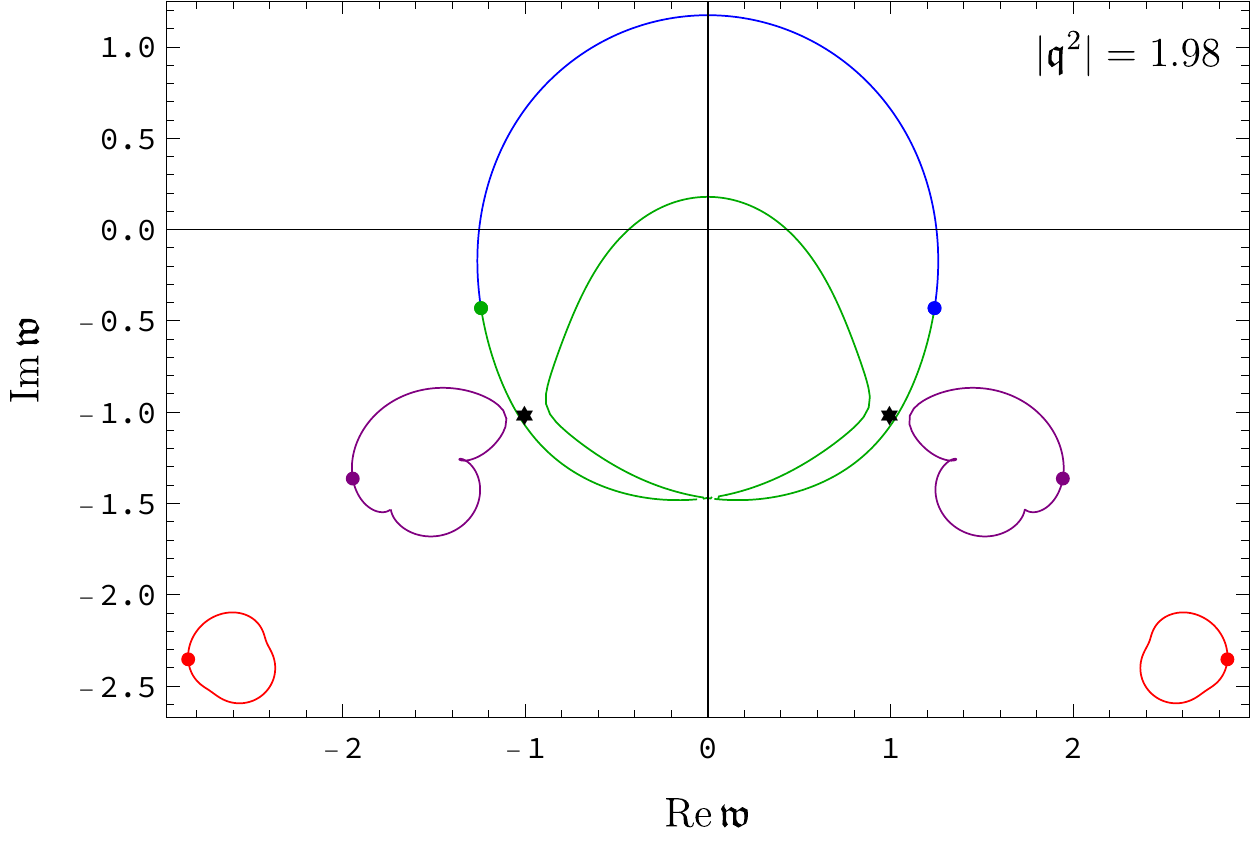}
\hspace{0.05\textwidth}
\includegraphics[width=0.45\textwidth]{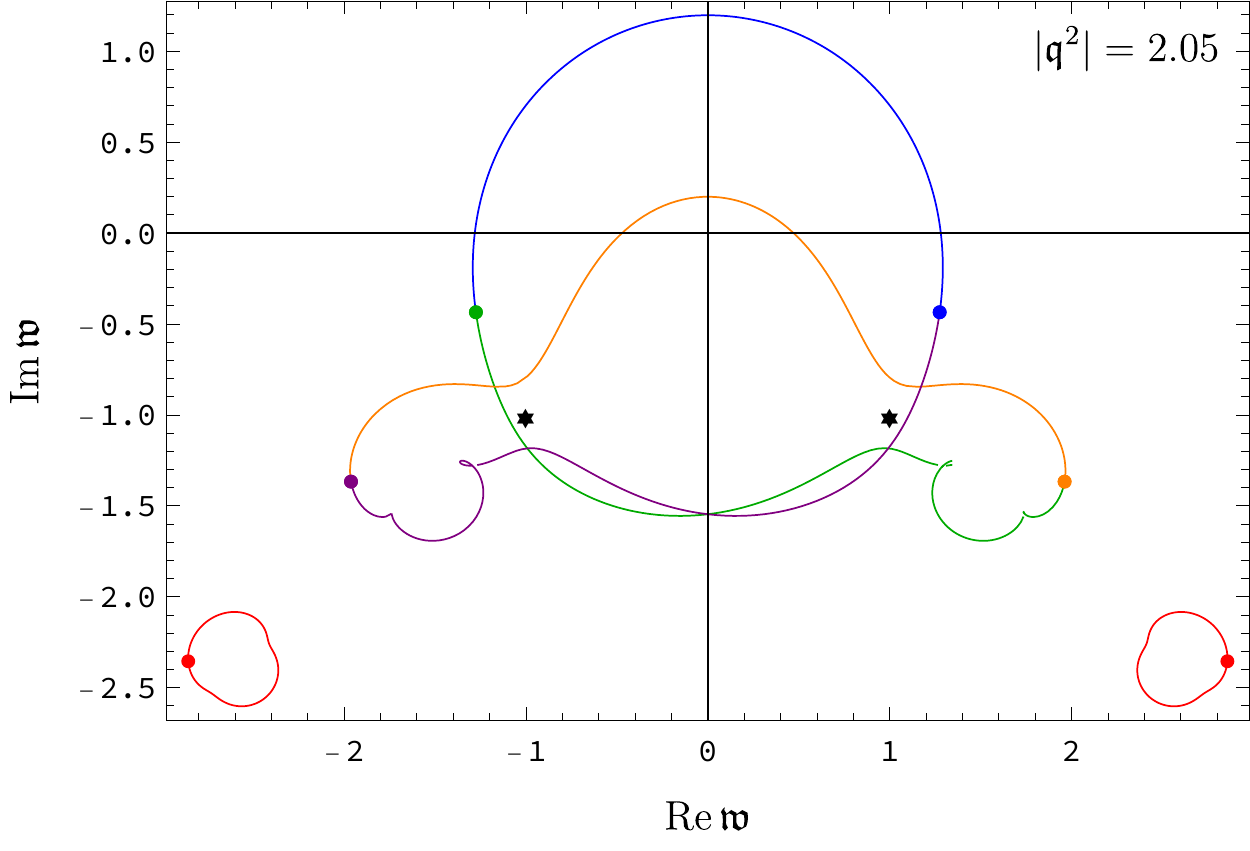}
\caption{
\label{fig:pole-collisions-complex-q} {\small Quasinormal spectrum 
(poles of the retarded energy-momentum tensor two-point function in  the ${\cal N}=4$ SYM theory)  in the  sound channel plotted in the complex $\wfr$-plane for different values of the complex momentum $\qfr^2 = |\qfr^2|e^{i \theta}$. Large dots in all four plots correspond to the location of the poles for purely real momentum, $\qfr^2$ (i.e. at $\theta=0$) \cite{Kovtun:2005ev}.  The hydrodynamic sound poles are the blue and the green poles closest to the real axis. As $\theta$ is tuned from $0$ to $2\pi$, each pole moves in a counter-clockwise direction and follows the trajectory of its colour. At $|\qfr^2|=1$ (top left panel), all poles follow a closed orbit. At $|\qfr^2|=1.95$ (top right panel), the trajectory of the two hydrodynamic sound poles comes close to the trajectories of the nearest gapped poles. With $|\qfr^2|$ further increasing to $|\qfr^2|=1.98$, the poles nearly collide at the positions marked by  asterisks (bottom left panel).  The actual collision occurs at the critical value of the momentum (\ref{eq:qw-crit-sound-x}), $|\qfr_{\rm c}^2| = 2$.  At $|\qfr^2|=2.05$ (bottom right panel), the orbits of the four uppermost poles are no longer closed: the hydrodynamic poles and the two gapped poles exchange their positions cyclically as the phase $\theta$ increases from $0$ to $2\pi$---again, a manifestation of the quasinormal mode level-crossing. The dispersion relation $\wfr(\qfr)$ therefore has branch cuts starting at  $\qfr_{\rm c}$.
}}
\end{figure*}

\subsection{Sound mode: full spectral curve}
\label{sound-lc}
\begin{figure*}[t]
\centering
\includegraphics[width=0.45\textwidth]{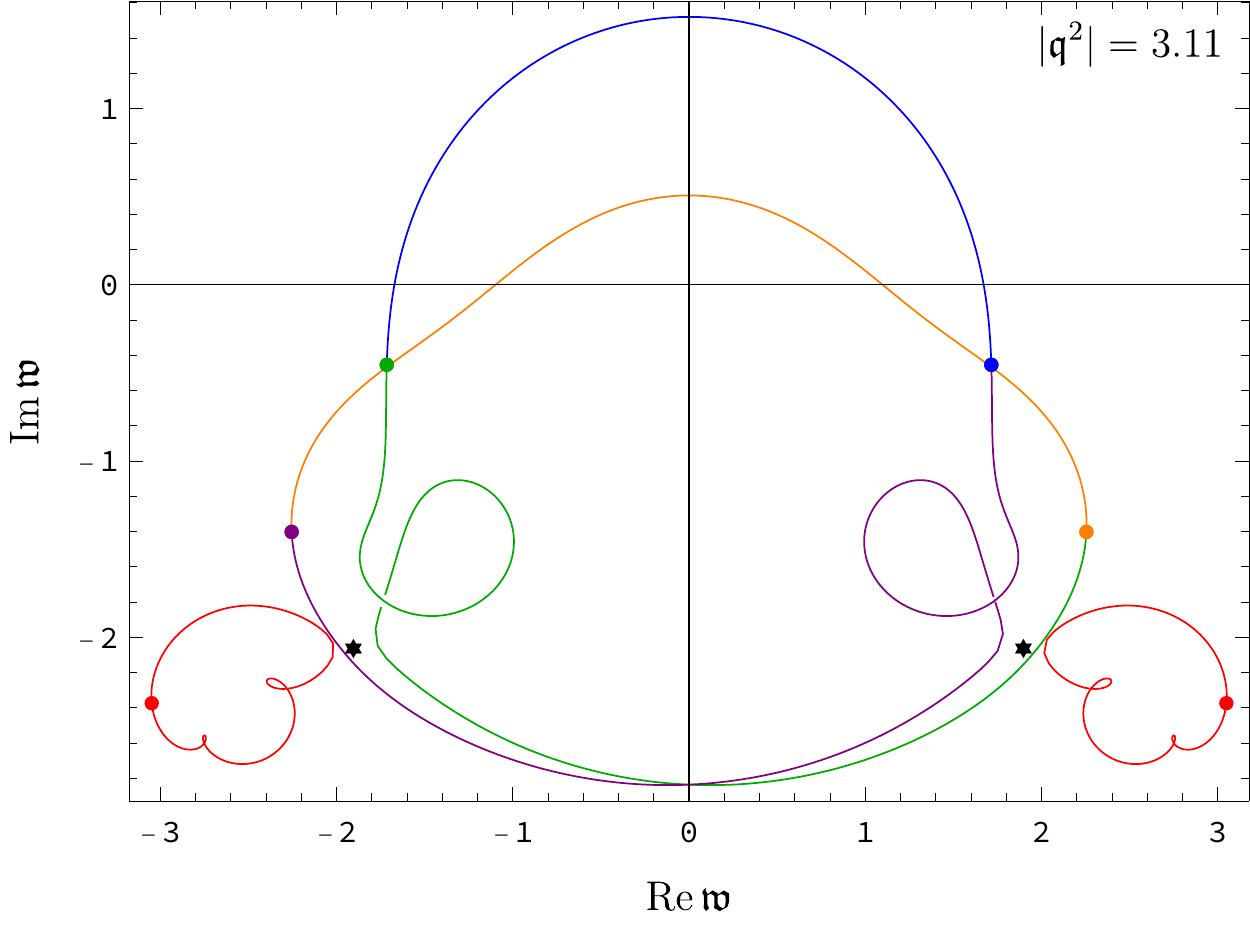}
\hspace{0.05\textwidth}
\includegraphics[width=0.45\textwidth]{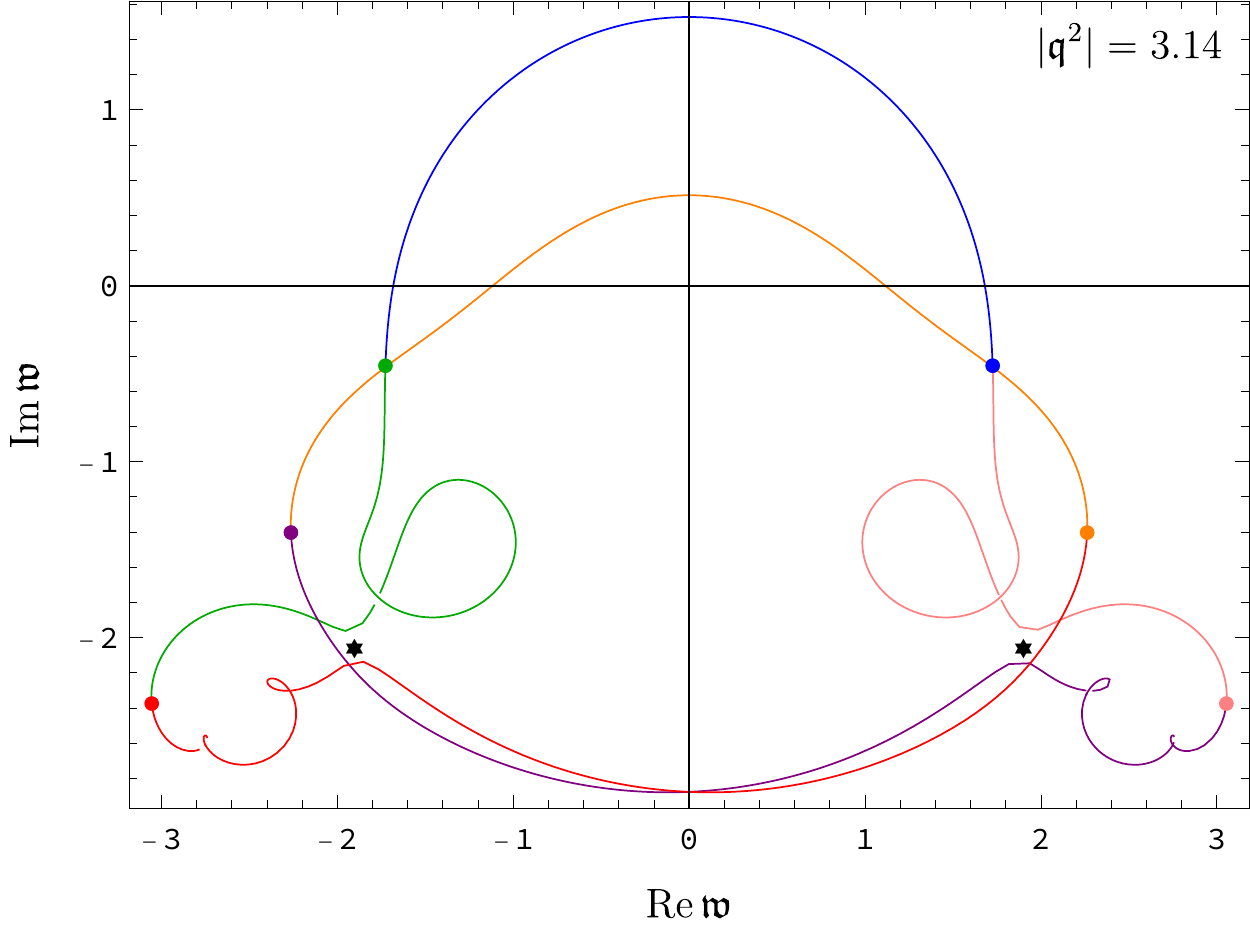}
\caption{
\label{fig:pole-collisions-complex-q-1} {\small Quasinormal spectrum 
(poles of the retarded energy-momentum tensor two-point function in  the ${\cal N}=4$ SYM theory)  in the  sound channel plotted in the complex $\wfr$-plane for different values of the complex momentum $\qfr^2 = |\qfr^2|e^{i \theta}$. Large dots in all four plots correspond to the location of the poles for purely real momentum, $\qfr^2$ (i.e. at $\theta=0$) \cite{Kovtun:2005ev}. What is shown is the level-crossing phenomenon for the critical points in eq.~\eqref{eq:qw-crit-sound-x-1x}.}}
\end{figure*}

\begin{figure}[t!]
\centering
\includegraphics[width=0.7\textwidth]{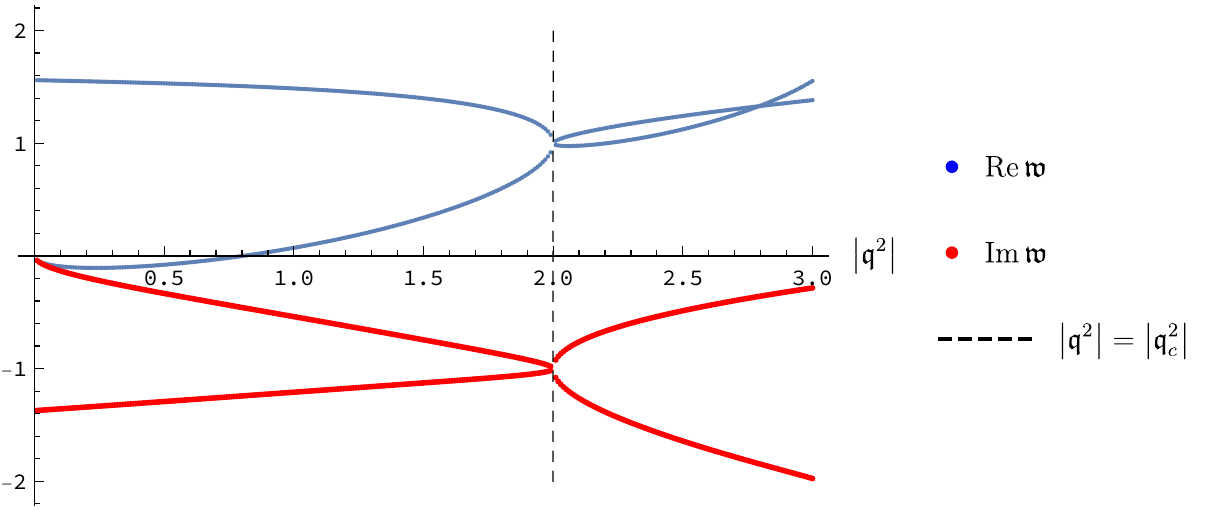}
\caption{{\small Quasinormal spectrum level-crossing in the sound channel: the real (blue curves) and the imaginary (red curves) parts of the hydrodynamic sound mode and the closest gapped quasinormal mode dispersion relations plotted as functions of $|\qfr^2|$ at the fixed phase 
$\theta = \pi/2$ of the complex momentum $\qfr^2 = |\qfr^2|e^{i \theta}$. At $|\qfr^2|=|\qfr_{\rm c}^2| =2$, the level-crossing occurs.}}
\label{qnm-lc-sound}
\end{figure}

As discussed in section \ref{sound-hydro-curve}, the origin \eqref{cps1} is a critical point of the sound mode dispersion relation $\wfr = \wfr(\qfr^2)$, 
as predicted by hydrodynamics (see section \ref{hydro-spectral-sound}). Proceeding  as in section \ref{shear-lc}, we find the first set of 
critical points nearest to the origin at 
\begin{align}
   \qfr_{\rm c}^2 &= \pm 2i \,, &  \wfr_{\rm c} &= \pm 1 - i \,, \label{eq:qw-crit-sound-x}
\end{align}
within the limits of our numerical accuracy. Curiously, although eq.~\eqref{eq:Z2eqn-x} looks rather complicated, one can check that with $\wfr$ and $\qfr^2$ given by \eqref{eq:qw-crit-sound-x}, a simple analytic solution satisfying the correct boundary conditions at $u=1$ and $u=0$ is available. Explicitly, the two solutions corresponding to the pair of critical points in \eqref{eq:qw-crit-sound-x} are 
\begin{align}
(\qfr_{\rm c}^2 = 2 i, \wfr_{\rm c} = 1- i ) : && Z_2 (u) &=  \mathcal{C}^+_2  \left(1-u\right)^{- \frac{i \left(1- i\right)}{2} } \left(1+ u\right)^{-\frac{1-i}{2}} u^2 \left(u - 3i  \right)  \,, \\
(\qfr_{\rm c}^2 = - 2 i, \wfr_{\rm c} = - 1- i ) : && Z_2 (u) &=  \mathcal{C}^-_2  \left(1-u\right)^{- \frac{i \left(-1- i\right)}{2} } \left(1+ u\right)^{\frac{-1-i}{2}} u^2 \left( u + 3 i \right)\,,
\end{align} 
where $\mathcal{C}_2^\pm$ are arbitrary constants. Although we were not able to show analytically that $\partial_\wfr Z(u=0) = 0$ at \eqref{eq:qw-crit-sound-x} as well, we have verified this numerically to high precision. The existence 
of the critical point \eqref{eq:qw-crit-sound-x} implies  that the convergence radius of the sound mode dispersion 
relation is given by $|\qfr_{\rm sound}^{\rm c}| = \sqrt{2} \approx 1.41421$. The next set of critical points is located at
\begin{align}
   \qfr_{\rm c}^2 &\approx - 0.01681 \pm 3.12967 i\,, &  \wfr_{\rm c} &\approx \pm 1.90134 - 2.04492 i \,. \label{eq:qw-crit-sound-x-1x}
\end{align}
The behaviour of poles in the complex frequency plane is shown in figs.~\ref{fig:pole-collisions-complex-q} and \ref{fig:pole-collisions-complex-q-1}, and the quasinormal level-crossing phenomenon is presented in fig.~\ref{qnm-lc-sound}.


\subsection{Scalar mode: full spectral curve}
\label{scalar-lc}
\begin{figure*}[t]
\centering
\includegraphics[width=0.45\textwidth]{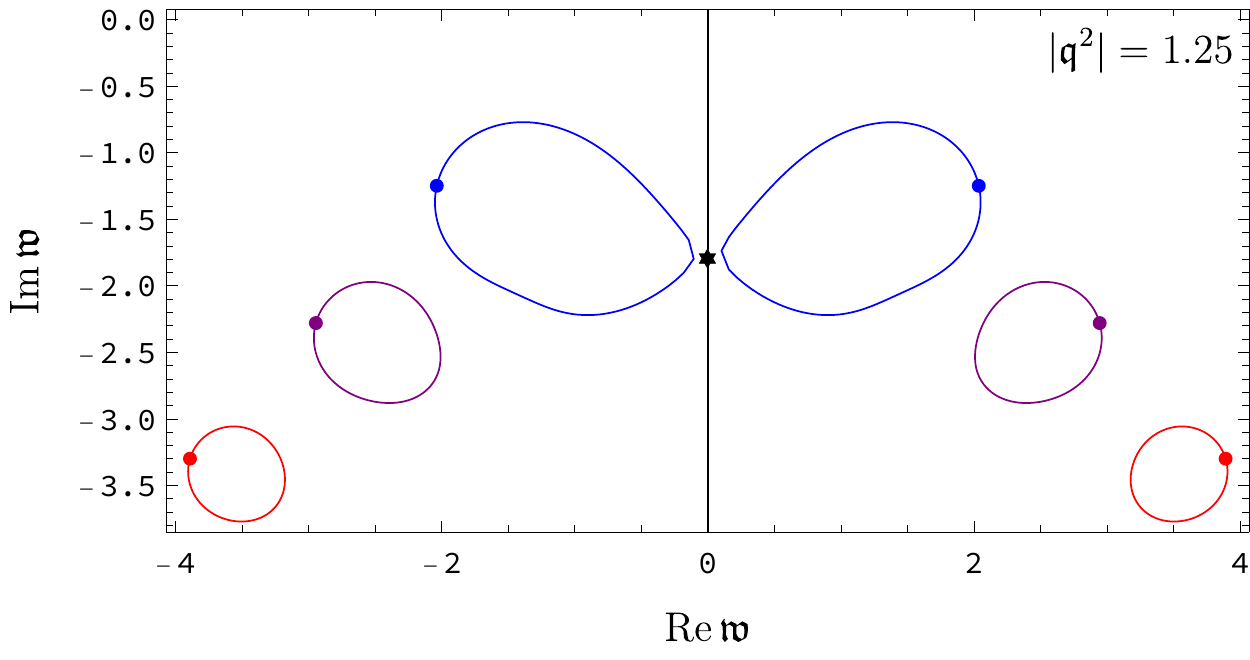}
\hspace{0.05\textwidth}
\includegraphics[width=0.45\textwidth]{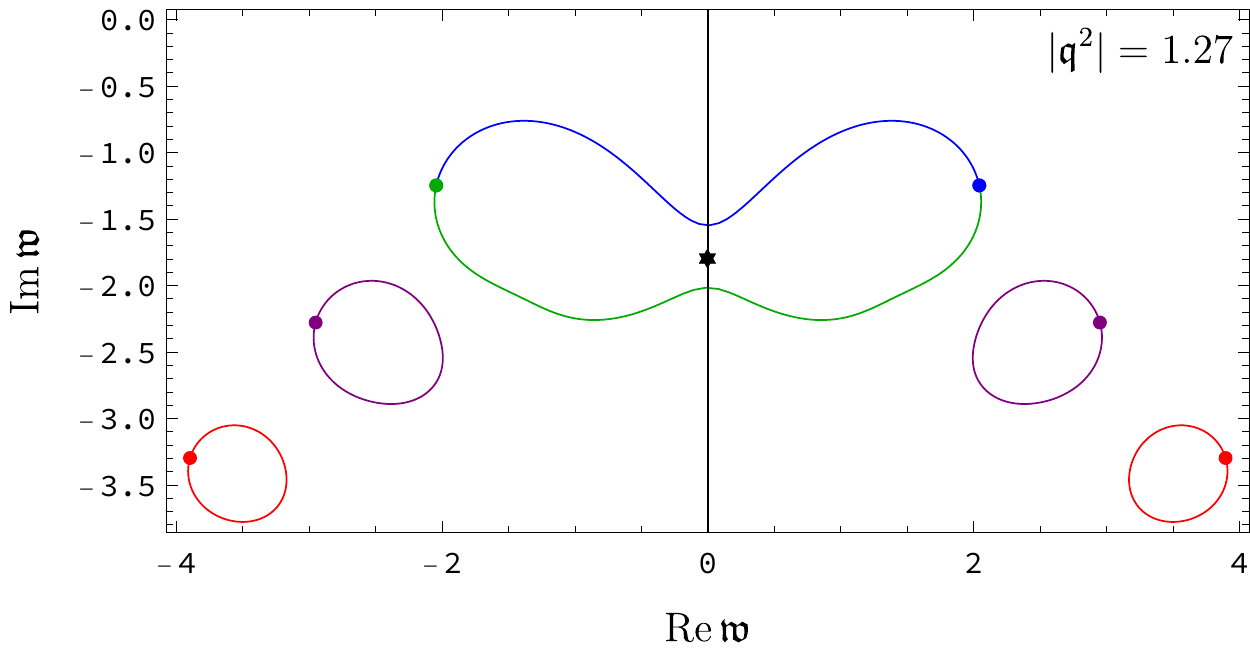}
\\
\includegraphics[width=0.45\textwidth]{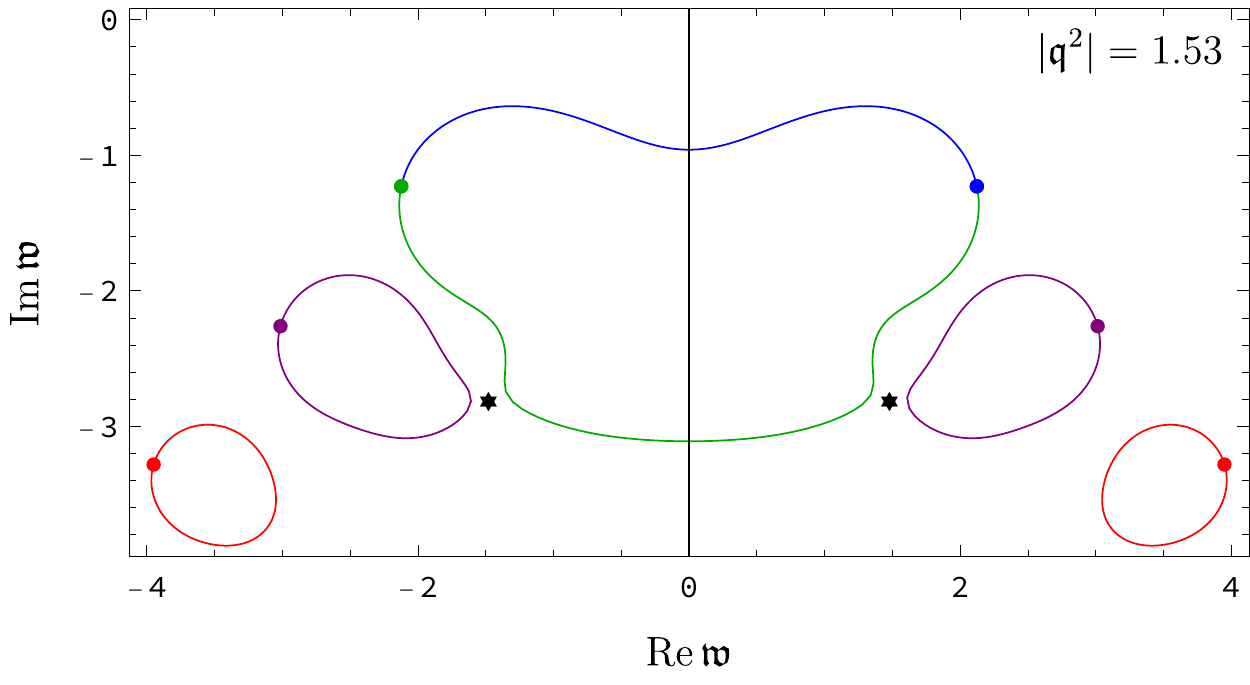}
\hspace{0.05\textwidth}
\includegraphics[width=0.45\textwidth]{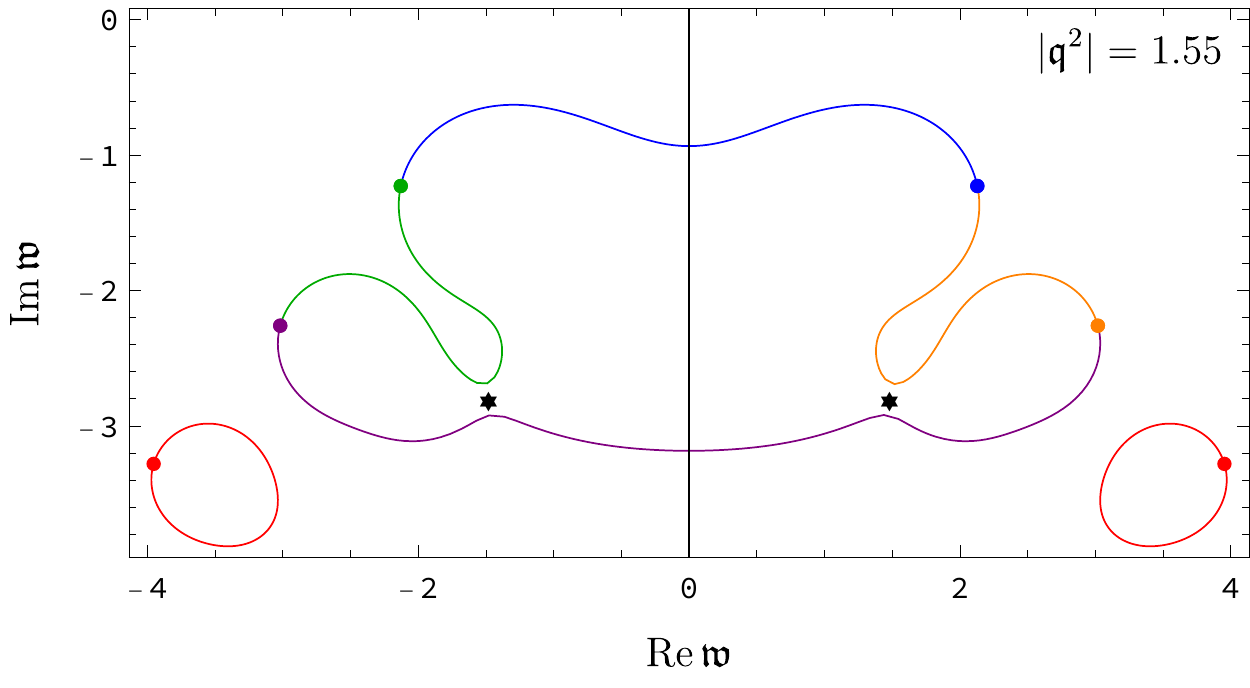}
\caption{
\label{fig:pole-collisions-complex-q-scal}
{\small Quasinormal spectrum 
(poles of the retarded energy-momentum tensor two-point function in  the ${\cal N}=4$ SYM theory)  in the  scalar channel plotted in the complex $\wfr$-plane for different values of the complex momentum $\qfr^2 = |\qfr^2|e^{i \theta}$. Large dots in all plots correspond to the location of the poles for purely real momentum, $\qfr^2$ (i.e. at $\theta=0$) \cite{Kovtun:2005ev}. There are no hydrodynamic modes in this channel, but the gapped modes exhibit the level-crossing phenomena at complex values of momenta given by eq.~\eqref{eq:qw-crit-scalar-x}.
}}
\end{figure*}

In the scalar channel, the gauge-invariant metric perturbation $Z_3(u)$  obeys the equation \cite{Kovtun:2005ev}
\begin{align}
  \,&  Z_3'' - 
    \frac{(1+u^2)}
         {u f } \, Z_3' 
    + \frac{\wfr^2-\qfr^2 f}
    {u f^2}\, Z_3
    = 0 \, .
\label{eq:Z3eqn-x}
\end{align}
The full spectral curve is constructed from the solution $Z_3 (u; \qfr^2,\wfr)$ obeying the incoming wave boundary conditions at the horizon and is given by 
\begin{align}
 F(\qfr^2,\wfr) =Z_3 (u=0; \qfr^2,\wfr) =0\,.
\label{full-scalar-spectral-curve}
\end{align}
The quasinormal spectrum in the scalar channel has no hydrodynamic modes, but it exhibits the phenomenon of level-crossing for the gapped modes, as shown in fig.~\ref{fig:pole-collisions-complex-q-scal}. The first two sets of critical points nearest to the origin are given by
\begin{subequations}
\label{eq:qw-crit-scalar-x}
\begin{align}
  \qfr_{\rm c}^2 &\approx -1.25309 \,, & \wfr_{\rm c} &\approx -1.76937 i\,,\\
  \qfr_{\rm c}^2 &\approx -1.49704 \pm  0.36674 i \, , & \wfr_{\rm c} &\approx \mp 1.47977 - 2.79404 i \,.
\end{align}
\end{subequations}
These are branch points in the scalar (gapped) dispersion relation $\wfr = \wfr (\qfr^2)$.

\subsection{Analytic structure of the hydrodynamic dispersion relations}\label{sec:AnalyticStructureN4}
\begin{figure}[t!]
\centering
\includegraphics[width=0.45\textwidth]{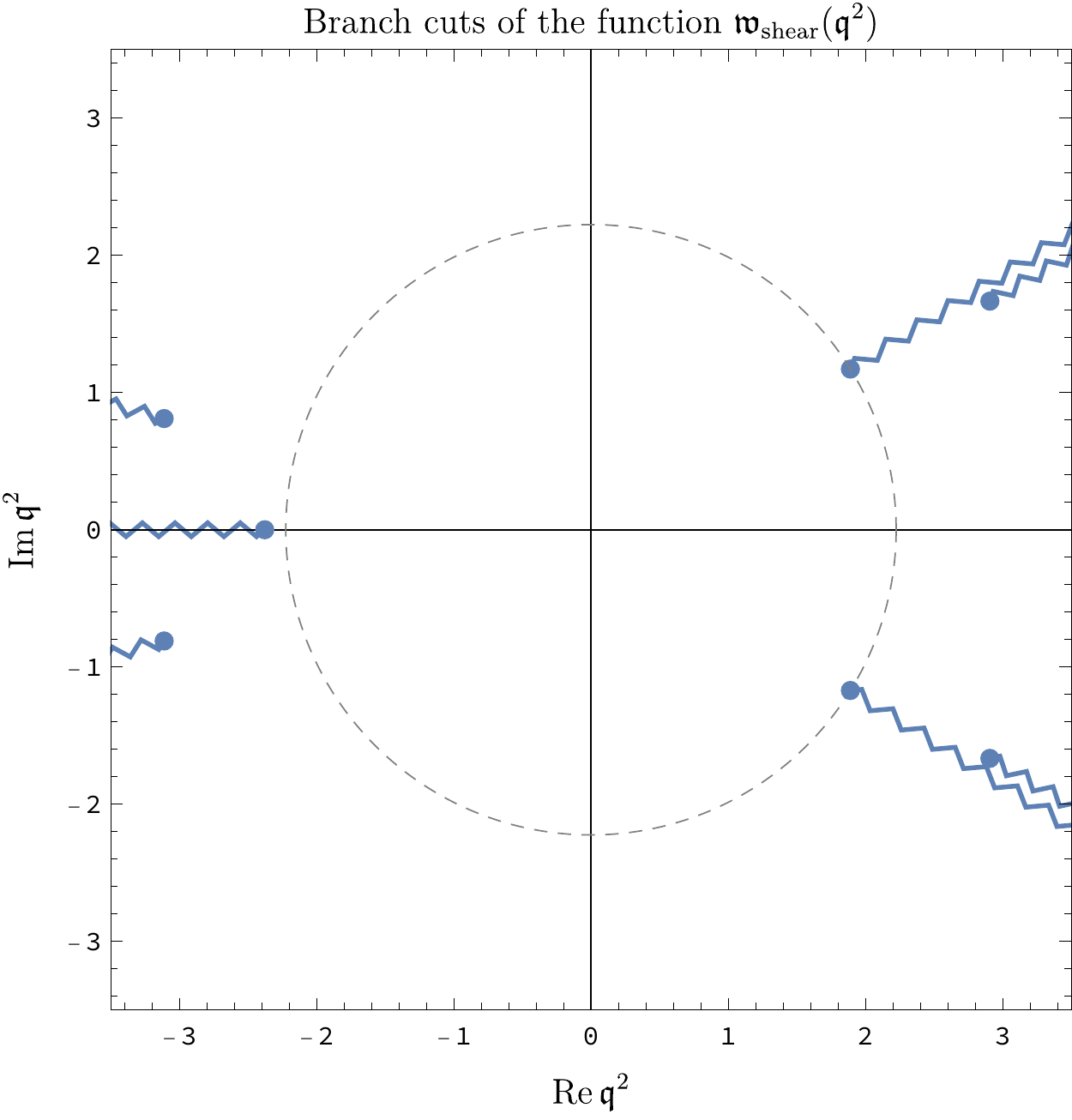}
\hspace{0.05\textwidth}
\includegraphics[width=0.45\textwidth]{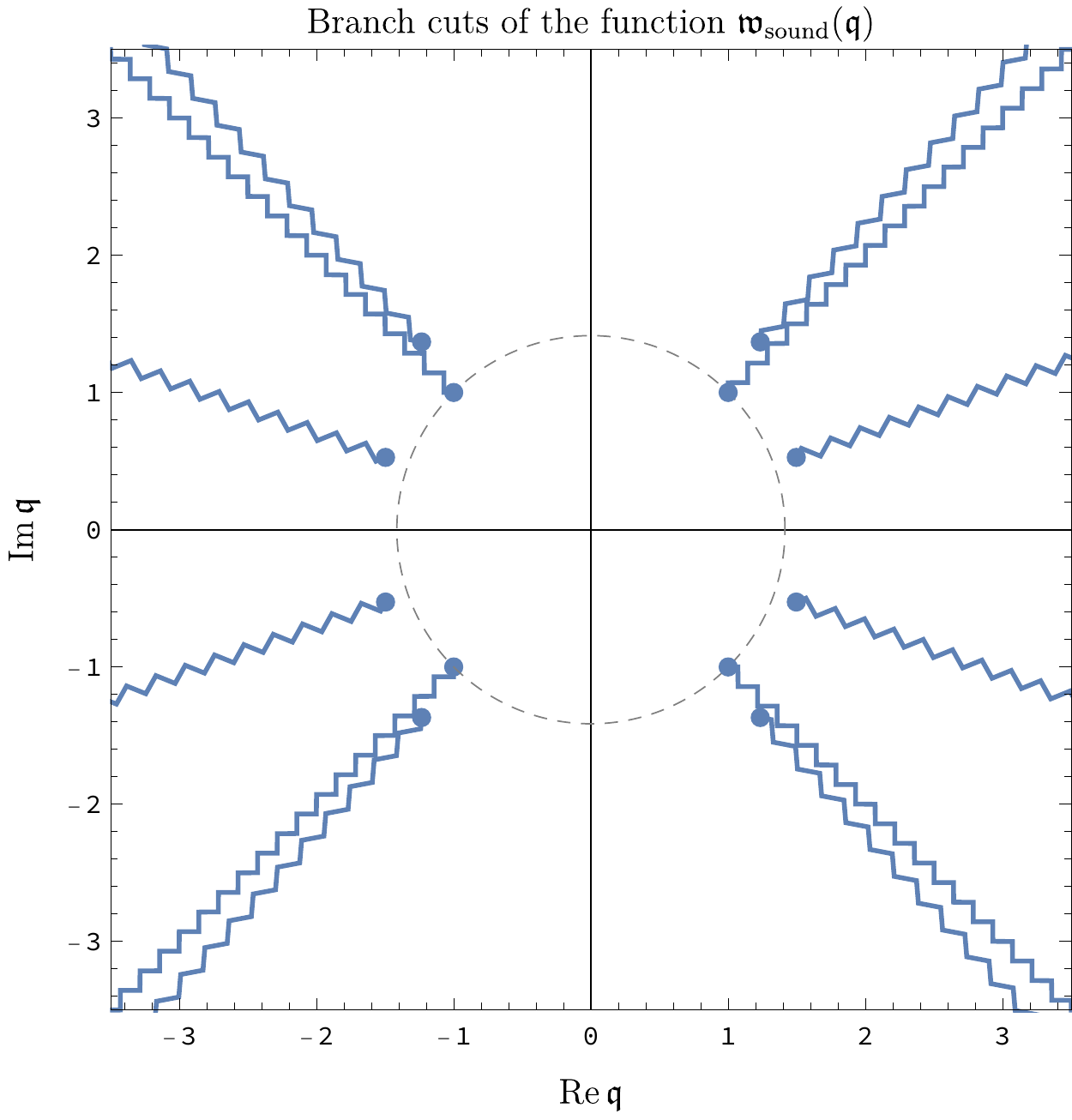}
\caption{{\small Branch point singularities, branch cuts and the domain of hydrodynamic series convergence for the shear mode in the complex $\qfr^2$-plane (left panel) and the sound mode in the complex $\qfr$-plane (right panel).  }
 \label{shear-c-x}
}
\label{fig:cuts-shear-sound}
\end{figure}

From the  discussion above, it is clear that the shear mode dispersion relation  $\wfr_{\rm shear}(\qfr^2)$ is an analytic function 
of complex $\qfr^2$ in the circle $|\qfr^2|<|\qfr_{\rm c}^2| \approx 2.224$. Since the appropriate second derivative of 
the spectral curve at the critical point is non-zero (i.e. $p=2$ in the Puiseux language of section \ref{hydro-series-convergence}; see eq.~\eqref{c-curve-critical-xp}) which corresponds to a collision of {\it two} quasinormal modes, the critical point is the branch point singularity of the square root type, with 
the Puiseux series in powers of $\pm \sqrt{\qfr^2-\qfr_{\rm c}^2}$ providing the extension  beyond the radius of convergence. We show the critical points, the radius of convergence and the appropriate branch cuts in the complex plane of $\qfr^2$ in fig.~\ref{fig:cuts-shear-sound} (left panel).

For the sound mode dispersion relation, considered as a  function of $\qfr^2\in \mathbb{C}$, the origin is a branch point, and 
the corresponding Puiseux series is given by eq.~\eqref{eq:sound-1x}. It will be more convenient to consider the sound dispersion relation 
 $\wfr_{\rm sound}(\qfr)$ as a function of complexified magnitude $\qfr \in \mathbb{C}$ of the wave-vector ${\bf q}$. The critical 
points, the radius of convergence and the appropriate branch cuts in the complex plane of $\qfr$ are shown 
in fig.~\ref{fig:cuts-shear-sound} (right panel).

\subsection{Pole-skipping in the full response functions}
\label{sec:Chaos}
As already discussed in the Introduction, analytically continued hydrodynamic modes appear to be connected 
to the parameters of an OTOC related to the microscopic many-body quantum chaos. The apparent connection 
is provided by the phenomenon of pole-skipping \cite{Grozdanov:2017ajz,Blake:2017ris,Blake:2018leo,Grozdanov:2018kkt}, whereby a pole and a zero of a two-point correlation  function collide for some $\wfr, \qfr \in \mathbb{C}$. In the sound channel of the energy-momentum tensor retarded two-point function, the pole-skipping has been studied  in the context of holography  \cite{Grozdanov:2017ajz,Blake:2018leo,Grozdanov:2018kkt}, effective field theory \cite{Blake:2017ris} and two-dimensional conformal field theory in the limit of large central charge \cite{Haehl:2018izb}. Here, we extend the discussion in refs.~\cite{Grozdanov:2017ajz,Blake:2018leo,Grozdanov:2018kkt} to show that pole-skipping also occurs in other channels.

Consider a retarded two-point function $G^R(\wfr,\qfr)$ of the energy-momentum tensor components at finite temperature. Schematically, and modulo tensor structure, the correlator can be written as 
\begin{align}
G^R(\wfr,\qfr) \sim \frac{b(\wfr,\qfr)}{a(\wfr,\qfr)}\,,
\end{align}
where $a(\wfr,\qfr)$ necessarily contains a gapless hydrodynamic mode $\wfr = \wfr(\qfr)$ (either shear or sound) in the appropriate channels \cite{Kovtun:2005ev}. More generally, let 
${ \cal Z}_d =\{ \wfr = \wfr(\qfr): a(\wfr(\qfr) ,\qfr)=0 \}$ and ${ \cal Z}_n =\{ \wfr = \wfr(\qfr): b(\wfr(\qfr) ,\qfr)=0 \}$, where we assume for simplicity that all zeros are simple. Then, pole-skipping occurs at generically complex $(\qfr_*, \wfr_*)\in \CP =  { \cal Z}_n \cap { \cal Z}_d$.

For theories with available gravity dual descriptions, the set $\CP$ can be determined directly either by computing ${ \cal Z}_n$ and ${ \cal Z}_d$ (the set ${ \cal Z}_d$ is nothing but the quasinormal spectrum) or from the dual gravity equations of motion (see below). In the case of energy-momentum tensor correlators of the ${\cal N}=4$ SYM theory in the limit of infinite $N_c$ and infinite `t Hooft coupling $\lambda$, pole-skipping in the three channels occurs at points $(\qfr_*, \wfr_*)$ given by
\begin{align}
{\rm Sound \, channel} :& &   \qfr_* &= \sqrt{\frac{3}{2}} i \,,  & \wfr_* &= i  \,, \label{pcsound} \\
{\rm Shear \, channel} :&   & \qfr_* &= \sqrt{\frac{3}{2}}  \,,  & \wfr_* &= - i \,,  \label{pcshear} \\
{\rm Scalar \, channel} :&  & \qfr_* &= \sqrt{\frac{3}{2}} i \,,  & \wfr_* &= - i  \,. \label{pcscalar} 
\end{align}
We observe that $|\qfr_*|=\sqrt{3/2}$, $|\wfr_*|=1$ in all three channels. The connection to the Lyapunov exponent $\lambda_L$ and the butterfly velocity $v_B$ is given by the formulae
\begin{align}
{\rm Sound \, channel} :& & \wfr_*(\qfr_*) &= \frac{i \lambda_L}{2\pi T} = i \Ofr_*\,, & \qfr_* &= i \kfr_* \,,  \label{PcSound} \\
{\rm Shear \, channel} :& & \wfr_*(\qfr_*) &= - \frac{i \lambda_L}{2\pi T} =-i \Ofr_*\,, & \qfr_* &=  \kfr_* \,, \label{PcShear} \\
{\rm Scalar \, channel} :& & \wfr_*(\qfr_*) &= - \frac{i \lambda_L}{2\pi T} =-i \Ofr_*\,, & \qfr_* &=  i \kfr_* \,, \label{PcScalar}
\end{align}
where $\kfr_* = \sqrt{3/2}$, $\Ofr_*=1$, and $v_B = \Ofr_*/\kfr_*$. It is clear from figs.~\ref{fig-chaos} and \ref{fig:disp-rel-n=4}  that the sound and the shear dispersion relations pass through their respective "chaos" points \eqref{pcsound}  or \eqref{pcshear}. In the scalar channel, which has no hydrodynamic modes, pole-skipping is exhibited by (one of the pair of) the lowest-lying gapped modes in the spectrum. This can be seen from fig. \ref{fig-chaos-scalar}. Thus, in the ${\cal N}=4$ SYM theory at infinite `t Hooft coupling, the values of $\lambda_L$ and  $v_B$ defined by pole-skipping in the complexified dispersion relations of the lowest-lying modes (either hydrodynamic or gapped) coincide with those obtained from the appropriate limit of 
the OTOC:\footnote{ To subleading order in the inverse 't Hooft coupling expansion, the butterfly velocity 
 is $v_B = \sqrt{2/3} \left( 1 + \frac{23\zeta(3)}{16} \lambda^{-3/2}  \right)$ while the relevant (long-distance) Lyapunov exponent remains uncorrected \cite{Grozdanov:2018kkt}.}
\begin{align}
\lambda_L = 2 \pi T\, ,&& v_B = \sqrt{\frac{2}{3}} \, , &&  \kfr_* = \frac{\lambda_L}{2\pi T v_B} = \sqrt{\frac{3}{2}} \, .
\end{align}
In fact, irrespectively of the channel in question, we can define the (maximal holographic) Lyapunov exponent and the butterfly velocity through the pole-skipping location exhibited by the mode closest to the origin in the complex plane of $\wfr$:
\begin{align}\label{GenDefPS}
\lambda_L = 2 \pi T \left|\wfr_*\right|\, , \qquad v_B = \left| \frac{\wfr_*}{\qfr_*} \right| = \frac{\left| \wfr_* \right|}{\kfr_*}\,.
\end{align}
 
Pole-skipping points $(\qfr_*, \wfr_*)$ can be found directly from the dual gravity equations of 
 motion \cite{Grozdanov:2017ajz,Blake:2018leo}. To show this explicitly for the $\CN = 4$ SYM theory, we follow the arguments of ref.~\cite{Blake:2018leo} and examine the horizon behaviour of Einstein's equations 
\begin{align}
E_{\mu\nu} \equiv R_{\mu\nu} - \frac{1}{2} g_{\mu\nu} R - 6 g_{\mu\nu} =0
\end{align}
in the infalling Eddington-Finkelstein (EF) coordinates $(v,r,x^i)$ with 
\begin{align}
v = t + r_*(r)\,, \qquad \frac{dr_*}{dr} = \frac{1}{r^2 f(r)} \, .
\end{align}  
We perturb the $5d$ AdS-Schwarzschild metric $ds^2 = g_{\mu\nu} dx^\mu dx^\nu = -r^2f(r) dv^2 + 2 dv dr + r^2 d\vec{x}^2$ to first order, $g_{\mu\nu} \to g_{\mu\nu} + \delta g_{\mu\nu} (r) e^{-i\omega v +i k z}$, and expand the (regular) metric fluctuations around the horizon $r = r_0$ as
\begin{align}
\delta g_{\mu\nu} (r) = \sum_{n=0}^{\infty} \delta g_{\mu\nu}^{(n)} (r-r_0)^n \, .
\end{align}
Similarly to what was observed in ref.~\cite{Blake:2018leo} for the sound channel, we find that in any channel,  
there  exists a linear combination of the components $E_{\mu\nu}$, which vanishes identically at the horizon 
$r\to r_0$ at the pole-skipping values of the parameters $(\qfr_*, \wfr_*)$. Explicitly, 
\begin{align}
{\rm Sound \, channel} &:  &\lim_{r\to r_0} E_{vv} &= 0\qquad  \mbox{at} & (\qfr_*,\wfr_*)  &= (\sqrt{3/2} i, i)  \,,\label{pcsoundeom} \\
{\rm Shear \, channel} &:   &\lim_{r\to r_0} \left(E_{vx} + i \sqrt{\frac{2}{3}} E_{xz}\right)   &= 0\qquad  \mbox{at}  &(\qfr_*,\wfr_*)  &= (\sqrt{3/2}, -i) \,,
 \label{pcsheareom} \\
{\rm Scalar \, channel} &: & \lim_{r\to r_0} E_{xy} &= 0\qquad  \mbox{at}  &(\qfr_*,\wfr_*)  &= (\sqrt{3/2}i, -i) \,.
 \label{pcscalareom}
\end{align}
In other words, pole-skipping occurs at values of the parameters $(\qfr_*, \wfr_*)$ for which the rank of the matrix $E_{\mu\nu}$ decreases at the horizon.

 We note also that for the $\CN = 4$ SYM theory, the chaos point $|\qfr_*^2|=3/2$ lies within the radius of convergence of the hydrodynamic series in both the shear ($|\qfr_{\rm c}^2|\approx 2.224$) and the sound ($|\qfr_{\rm c}^2| = 2$) channels.

\section{Pole-skipping and level-crossing in $2d$ thermal CFT correlators}
\label{sec:pskk}
 In a $2d$ CFT, the (equilibrium) retarded finite-temperature two-point correlation function of an operator of non-integer scaling dimension $\Delta$ and spin zero in momentum space is given by the expression\footnote{ In ref.~\cite{Son:2002sd}, 
 the expression for  $G^{R}(\omega , q)$ was derived holographically from dual gravity. For integer $\Delta$, it was further  checked in ref.~\cite{Son:2002sd} that thus obtained formula coincides (up to normalisation) with the retarded correlator obtained from $2d$ CFT.}\textsuperscript{,}\footnote{ The expression for $G^{R}(\omega , q)$ in the form (4.16) of ref.~\cite{Son:2002sd} 
 assumes $\wfr, \qfr \in \mathbb{R}$. To be valid for generic $\wfr, \qfr \in \mathbb{C}$, it has to be rewritten in the form 
 \eqref{full_green_ni}. We would like to thank D.~Vegh for pointing this out.} \cite{Son:2002sd}
\begin{equation}
\begin{split}
 G^{R}(\wfr , \qfr) &= C_\Delta \, 
\Gamma \left( \frac{\Delta}{2}+  {i (\wfr - \qfr)\over 2 }\right) \Gamma \left(   \frac{\Delta}{2}+  {i (\wfr + \qfr )\over 2 }      \right) \Gamma \left( \frac{\Delta}{2}-  {i (\wfr - \qfr)\over 2 }\right)
  \\
  & \times \Gamma \left(   \frac{\Delta}{2}-  {i (\wfr + \qfr )\over 2 }      \right) \biggr[ 
\cosh{ \left(\pi \qfr\right)  } -\cos{\left(\pi\Delta\right)}
\cosh{ \left( \pi \wfr\right)  }  + i \sin{ \left(\pi\Delta\right)}
\sinh{\left( \pi \wfr\right) } 
\biggr]\,,
\label{full_green_ni}
\end{split}
\end{equation}
where $C_\Delta$ is the normalisation constant, and we put $T_L=T_R=T$. Note also that 
here, in $1+1$ dimensions, the symbol $\qfr$ denotes $q/2\pi T$, rather than $|q|/2\pi T$. Similar formulae can be written for integer 
 $\Delta$ \cite{Son:2002sd}. The correlator \eqref{full_green_ni} has a sequence of poles at
\begin{equation}
\wfr_n (\qfr)  = \pm \qfr - i \left( 2 n +\Delta\right)\,,
\label{btz-qnm}
\end{equation}
where $n=0,1,2,\ldots$. These are precisely the quasinormal frequencies of the dual BTZ black hole \cite{Birmingham:2001pj}, \cite{Son:2002sd}. In this section, we shall examine these correlators for their pole-skipping and level-crossing properties.\footnote{ Similar issues have been recently independently studied in  ref.~\cite{Blake:2019otz}. The results of  ref.~\cite{Blake:2019otz} agree with ours whenever they overlap.}\textsuperscript{,}\footnote{ Here, we only consider correlation functions of $2d$ CFT operators with scaling dimension $\Delta$ and spin $s=0$. The energy-momentum, having $\Delta=2$ and $s=2$, is not of this type. Its finite-temperature two-point function 
(see e.g.\ refs.~\cite{Haehl:2018izb}, \cite{Datta:2019jeo}) in momentum space has a pole corresponding to a mode propagating on the light cone. The corresponding dispersion relation line passes through the ``chaos'' point of that correaltor \cite{Haehl:2018izb}, just as it does in $4d$.}
\subsection{Pole-skipping in the full response functions}
\label{sec:BTZ-PS}
The zeros of  the correlator \eqref{full_green_ni}  come from the zeros of the expression in the square brackets,
\begin{align}
\,& \cosh{\left( \pi \qfr\right)  } -\cos{ \left(\pi\Delta\right)} \cosh{ \left( \pi \wfr \right) }  + i \sin{ \left(\pi\Delta\right)}
\sinh{ \left(\pi \wfr\right) } \nonumber \\
\,& =2 \sin{\left[\frac{\pi}{2} (\Delta + i\wfr - i \qfr ) \right]}  \sin{\left[\frac{\pi}{2} (\Delta + i\wfr + i \qfr ) \right]}\,,   
\label{trigid}
\end{align}
and are given by
\begin{align}
 \,& \Delta + i \wfr - i \qfr = 2 n_1\,, \label{zc1}  \\
  \,& \Delta + i \wfr + i \qfr = 2 n_2\,, \label{zc2}
\end{align}
where $n_1, n_2 = 1,2,3,\ldots$. Note that the zeros of eq.~\eqref{trigid} with $n_1, n_2=0,-1,-2,\ldots$ are not  among the zeros of the correlator since they are identically (i.e. for arbitrary $\wfr$, $\qfr$) cancelled by the poles of the first two Gamma-functions in eq.~\eqref{full_green_ni}. 

The pole-skipping phenomenon in $G^R$ occures for $\wfr$ and $\qfr$ simultaneously 
satisfying the conditions \eqref{btz-qnm} and \eqref{zc1}, \eqref{zc2}, i.e. for 
\begin{align}
\qfr_* &= \pm i (\Delta + n-n_*)\,, \\ 
 \wfr_* &= -i (n+n_*)\,,
\label{psqw}
\end{align}
where $n=0,1,2,\ldots$ and $n_*=1,2,\ldots$ (here $n_*$ denotes either $n_1$ or $n_2$), and $\Delta$ is not an
 integer.\footnote{ Introducing $Q=n_*$ and $N=n+n_*$, the pole-skipping condition can be written as 
$\qfr_* = \pm i (\Delta + N - 2 Q)$, $\wfr_* = -i N$, where $N=1,2,...$ and $Q=1,\ldots,N$. This coincides with the results in ref.~\cite{Blake:2019otz}. We would like to thank R.~Davison for pointing out an error in eqs. (5.6), (5.7) in
 the first version of this paper.}

 We also note that the Euler reflection formula, $\Gamma(z)\Gamma(1-z)=\pi/\sin{\pi z}$, can be used to rewrite the correlator  \eqref{full_green_ni} in the form
\begin{equation}
 G^{R}(\wfr , \qfr) =  C_\Delta \, 
\frac{2 \pi^2 \Gamma \left( \frac{\Delta}{2}-  {i (\wfr - \qfr)\over 2 }\right)
  \Gamma \left(   \frac{\Delta}{2}-  {i (\wfr + \qfr )\over 2 }      \right) }{ 
  \Gamma \left( 1- \frac{ \Delta}{2}-  {i (\wfr - \qfr)\over 2 }\right)
  \Gamma \left(  1-  \frac{\Delta}{2}-  {i (\wfr + \qfr )\over 2 }      \right)  }\,.
\label{full_green_ni_alt}
\end{equation}

\begin{figure}[h!]
\centering
\includegraphics[width=0.7\textwidth]{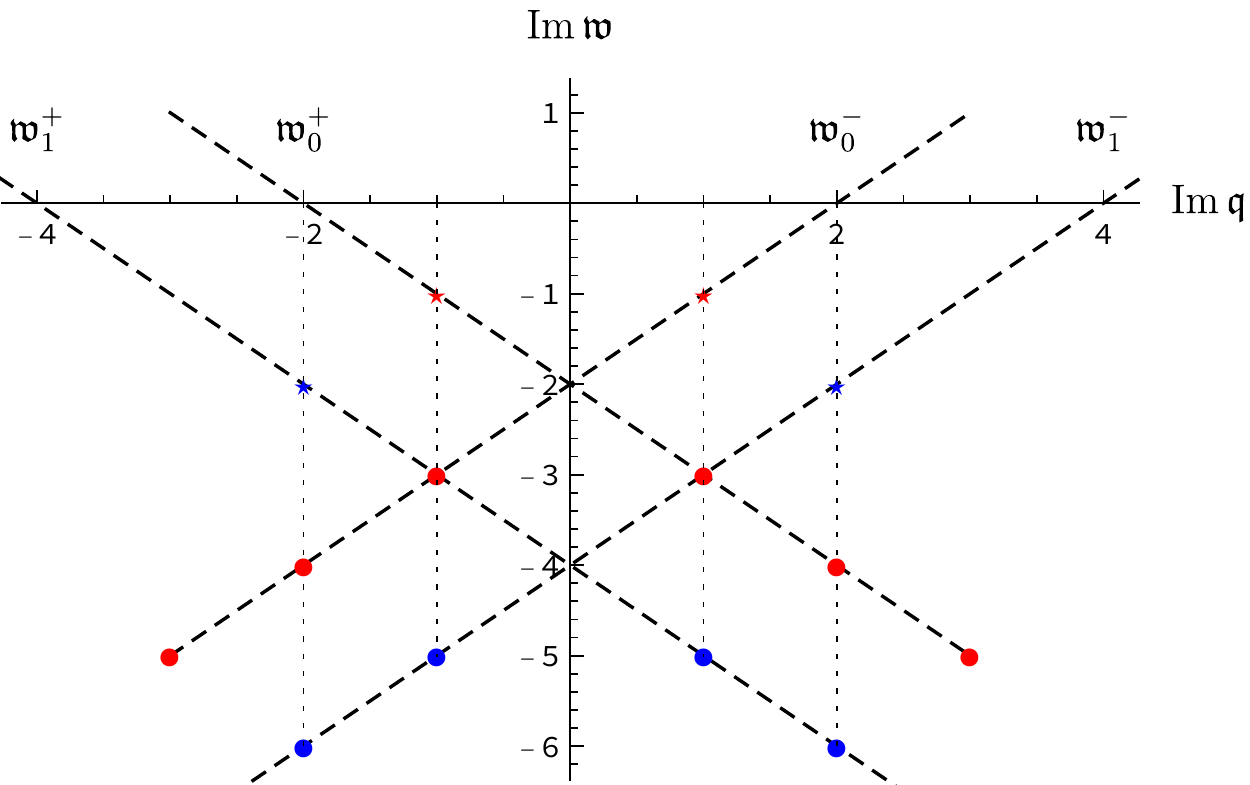}
\caption{{\small  Pole-skipping and level-crossing points in a $2d$ CFT correlator of 
operators with conformal dimension $\Delta=2$, for the smallest in magnitude poles $\wfr_0^\pm$ and $\wfr_1^\pm$. The red stars at $\mbox{Im}\, \qfr =\pm 1$ correspond to the pole-skipping points and the red dots label the critical points of level-crossings of $\wfr_0^\pm$. The blue stars at $\mbox{Im}\, \qfr =\pm 2$ correspond to the pole-skipping points and the blue dots label the critical points of level-crossings of $\wfr_1^\pm$.} }
\label{fig-chaos-btz-combo}
\end{figure}

For integer $\Delta$, the poles of $G^R$ are still given by eq.~\eqref{btz-qnm}, but the functional form of the 
correlators is somewhat different from  \eqref{full_green_ni} (see ref.~\cite{Son:2002sd}). Here, we focus on the case of $\Delta =2$. For $\Delta =2$, one has \cite{Son:2002sd}
\begin{align}\label{G-btz-Integer}
G^R \sim \left( \wfr^2 - \qfr^2\right) \left[  \psi \left( 1 - \frac{i}{2} \left( \wfr - \qfr\right)\right) + \psi \left( 1 - \frac{i}{2} \left( \wfr + \qfr\right)\right) \right]\, ,
\end{align}
 where $\psi(x)=\Gamma'(x)/\Gamma(x)$. The singularities of the correlator \ref{G-btz-Integer} 
 are simple poles located at 
\begin{equation}
\wfr_n (\qfr)  = \pm \qfr - 2 i \left( n + 1\right)\,, \qquad n=0,1,2,\ldots\,.
\label{btz-qnm-d2}
\end{equation}
In the case of pole-skipping, they are cancelled by the zeros coming from the numerator $\wfr^2-\qfr^2$, which occur when $\wfr_n = \mp \qfr$. The discrete set of momenta $\qfr$ that satisfies this condition is
\begin{equation}
\qfr = \pm i (n + 1)\,, \qquad n=0,1,2,\ldots\,.
\label{psqi}
\end{equation}
Therefore, the pole-skipping points for $\Delta=2$ are given by
\begin{align}
\qfr_* &= \pm i (n+1)\,, \\
\wfr_* &= -i (n+1)\,,
\label{psqw}
\end{align}
where $n=0,1,2,\ldots$. In particular, for the pair of poles that lies closest to the origin in the complex $\wfr$ plane (ones with $n=0$ among those in eq. \eqref{btz-qnm-d2}),
\begin{equation}
\wfr_0^\pm (\qfr)  = \pm \qfr - 2 i\,,
\label{btz-qnm-2c}
\end{equation}
the branch $\wfr_0^+$ passes through the (lowest-lying) pole-skipping point $\qfr_*=i$, $\wfr_*=-i$, whereas the branch $\wfr_0^-$ does not. The branch $\wfr_0^-$ passes through the pole-skipping point $\qfr_*=-i$, $\wfr_*=-i$, whereas the branch $\wfr_0^+$ does not (see fig.~\ref{fig-chaos-btz-combo}). Finally, we note that for  $\Delta=1$, the correlator is directly proportional to the sum of two $\psi-$functions  \cite{Son:2002sd}, and there is no pole-skipping.

\subsection{BTZ spectrum level-crossing}
\begin{figure}[h!]
\centering
\includegraphics[width=0.5\textwidth]{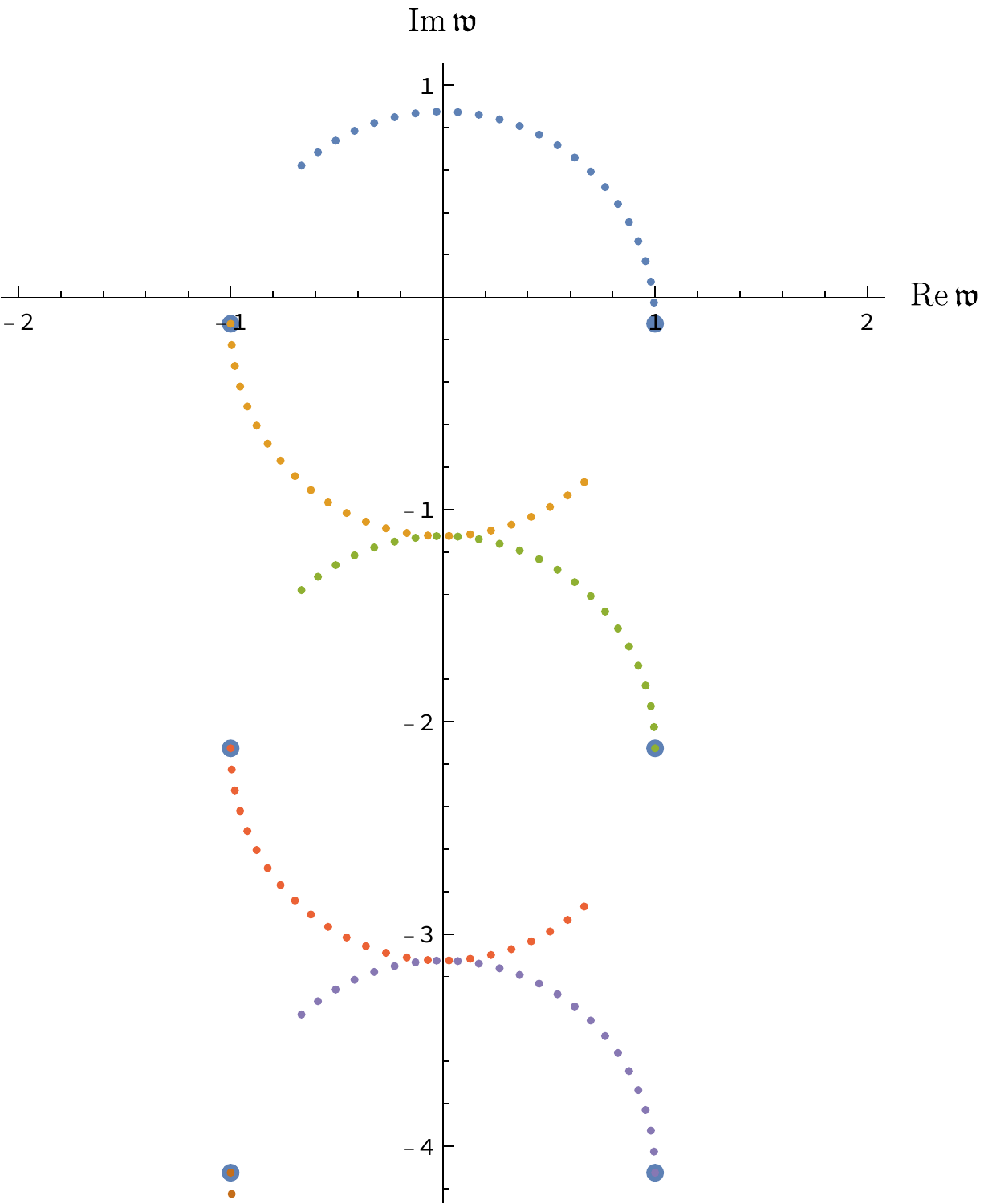}
\caption{{\small BTZ quasinormal spectrum for $\Delta=1/8$ in the complex $\wfr$-plane at complex momentum $\qfr = |\qfr|e^{i \theta}$, 
with $|\qfr|=1$ and $\theta$ changing from $0$ to $3\pi/4$. The spectrum at $\theta=0$ is shown by large dots.}}
\label{fig-chaos-btz}
\end{figure}

 Since the quasinormal spectrum $\wfr_n^{\pm} (\qfr)$ is known explicitly (see eq.~\eqref{btz-qnm}), the level-crossing points can be found directly. Such level-crossing points were used above in theories with gapless excitations to determine the radius of convergence of their hydrodynamic series. Considering complex $\qfr = |\qfr| e^{i\theta}$, we have
\begin{align}
\mbox{Re}\,  \wfr_n^\pm &= \pm |\qfr| \cos{\theta} \equiv X \,,\\
\mbox{Im}\, \wfr_n^\pm &= \pm |\qfr| \sin{\theta} - 2 n -\Delta \equiv Y\,,
\end{align}
and thus the orbits followed by the poles  in the complex $\wfr$ plane when the phase $\theta$ changes from  $0$ to $2\pi$ are circles
\begin{align}
X^2 +\left( Y + 2n +\Delta \right)^2 = |\qfr|^2\,, \qquad n=0,1,2, \ldots \,.
\end{align}
The poles move counter-clockwise and collide on the imaginary axis of $\wfr$ (moreover, at integer values of $|\wfr|$ if $\Delta$ is an integer).  More precisely, two poles collide when $\wfr_n^-=\wfr_m^+$, $m\neq n$, i.e. when (the case $n=0$ or $m=0$  is treated separately below) 
\begin{align}
\qfr_{\rm c} &= i \left( m-n\right)\,, \label{cp1} \\
\wfr_{\rm c} &= -  i \left( m+n + \Delta \right)\,, \label{cp2} \\
m,n &= 1,2,\ldots,\, \text{with}~ m\neq n\,,
\end{align}
with the first collision occurring for $m=n\pm1$, i.e. for $\qfr_{\rm c}=\pm i$, $\wfr_{\rm c} = - i \left(  2 n + \Delta \pm 1\right)$, $n=1,2,\ldots$. The mode with $n=0$, i.e. $\wfr_0^\pm = \pm \qfr - i \Delta$, has only one neighbour, and the critical points correspond to $\wfr_0^-=\wfr_n^+$ (i.e. $\qfr_{\rm c}=i n$ and $\wfr_{\rm c}=-i (n+\Delta)$) or $\wfr_n^-=\wfr_0^+$ (i.e. $\qfr_{\rm c}=-i n$ and $\wfr_{\rm c}=-i (n+\Delta)$, $n=1,2,\ldots$). Thus, the zero mode critical points are
\begin{align}
\qfr_{\rm c} &= \pm i n \,, \label{cp1-0} \\
\wfr_{\rm c} &= -i \left(n+\Delta\right)\,, \label{cp2-0} \\
n&=1,2,\ldots\,.
\end{align}
The motion of poles in the complex frequency plane and their level-crossings are illustrated for  $\Delta=1/8$ and $|\qfr|=1$ in fig.~\ref{fig-chaos-btz}.

For larger $|\qfr|$, the trajectories intersect but the poles miss each other, so there is no phenomenon of one trajectory crossing into and continuing as the other. In a sense, here, we have ``level-touching'' rather than ``level-crossing''.  The nearest critical points are thus 
\begin{align}
\qfr_{\rm c} &= \pm i\,, & \wfr_{\rm c} &= - i \left(  2 n + \Delta \pm 1\right)\,, & n&=1,2,\ldots\,, \\
\qfr_{\rm c} &= \pm i\,, & \wfr_{\rm c} &= - i \left(  \Delta + 1\right)\,, & n&=0\,.
\end{align}
In particular, for $\Delta=2$, we have\footnote{ These critical points are single poles: the collisions may occur at the same point $\wfr_{\rm c}$, but the phases are different for different modes.}
\begin{align}
 \qfr_{\rm c} &=  i\,, & \wfr_{\rm c} &= -3i, -5 i, -7i, -9 i \ldots\,, \\
 \qfr_{\rm c} &= - i\,, &\wfr_{\rm c} &=  -3i,  -3i, -5 i, -7 i,\ldots \,.
\end{align}
 We observe that for non-integer $\Delta$, the values of $\qfr$ corresponding to the pole-skipping and the level-crossing do not coincide. The same is (trivially) true for $\Delta=1$ as well, where there is no pole-skipping in the correlator at all. For $\Delta=2$, however, a curious picture emerges.
Consider again the ``sound'' mode $\wfr_0^\pm$ \eqref{btz-qnm-2c}. The imaginary parts of the two branches obey
\begin{align}
\,& \mbox{Im}\, \wfr_0^\pm = \pm\mbox{Im}\, \qfr - 2\,.
\end{align}
The pole-skipping points found in section \ref{sec:BTZ-PS} are $\qfr_*=i$, $\wfr_*=-i$ and $\qfr_*=-i$, $\wfr_*=-i$, and the critical points are $\qfr_{\rm c}=\pm i n$, $\wfr_{\rm c}=-i (n+2)$, $n=1,2,\ldots$. For $\Delta=2$, and the mode with $n=1$,
\begin{equation}
\wfr_1^\pm (\qfr)  = \pm \qfr - 4 i\,,
\label{btz-qnm-4c}
\end{equation}
the pole-skipping occurs at $\qfr=\pm2 i$ and $\wfr=-2i$. The critical points are located at $\qfr_{\rm c}=\pm i (m-1)$, $\wfr_{\rm c}=-i (m+3)$, $m=2,3,\ldots$. This is illustrated in fig.~\ref{fig-chaos-btz-combo}.

\section{Discussion}
\label{disc}
In this paper, we introduced spectral curves as a useful tool for investigating analytic properties of gapless collective excitations in classical hydrodynamics.\footnote{ See footnote \ref{fn1}.} We showed that the dispersion relations of hydrodynamic modes, such as shear and sound modes, are generically given by Puiseux series expansions in rational powers of the spatial momentum squared. These series are guaranteed to converge in some vicinity of the origin (the point with zero frequency $\omega = 0 $ and zero spatial momentum $\q^2 =0 $ in the $(\omega,\qfr^2) \in \mathbb{C}^2$ space), so long as the analyticity of the spectral curve at the origin can be proven (e.g. by holographic or other means). Thus, given the analyticity of the spectral curve, the asymptotic nature of the series for hydrodynamic modes in momentum space can be automatically ruled out.   The radius of convergence of the series is given by the distance from the origin to the critical point of the spectral curve nearest to the origin. After developing the general theory, we then used holography as a theoretical laboratory where all these of features can be seen and analysed explicitly. Before focusing on the main example of the strongly coupled ${\cal N}=4$ supersymmetric Yang-Mills theory at finite temperature, to illustrate our method, we first considered the  holographic model with broken translation symmetry, where an exact spectral curve is available. We have shown that the critical points of the spectral curves can be found by studying quasinormal spectra at complex values of spatial momentum: the critical points correspond to the collisions of quasinormal frequencies (poles of dual correlators) in the complex frequency plane at critical (generically, complex) values of spatial momentum. These values also set the radii of convergence for the dispersion relation.   We call this phenomenon the quasinormal mode level-crossing, in analogy with the well-known phenomenon of level-crossing for eigenvalues of Hermitian operators.

Applying these methods to the strongly coupled ${\cal N}=4$ supersymmetric Yang-Mills theory, we found that the gradient expansions for the hydrodynamic shear and sound modes have finite radii of convergence given by $q_{\rm sound}^{\rm c} = \sqrt{2}\, \omega_0$ for the sound mode and by $q_{\rm shear}^{\rm c}\approx 1.49\, \omega_0$ for the shear mode, where $\omega_0= 2 \pi T$ is the fundamental Matsubara frequency. Thus, in both channels, the hydrodynamic series converge up to the order of $|\q|/ T \sim O(10)$, which is a vast improvement over the naive expectation that $|\q|/T \ll 1$ provides a natural expansion parameter for hydrodynamic dispersion relations. The obstruction to convergence in the example of  the ${\cal N}=4$ SYM theory comes from the collision of poles of the two-point correlation function of the energy-momentum tensor at complex $q^2$. 

As mentioned in the Introduction, the problem of extending the hydrodynamic modes in the complex momentum plane beyond the branch point singularity was recently investigated by Withers \cite{Withers:2018srf} in the context of a holographic model in $2+1$ dimensions with finite chemical potential. The shear-diffusion mode series could be Pad\'{e}-resummed and extended beyond the branch point singularity, which was in that case located at an imaginary value of momentum. The main focus of ref.~\cite{Withers:2018srf} was on the possibility of recovering the full spectrum from the hydrodynamic derivative expansion, similar to recovering the non-hydrodynamic modes from asymptotic series via Borel resummation and resurgence \cite{Heller:2013fn,Heller:2015dha,Casalderrey-Solana:2017zyh}. The quasinormal spectrum in the holographic models with finite temperature $T$ and non-vanishing 
chemical potential $\mu$ such as the one considered in ref.~\cite{Withers:2018srf} is rather complicated and changes qualitatively as the parameter $T/\mu$ is varied (in particular, it involves pole collisions even at real values of the momentum) \cite{Edalati:2010hk,Edalati:2010pn,Brattan:2010pq,Davison:2011uk,Davison:2011ek,Gushterov:2018spg,Gushterov:2018nht}. In ref.~\cite{Withers:2018srf}, the shear-diffusion mode was found to have a radius of convergence inversely proportional to the chemical potential. Naively, this would imply infinite radius of convergence in the limit of vanishing $\mu$, in apparent contradiction with our results. However, the result of ref.~\cite{Withers:2018srf} was obtained at a specific fixed value of $T/\mu$, and we expect it to change when the complex momentum behaviour of other gapped modes in the model is taken into account with $T/\mu$ increasing. This will require further study. It would be also interesting to extend the results of the present work to the sound channel (not considered in  ref.~\cite{Withers:2018srf}) as well as to other holographic models with finite chemical potential such as the STU model \cite{Son:2006em}, and other models \cite{Anantua:2012nj}, \cite{Betzios:2017dol,Betzios:2018kwn}, including those in the large $D$ limit \cite{Casalderrey-Solana:2018uag}.

Pole collisions in the correlation functions appear in holographic models in different contexts \cite{Davison:2011ek,Davison:2014lua,Grozdanov:2016vgg,Grozdanov:2016fkt,Grozdanov:2018ewh,Baggioli:2018nnp,Baggioli:2018vfc,Grozdanov:2018fic,Grozdanov:2018gfx,Gushterov:2018spg,Davison:2018ofp,Baggioli:2019jcm}. No less interesting are collisions among poles and zeros of the correlators known as pole-skipping \cite{Grozdanov:2017ajz,Blake:2017ris,Blake:2018leo,Grozdanov:2018kkt}. What we have shown here is that this phenomenon, known to exist in the sound channel of strongly coupled ${\cal N}=4$ SYM theory \cite{Grozdanov:2017ajz}, exists also in the shear and scalar channels of the energy-momentum correlators. The  conjectured connection to the OTOC thus allows one to determine the parameters quantifying microscopic many-body chaos (scrambling time and butterfly velocity) by considering the complexified behaviour of the lowest-lying modes (those with the smallest $|\omega|$ in the spectrum) in any channel, be it a channel with or without hydrodynamic modes. In general, the critical points and the pole-skipping points are different. We have analysed the $2d$ CFT finite-temperature correlators and the spectra of the dual BTZ black hole to demonstrate this explicitly. What this implies for the relation between chaos and hydrodynamics is that the ``chaos'' (or pole-skipping) points can lie within or outside of the radius of convergence of the hydrodynamic series. In particular, while this is not the case in the  holographic model with broken translation symmetry considered in section \ref{DG-model}, we found that in the ${\cal N}=4$ SYM theory, pole-skipping points for both of the two hydrodynamic modes lie within the radius of convergence of the corresponding dispersion relations. This provides an explanation for the observation of the fast convergence of the hydrodynamic series to the exact chaos point in ref. \cite{Grozdanov:2017ajz}.

Can finiteness of the convergence radius of the hydrodynamic modes dispersion relations expansion be taken as a criterion for validity of hydrodynamics? By analogy, one may think of a free particle whose dispersion relation $\omega = \sqrt{p^2+m^2} -m = p^2/2m + \ldots$ has branch points located at $p=\pm i m$, and for which the failure of the convergence of the gradient expansion 
corresponds to the breakdown of the non-relativistic approximation. We hope 
our results may be of interest for  studies of higher-order 
hydrodynamics necessary for improving the precision of hydrodynamic predictions and 
also for justifying the construction of the effective field theory of hydrodynamics 
formulated as a gradient expansion \cite{Dubovsky:2011sj,Grozdanov:2013dba,Kovtun:2014hpa,
Haehl:2015foa,Crossley:2015evo,Haehl:2015uoc,Glorioso:2016gsa,Jensen:2017kzi,
Glorioso:2018wxw,Banks:2018aob,Chen-Lin:2018kfl}. As already mentioned in ref.~\cite{Grozdanov:2019kge} in the context of the discussion of
 the ``unreasonable effectiveness''  of hydrodynamics as an effective theory, many previous studies have reported the divergence of 
 the derivative expansion in relativistic hydrodynamics~\cite{Heller:2013fn,Heller:2015dha,Buchel:2016cbj,Casalderrey-Solana:2017zyh,Heller:2015dha}. Possibly, the asymptotic nature of the expansion appearing in those publications should be viewed as a reflection of the singular nature of the state about which this expansion is performed, rather than a generic property of the hydrodynamic gradient expansion itself. On the other hand, even for a free particle, the momentum space and position space pictures look different in this respect: the small-momentum expansion of the corresponding dispersion relation has a finite radius of convergence, whereas e.g. for the position space propagator, the large-time expansion is only 
asymptotic.\footnote{ We thank Hong Liu for pointing out this example to us.} This issue needs further investigation. The role of the non-hydrodynamic degrees of freedom is the common feature of the mentioned works and the present paper.

Of special interest is the dependence of the radii of convergence on coupling. In ref.~\cite{Grozdanov:2019kge}, using eq.~\eqref{rad-sound-1} as a crude approximation and the results of refs.~\cite{Buchel:2004di,Buchel:2008sh}, we argued that in the ${\cal N}=4$ SYM theory, the radius of convergence is smaller at weaker coupling. This, of course, requires the actual study of the spectrum at finite coupling. More generally, in the context of the problem of interpolating between weak and strong coupling regimes of the same theory at finite temperature  \cite{Grozdanov:2016vgg,Grozdanov:2016zjj}, one may note\footnote{ We would like to thank J.~Noronha for bringing
 ref.~\cite{mclennan} to our attention.} that the problem of convergence of hydrodynamic series has been raised and partially investigated in the 1960s in kinetic theory \cite{mclennan}. This approach, together with recent studies of relevant issues at weak coupling \cite{Romatschke:2015gic,Kurkela:2017xis,Moore:2018mma,Grozdanov:2018atb}, may deserve more attention in the context of the problem of the validity of the hydrodynamic description at finite coupling.

\acknowledgments{\small S.G. and A.O.S. would like to thank the organisers of  the programme ``Bounding transport and chaos in condensed matter and holography'' at Nordita, where part of this work was initiated. S.G. is supported by the U.S. Department of Energy under grant Contract Number DE-SC0011090. S.G. would also like to thank H.~Liu, D.~Sustretov and W.~Taylor for stimulating and illuminating discussions, and for suggestions on the relevant algebraic geometry literature. P.K. would like to thank the Rudolph Peierls Centre for Theoretical Physics at the University of Oxford for hospitality during the initial stage of this work, and the organisers of the KITP programme ``Chaos and Order'', where part of this work was completed. P.K.'s work was supported in part by NSERC of Canada. A.O.S. would like to thank Moscow State University and especially A.V.~Borisov, as well as the Kadanoff Center for Theoretical Physics at the University of Chicago for hospitality during the final stage of this work. He also thanks P.~Glorioso, D.T.~Son, M.~Stephanov, P.B.~Wiegmann and F.~Essler, J.~March-Russell, S.~Parameswaran for discussions in Chicago and Oxford, respectively, and the participants of the seminars at the Institute for Nuclear Research, Steklov Mathematical Institute and Lebedev Physical Institute of the Russian Academy of Sciences for critical questions and useful suggestions. The work of P.T. is supported by an Ussher Fellowship from Trinity College Dublin. We would like to thank A.~Buchel, M.~Heller, A.~Kurkela and J.~Noronha  for illuminating correspondence. We also would like to thank B.~Withers for correcting our inadvertent partial misrepresentation of his results in the first version of our preprint \cite{Grozdanov:2019kge}.}

\appendix

\section{Analytic implicit function theorem and Puiseux series}
\label{puiseux-app}
Here, we collect the necessary information from complex analysis regarding the following problem. Given an implicit  function $f(x,y)=0$, where $x,y \in  \mathbb{C}$, we would like to find  explicit solution(s) in the form $y = y(x)$, at least locally in the vicinity of some point $(x_0,y_0)$, where $y(x)$ may be represented by a finite or infinite series in $x$. We would like to determine, furthermore, what sets the radius of convergence of such series.  

A simple example is provided by the function $f(x,y)=x^2+y^2-1=0$. Since $f(x,y)$ is a polynomial, it determines a complex algebraic curve. Singular  points of $f(x,y)$ are determined by the simultaneous solution of the equations $f(x,y)=0$, $f_{,x} (x,y)=0$, $f_{,y}(x,y)=0$, where the comma subscript denotes the partial derivative with respect to the argument after the comma. Clearly, this particular curve has no singular points. It does, however, have the so-called ``points of multiplicity 1'' or ``one-fold points'', where $f_{,x} (x,y)=0$ or $f_{,y} (x,y)=0$ (but not both simultaneously). These are sometimes called critical points. We are interested in the local behaviour of $y=y(x)$ near a critical point defined by the conditions $f(x,y)=0$, $f_{,y}(x,y)=0$. In our example, there are two such points: $(x,y)=(\pm 1,0)$. The series representation $y=y(x)$ in the vicinity of e.g. 
$(x,y)=(1,0)$ has two branches:
\begin{align}
\,& y=  y_1(x) = i \sqrt{2}  (x-1)^{\frac{1}{2}} + i 2^{-\frac{3}{2}} (x-1)^{\frac{3}{2}} + \cdots\,,  \\
\,& y=y_2(x) = -i \sqrt{2}  (x-1)^{\frac{1}{2}} - i 2^{-\frac{3}{2}} (x-1)^{\frac{3}{2}} + \cdots\,.
\end{align}
This is an example of the Puiseux series, i.e. the power series with fractional exponents. These series converge in the circle with the centre at $(x,y)=(1,0)$ and radius $R=2$ which is the distance from $(1,0)$ to the nearest critical point, $(x,y)=(-1,0)$.

One may be interest in the behaviour $y=y(x)$ in the vicinity of a regular point, where $f_{,y}(x,y)\neq 0$, for example, near $(x,y)=(0,1)$ in our example. Here, since $f_{,y}(x,y)\neq 0$, the implicit function theorem guarantees that we can compute the derivatives $y'(x)$, $y''(x)$ and so on, and represent $y(x)$ by the Taylor series in the vicinity of $x=0$,
\begin{align}
y = y(x) = 1 - \frac{x^2}{2}  - \frac{x^4}{8} + \cdots\,.
\end{align}
This series is convergent in the circle of radius $R=1$, determined by the distance from  the point $x=0$ to the nearest critical point(s) at $x=\pm 1$.

In general, for an implicit function given by the equation $f(x,y)=0$, where $f(x,y)$ is either a finite polynomial in $x$ and $y$, or an analytic function at a point $(x,y)$ (i.e. a polynomial of an infinite degree), the behaviour at a regular point is governed by the analytic implicit function theorem \cite{implicit-function-theorem}, and the behaviour in the vicinity of a critical point is determined by the Puiseux theorem. In the former case, $y=y(x)$ is represented by a Taylor series converging in some vicinity of a regular point. In the latter case, it is represented by a Puiseux series converging in some vicinity of a critical point. We now recall some facts from complex analysis \cite{gunning-rossi} and explain the Puiseux construction \cite{walker}, \cite{wall}. 

\noindent
{\bf Definition:} {\it A function, $f(x,y)$, from a neighbourhood of $(x_0,y_0) \in  \mathbb{C}^2$  to $ \mathbb{C}$ is called analytic at $(x_0,y_0)$ if near $(x_0,y_0)$ it is given by the uniformly convergent power series
\begin{align}
f(x,y) = \sum\limits_{n,m=0}^\infty a_{nm} (x-x_0)^n (y-y_0)^m\,.
\end{align}
}

\noindent
{\bf Analytic implicit function theorem:} {\it 
If $f(x_0,y_0)=0$ and $f_{,y}(x_0,y_0)\neq 0$, there exist $\epsilon >0$ and $\delta >0$ so that $\mathbb{D}_\epsilon (x_0)  \times 
\mathbb{D}_\delta (y_0)$ is in the neighbourhood where $f$ is defined, and $g$ is a map of $\mathbb{D}_\epsilon (x_0)$ into $\mathbb{D}_\delta (y_0)$
so that $f(x,g(x))=0$ and for each $x\in  \mathbb{D}_\epsilon (x_0)$, $g(x)$ is the unique solution of $f(x,g(x))=0$ with $g(x)\in \mathbb{D}_\delta (y_0)$. Moreover, $g(x)$ is analytic in $\mathbb{D}_\epsilon (x_0)$ and 
\begin{align}
g'(x) = - \frac{\frac{\partial f}{\partial x}(x,g(x))}{\frac{\partial f}{\partial y}(x,g(x))} \,.
\end{align}
}Similarly, one can compute higher derivatives of $g(x)$ and represent it by a Taylor series around $x=x_0$ convergent in $\mathbb{D}_\epsilon (x_0)$. Note that the statements of the theorem are {\it local}, e.g. the size of the domain $\mathbb{D}_\epsilon (x_0)$ is unspecified, it is only known that it exists for some $\epsilon>0$. In other words, we know that the radius of convergence of the series of $g(x)$ around $x=x_0$ is non-zero but its value is left unspecified. In the 
example above, we saw that the value of the radius of convergence is determined by the distance from the centre of the expansion $x_0$ to the nearest critical point of $f(x,y)$. Note also that the statements of the theorem depend crucially on $f(x,y)$ being an analytic function at $(x_0,y_0)$ (in particular, a finite order polynomial in $x$ and~$y$). 

Now we return to the original problem: for $f(x,y)=0$, where $x,y \in  \mathbb{C}$, find  explicit solution(s) in the form $y = y(x)$, at least locally in the vicinity of some point $(x_0,y_0)$, where $y(x)$ may be represented by series (possibly infinite) at $x=x_0$. For simplicity, we set $x_0=0$. First, we check $f(0,y)$. If this is a 
polynomial in $y$ of degree $n$, then the equation $f(0,y)=0$ has $n$ roots $y_i$, $i=1,\ldots,n$. There are  two possibilities.

\noindent
{\bf Local behaviour at regular points:} all the roots $y_i$, $i=1,\ldots,n$ of 
the equation   $f(0,y)=0$ are distinct.  Then $f_{,y}(0,y_i)\neq 0$, and the analytic implicit function theorem guarantees the existence of a unique Taylor expansion $y=y(x)$ at $x=0$.

\noindent
{\bf Local behaviour at critical points:}  the equation $f(0,y)=0$ has multiple roots. For simplicity, let $y_0=0$ be such a root. Then we have $f(0,0)=0$, $f_{,y}(0,0)=0$, $(\partial^2 \!f/\partial y^2)(0,0)=0,\ldots$ $(\partial^p\! f/\partial y^p)(0,0)\neq 0$, if $y=0$ is a zero of $f(0,y)=0$ of order~$p$. We expect $p$ branches of the solutions $y=Y_j(x)$, $j=1,\ldots,p$,  at $x=0$. They are given by Puiseux series
\begin{align}
y =Y_j (x)  = \sum\limits_{k\geq k_0}^{\infty} a_k x^{\frac{k}{m_j}}\,, \qquad j=1,\ldots, p\,,
\label{puiseux-1}
\end{align}
where $m_j$ are positive integers, and $k_0$ is a non-negative integer which in general depends on $j$. The exponents $k_0/m_j$, $(k_0+1)/m_j$, etc, and the coefficients $a_{k_0}, a_{k_0+1}$, etc, can be determined by the Newton polygon method (1671), as described e.g. in refs.~\cite{walker}, \cite{wall}. The Puiseux series are converging in some vicinity of the point $x=0$ provided $f(x,y)$ is an analytic function at $(x,y)=(0,0)$ (or a finite polynomial). If some $m_j>1$, we necessarily have among those $p$ branches a family of $m_j$ solutions of the form 
\begin{align}
y =Y_l (x)  = \sum\limits_{k\geq k_0}^{\infty} a_k \left( e^{\frac{2\pi i l}{m_j}}\right)^k \, x^{\frac{k}{m_j}}\,, \qquad l=0,1,\ldots, m_j-1\,.
\label{puiseux-2}
\end{align}

As an example, consider the  algebraic curve \cite{sendra}
\begin{align}
f(x,y)=y^5 - 4 y^4 + 4 y^3 + 2 x^2 y^2 - x y^2+ 2 x^2 y + 2 x y + x^4 + x^3 =0\,.
\label{puiseux-3}
\end{align}
Since $f(0,y)=y^3(y-2)^2$, the points $(0,0)$ and $(0,2)$ are critical points with multiplicities $3$ and $2$, respectively. We expect $y=y(x)$ to be given by 3 branches of Puiseux series at $(0,0)$ and by 2 branches at $(0,2)$. Applying the Newton polygon method \cite{walker}, \cite{wall}, \cite{sendra} at the point $(0,0)$, we find $m_1=m_2=2$ and $k_0=1$, $m_3=1$ and $k_0=2$, with appropriate coefficients:
\begin{align}
\,& y =Y_1 (x)  = i \frac{\sqrt{2}}{2} x^{\frac{1}{2}} - \frac{1}{8}\, x + i \frac{27 \sqrt{2}}{128} x^{\frac{3}{2}} - \frac{7}{32}\, x^2 + \cdots \,,\\
\,& y =Y_2 (x)  = - i \frac{\sqrt{2}}{2} x^{\frac{1}{2}} - \frac{1}{8}\, x - i \frac{27 \sqrt{2}}{128} x^{\frac{3}{2}} - \frac{7}{32}\, x^2 + \cdots \,, \\
\,& 
y =Y_3 (x)  =  - \frac{1}{2}\, x^2  +\frac{1}{8}\, x^4 - \frac{1}{8}\, x^5 + \frac{1}{16}\, x^6 + \cdots \,.
\label{puiseux-4}
\end{align}
At the point $(0,2)$, we have 2 branches, as expected:
\begin{align}
\,& y =Y_4 (x)  = 2 + \frac{1+i \sqrt{95}}{8}\, x + \cdots \,,\\
\,& y =Y_5 (x)  = 2 + \frac{1-i \sqrt{95}}{8}\, x + \cdots\,.
\label{puiseux-5}
\end{align}

\section{Perturbative solution of eq.~\eqref{shear-old-r}}
\label{app-cc}
Here, we list the explicit expressions for the components of the perturbative solution of eq.~\eqref{shear-old-r},
\begin{align}
G_0 (u) \,&= u\,, \nonumber \\
 G_1 (u) \,&=\left( \frac{\qfr^2}{2} - i \wfr\right) \left( 1- u\right) + i \wfr  \frac{u}{2} \ln{\frac{1+u}{2}} \nonumber \\
\,& = 
\left( \frac{\qfr^2}{2} - i \wfr\right) \left( 1- u\right) - i \wfr \, \frac{u}{2}\,  \mbox{Li}_1 \left(\frac{1-u}{2}\right)\,,  \nonumber \\
 G_2 (u) \,&= \wfr^2 \left[ \frac{u}{2} \, \mbox{Li}_2 \left( \frac{1-u}{2}\right) + 
 \frac{u}{8} \, \mbox{Li}_1^2 \left( \frac{1-u}{2}\right) +  \frac{1+u}{2}  \,  \mbox{Li}_1 \left( \frac{1-u}{2}\right)\right]
 \nonumber \\
 \,& +\qfr^2 \, \left(\qfr^2   - \frac{3  i \wfr }{2} \right) \frac{u}{2}\, \mbox{Li}_1 \left( \frac{1-u}{2}\right) - \frac{i \wfr \qfr^2}{4} \,  \mbox{Li}_1 \left( \frac{1-u}{2}\right) \nonumber \\
 \,& + \qfr^2 \left( \frac{\qfr^2 }{2} - i \wfr   \right) u \ln{u} + \frac{\qfr^4}{4} \left( 1- u \right)\,, \nonumber
\end{align}
as well as the appropriate boundary values, 
\begin{align}
G_0 (0) \,&= 0\,, \nonumber \\
G_1 (0) \,& = - i \wfr +\frac{\qfr^2}{2}\,, \nonumber \\
G_2 (0) \,& = \frac{\qfr^4}{4} - \frac{i \wfr \qfr^2 \ln{2}}{4} + \frac{\wfr^2 \ln{2}}{2}\,, \nonumber \\
G_3 (0) \,& = i \wfr^3 \left( \frac{\pi^2}{24} +\ln{2} - \frac{3}{8} \ln^2 2\right) +\qfr^6 \left( \frac{\ln{2}}{4} -\frac{1}{8}\right)  + i \wfr \qfr^4 
\left( \frac{1}{4}  - \frac{\ln 2}{8}\right) \nonumber \\
\,& +   \qfr^2 \wfr^2 \left( \frac{\pi^2}{48} -\frac{\ln{2}}{2}  - \frac{\ln^2 2}{16} \right)\,, \nonumber 
\\
G_4 (0) \,&= \qfr ^8 \left(-\frac{1}{16}+\frac{\pi^2}{64}-\frac{\ln{2}}{8}\right)-\qfr^4 \wfr^2 \left(\frac{\pi^2}{96} + \left(12 -7\ln{2}\right)\frac{\ln{8}}{96}\right)\nonumber \\ \,& -i \qfr^6 \wfr\left(\frac{\pi^2}{96} +\left(-5+\ln{4}\right)\frac{\ln{64}}{96}\right) \nonumber \\ \,& + \wfr^{4}\left(\left(24-5\ln{2}\right)\frac{\ln^{2}{2}}{48}+\frac{\pi^2}{48}\left(-4+\ln{8}\right)-\frac{1}{2}\zeta(3)\right) \nonumber \\ \,& +i \qfr^{2}\wfr^{3}\left(-\frac{\pi^{2}\ln{2}}{96}+\left(-24+\ln{2}\right)\frac{\ln^{2} {2}}{96}+\frac{3}{16}\zeta(3)\right)\,. \nonumber 
\end{align}

\section{Kepler's equation at complex eccentricity}
\label{kepler-app}
The connection between algebraic curves, their critical points  and non-analyticity of associated integrals has an interesting history \cite{arnold-nh}. Newton proved in ``Principia'' that every algebraically integrable oval must have singular points: all smooth ovals are algebraically non-integrable (hence the non-analyticity $T\propto a^{3/2}$ in the Kepler's third law). Moreover, the radius of convergence of the series solving Kepler's equation is determined by the critical points in the complex eccentricity plane.

\begin{figure}[t!]
\centering
\includegraphics[width=0.45\textwidth]{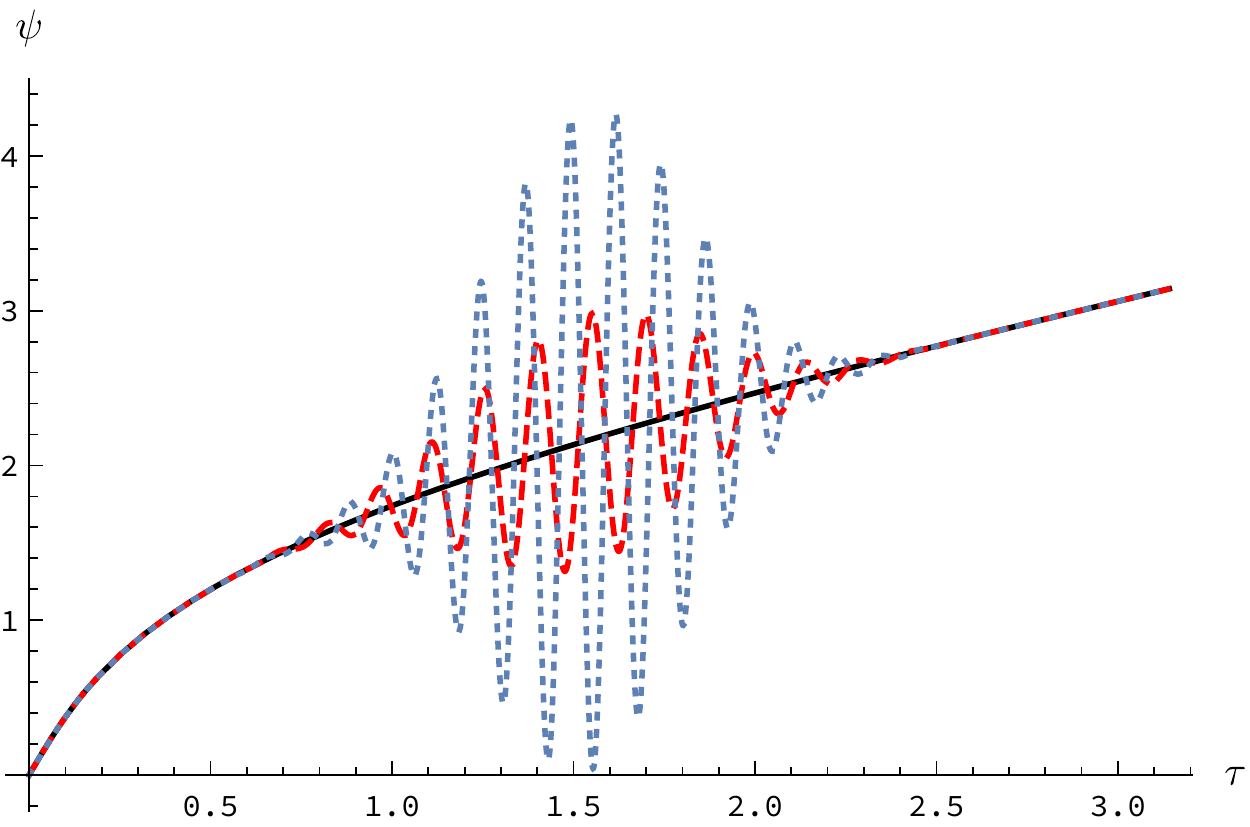}
\hspace{0.05\textwidth}
\includegraphics[width=0.45\textwidth]{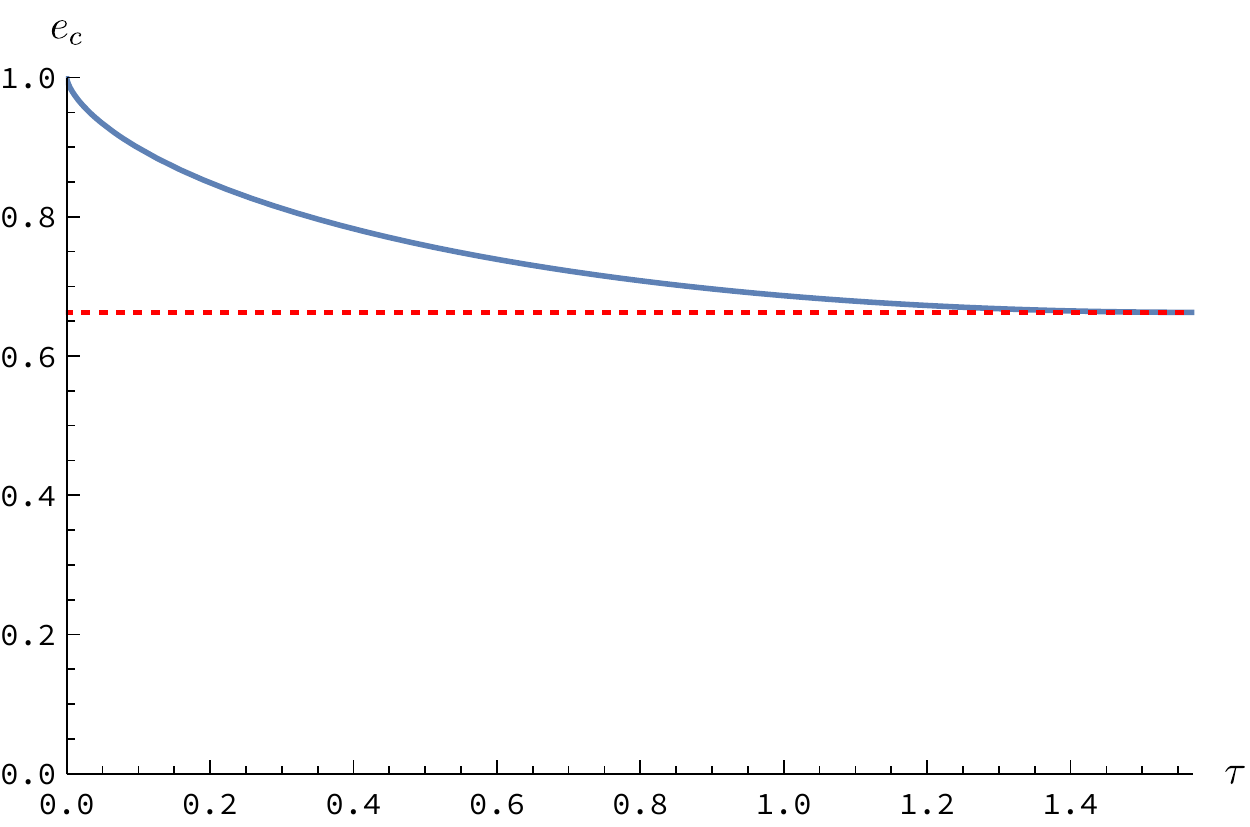}
\caption{{\small Left panel: Solutions $\psi (\tau)$ of Kepler's equation  \eqref{kepler-4}  at $e=0.75$ for $\tau \in [0,\pi]$ (half a period): ``exact'' numerical solution of \eqref{kepler-4} (solid black line), series solution \eqref{kepler-5} truncated at 50 terms (dashed red line) and 60 terms (dotted blue line). The rate of convergence is maximal at $\tau=\pi/2$. Right panel: The radius of convergence $e_c(\tau)$ of the series  \eqref{kepler-5} as a function of $\tau$ for $\tau \in [0,\pi/2]$. The red dotted line coresponds to Laplace's value $e_L$. For $0<e<e_L$, the series converges for all $\tau \in [0,2\pi]$.}
}
\label{fig:kepler-series}
\end{figure}
\begin{figure}[t!]
\centering
\includegraphics[width=0.75\textwidth]{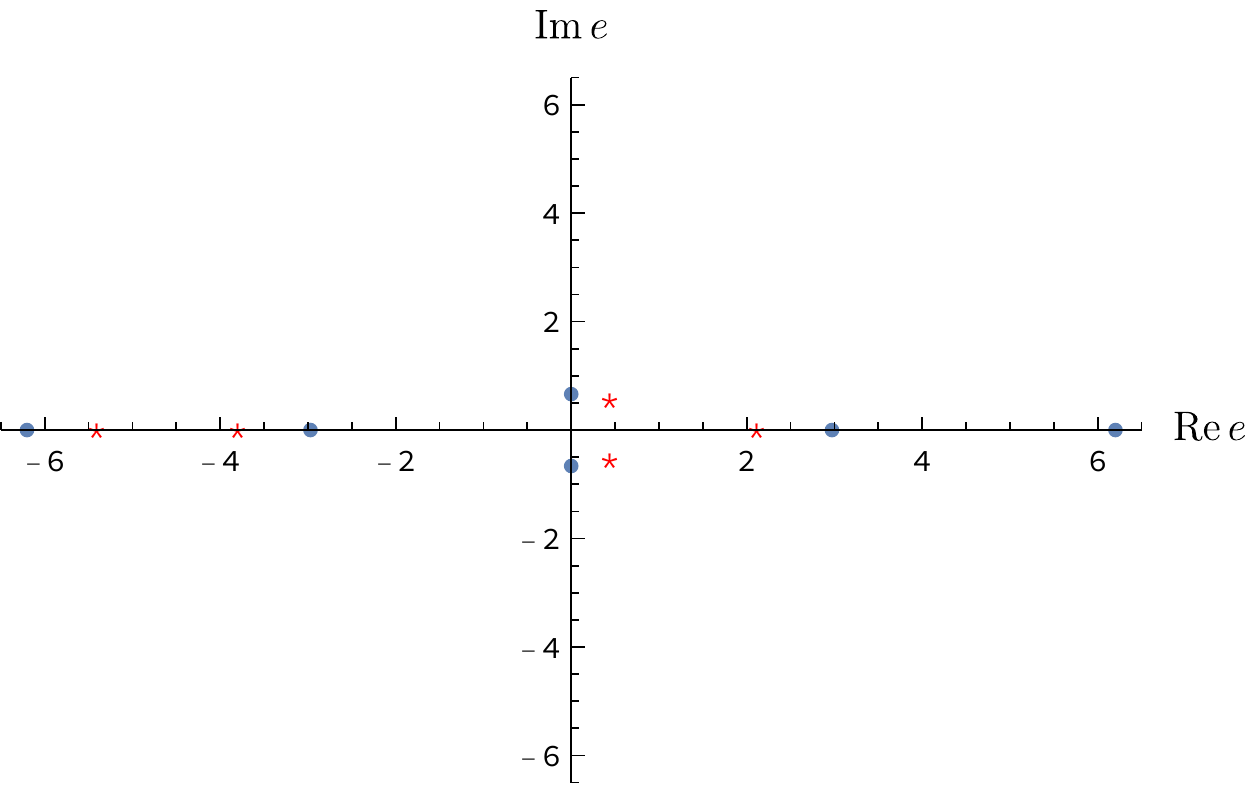}
\caption{{\small Critical points of the Kepler's curve  \eqref{kepler-4} in the complex eccentricity plane. The blue dots are the critical points at $\tau=\pi/2$. The points closest to the origin are located at $e\approx \pm 0.662743 i$. They determine the radius of convergence $e_L$ of the series  \eqref{kepler-5}. For $\tau < \pi/2$, the critical points are located at a larger distance from the origin. For example, the critical points at $\tau = \pi/4$ are shown by red asterisks. They determine the radius of convergence $e_c(\tau)$, shown in the right panel of fig.~\ref{fig:kepler-series}. With $\tau\rightarrow 0$, the three critical points merge at $e=1$.}}
\label{kepler-complex-e-plane}
\end{figure}

Kepler's law of motion of a planet in an elliptical orbit with eccentricity $e$, $0<e<1$, is usually expressed in a parametric form \cite{landau-mechanics}
\begin{align}
\,& r = a \left( 1 - e \cos{\psi}\right)\,, \label{kepler-1}\\
\,&  t = \frac{T}{2\pi} \left( \psi - e \sin{\psi}\right)\,,
\label{kepler-2}
\end{align}
where $r$ is the magnitude of the radius-vector from the centre of the force to the planet, $a$ is the major semi-axis of the ellipse, $T$ is the period of revolution and $\psi\in[0,2\pi]$ is the parameter known in astronomy as the {\it eccentric anomaly}. Knowing $\psi(t)$, one can find the position of the planet in the polar coordinates $(r(t),\varphi (t))$ as a function of time using 
Eq.~\eqref{kepler-1} and the equation
\begin{align}
\tan{\frac{\varphi}{2}} = \sqrt{\frac{1+e}{1-e}} \tan{\frac{\psi}{2}}\,.
 \label{kepler-3}
\end{align}
Introducing $\tau \equiv 2\pi t/T$, we rewrite Eq.~\eqref{kepler-2} as
\begin{align}
K = \tau  -  \psi + e \sin{\psi} =0\,.
\label{kepler-4}
\end{align}
Eq.~\eqref{kepler-4} is known as Kepler's equation. The task of finding a solution $\psi=\psi(\tau)$ preoccupied Kepler,
Newton, Lagrange, Laplace, Bessel, Cauchy and other great minds and led to progress in various mathematical disciplines. To quote V.I. Arnold \cite{arnold-nh}: {\it ``This equation plays an important role in the history of mathematics. From the time of Newton, the solution has been sought in the form of a series in powers of the eccentricity $e$. The series converges when $|e|\leq 0.662743...$. The investigation of the origin of this mysterious constant led Cauchy to the creation of complex analysis. Such fundamental mathematical concepts and results as Bessel functions, Fourier series, the topological index of a vector field, and the ``principle of the argument'' of the theory of functions of a complex variable also first appeared in the investigation of Kepler's equation''}.

A formal series solution of Kepler's equation was found by Lagrange \cite{Lagrange-1771} who apparently was not concerned with the series convergence (more details can be found in the book \cite{colwell})
\begin{align}
\psi (\tau,e) = \tau + \sum\limits_{n=1}^\infty\, a_n(\tau) \frac{e^n}{n!}\,,
\label{kepler-5}
\end{align}
where 
\begin{align}
a_n = \frac{d^{n-1}( \sin^n \tau )}{d \tau^{n-1}}\,.
\label{kepler-6}
\end{align}
As pointed out by Laplace \cite{Laplace-1823},  the series \eqref{kepler-5} converges for all $\tau \in [0,2\pi]$ as long as $|e|\leq e_L \approx 0.662743...$. For $e>e_L$, the series diverges for some values of $\tau$, in a rather peculiar manner (see fig.~\ref{fig:kepler-series}, left panel). 

What determines the radius of convergence $e_c(\tau)$ of the series  \eqref{kepler-5}? This problem was investigated by Cauchy, Puiseux and Serret in a series of papers in 1849-1859 \cite{colwell}. In modern language, the answer is the following. Treat Kepler's equation \eqref{kepler-4} as a complex analytic curve in the space of $\psi \in \mathbb{C}$, $e\in \mathbb{C}$, with $\tau$ remaining  a real parameter. The critical points of the curve $K=0$ 
obey  Eq.~\eqref{kepler-4} as well as the equation 
\begin{align}
\frac{\partial K}{\partial \psi} = e \cos{\psi} - 1 = 0\,.
\label{kepler-7}
\end{align}
The critical points closest to the origin in the complex eccentricity plane are shown in fig.~\ref{kepler-complex-e-plane}. Their location is parametrised by $\tau$. The radius of convergence $e_c(\tau)$ is given by the distance from the origin to the nearest singularity. This distance is a monotonic function of $\tau$ in the interval $[0,\pi]$ (half a period), with the minimum at $\tau=\pi/2$ given by $e_c(\frac{\pi}{2})=e_L$. Thus, for $0<e<e_L$, the series  \eqref{kepler-5} converges for all $\tau \in [0,2\pi]$. The dependence of the radius of convergence  on $\tau$ is shown in fig.~\ref{fig:kepler-series} (right panel).

\bibliographystyle{JHEP}
\bibliography{Genbib}{}

\end{document}